\documentclass[aps,prd,twocolumn,groupedaddress]{revtex4-2}

\usepackage{physics}
\usepackage{amsfonts}
\usepackage{amssymb}
\usepackage{graphicx}
\usepackage[english]{babel} 			
\usepackage[svgnames]{xcolor} 		
\usepackage{slashed}				
\usepackage{dsfont} 					
\usepackage{lipsum}
\usepackage{bm}
\usepackage{xfrac}
\usepackage{cancel}

\usepackage{hyperref}
\usepackage{cleveref}
\crefformat{figure}{fig.~(#2#1#3)}
\crefmultiformat{figure}{figs.~(#2#1#3)}%
       { and~(#2#1#3)}{, (#2#1#3)}{ and~(#2#1#3)}

\usepackage{feynmp}

\makeatletter
\def\endfmffile{%
    \fmfcmd{\p@rcent\space the end.^^J%
        end.^^J%
        endinput;
    }%
    \if@fmfio
    \immediate\closeout\@outfmf
    \fi
    \ifnum\pdfshellescape=\@ne
    \immediate\write18{mpost \thefmffile}%
    \fi}
\makeatother

\usepackage{gomezmacros}

\begin{document}


\title{Massive spinor helicity amplitudes, cross sections, and coalescence}


\author{Camille G\'omez-Laberge}
\email{c.gomez-laberge@northeastern.edu}
\homepage{https://gomezlaberge.org}
\affiliation{Department of Physics, Northeastern University, Boston, Massachusetts 02115, USA}


\date{\today}

\begin{abstract}
We examine recent advancements of the spinor helicity formalism of massive particles.
Technical aspects about the formulation of massive helicity spinors are presented in detail to analyze the projective-geometry kinematics of helicity spinors as well as the diagrammatical and analytical structure of their interactions.
Two new methods for calculating massive cross sections are derived and tested on Bhabha and Compton processes: a quasi-high-energy limit and an assembly of partial cross sections.
The acquisition of mass, where ultrarelativistic amplitudes coalesce at low energy, is given a physical interpretation as the localization of particle worldlines in twistor theory.
By subsuming the spinor helicity formalism in this way, both spacetime and particle content can emerge from null, lightlike, and timelike twistors.
\end{abstract}

\maketitle


\section{Introduction \label{sec_intro}}

Theoretical particle physics is undergoing a remarkable period of growth due to recent developments of the on-shell formulation of scattering amplitudes.
This alternative approach to scattering theory distinguishes itself from quantum field theory (QFT) already at the kinematical level by representing particles using helicity spinors,
objects that quantify momentum as a complex bi-spinor and transform under the little group without superfluous degrees of freedom.
Centering the \emph{particle} as the primary object of the theory, rather than as a subordinate embedding within the Lorentz 4-vector \emph{field}, has proven beneficial for allowing us to cut through much of the computational complexity of QFT.
The simplicity of the formalism entices us to study it and perhaps glean new insights into the nature of scattering and spacetime.

The last decade or so saw the first efforts to represent massive particles as helicity spinors \cite{Arkani-Hamed:2021a,Conde:2016vxs,Dittmaier:1998nn}.
Many applications have since been reported, for example, in the Standard Model \cite{Ochirov:2018,Alves:2022}, black hole physics \cite{Guevara:2019}, supersymmetry \cite{Herderschee:2019}, supergravity \cite{Chiodaroli:2023}, and twistor space \cite{Kim:2021,Albonico:2024,Kim:2025}.
Progress also continues in advancing our understanding the formalism itself and its utility, which are the twin foci of this paper.

We see a pedagogical opportunity to examine the recent description of massive particles in terms of helicity spinors and to further see how we can interpret these objects within the projective geometry of twistor theory.
It will be both practical and insightful to compare our calculations of scattering amplitudes and their cross sections to those obtained from QFT.
In doing so, we shall come to see each massive spinor helicity amplitude as the unique object that is consistent with all helicity configurations of a corresponding ultrarelativistic scattering process.
This \emph{coalescence} of high-energy configurations into a single massive amplitude seems to be of central importance for extending the spectrum of particles represented by the spinor helicity formalism, and we shall see its manifestations throughout this paper, which is organized as follows.

Section \ref{sec_mhs} begins by setting notation and examining the basic properties of the massive helicity spinor.
Scattering amplitudes and their on-shell diagrams are constructed in section \ref{sec_amp} with examinations of their transformation properties, singularity structures, and crossing symmetry relations.
New methodology for calculating cross sections are presented in section \ref{sec_cross}, and direct comparisons with QFT are made for Bhabha and Compton cross sections.
In section \ref{sec_mass}, we explore the physics of mass acquisition in terms of coalescence and worldline parametrization, given their importance for calculating massive amplitudes and cross sections.
Finally, section \ref{sec_twistor} pursues a more fundamental picture by subsuming the spinor helicity formalism within twistor theory, whose program also aims to derive representations for particles of any mass and spin---as well as the spacetime they occupy---from a more primitive object.

\section{Massive helicity spinors \label{sec_mhs}}

We begin by reviewing how massive particles are represented by spinor helicity variables.
The notation and conventions are those found in the textbook by Elvang \& Huang \cite{Elvang:2015} and the groundbreaking approach for massive helicity spinors by Arkani-Hamed, Huang, and Huang \cite{Arkani-Hamed:2021a}.
A student-friendly introduction to the spinor helicity formalism appears in the textbook by Schwartz \cite{Schwartz:2013} and in the TASI conference proceedings by Dixon \cite{Dixon:2014}.

\subsection{Bi-spinor representations with positive energy \label{sec_mhs_1}}

The on-shell program aims to construct amplitudes exclusively from the asymptotic state of each external particle $\ket{p,\sigma}$ with 4-momentum $p^\mu$ and spin state $\sigma$.
The spinor helicity formalism leverages the fact that 4-momenta transform according to the Lorentz group $\SO(1,3)$ which can be decomposed into two subalgebras representing the commuting generators of 3D rotations $J^+$ and $J^-$.
Therefore, $\so(1,3) = \su(2) \oplus \su(2)$, and $p^\mu$ has a bi-spinor representation of the Lorentz group $\left( \tfrac12, 0 \right) \oplus \left( 0, \tfrac12 \right)$ given by the direct sum of left- and right-handed helicity spinors $\la$ and $\lat$, respectively.
The spinors both transform under the particle's little group and must do so inversely with respect to the other so to ensure the overall invariance of $p^\mu$.
Hence, the Lorentz index $\mu$ is replaced by a pair of spinorial indices $\a$ and $\ad$, which identify the elements of the $2\times 2$ Pauli matrices $\sigma_\mu \equiv ( \id \;\; \vec\sigma)$ that provide passage between representations: $p^{\a\ad} = \sigma^{\a\ad}_\mu p^\mu$ or alternatively $p_{\ad\a} = \bar{\sigma}_{\ad \a}^\mu p_\mu$ with $\bar{\sigma}_{\ad \a}^\mu \equiv ( \id, \; -\vec\sigma)$. 

To see energy positivity between representations, first consider the equivalence between $\la \lat$ and $\sigma_\mu p^\mu$.
Following Schwartz \cite{Schwartz:2013}, massless particles with little group $\U(1)$ are represented by Weyl spinors $\la^\a$ and $\lat^\ad$ with complex-valued weight $z$
\ealin{
\la^\a &= \frac{z}{\sqrt{p_0 - p_3}} \mqty( p_0 - p_3 \\ -p_1 - \ii p_2), \nonumber \\
\lat^{\ad} &= \frac{z^{-1}}{\sqrt{p_0 - p_3}} \mqty( p_0 - p_3 & \ -p_1 + \ii p_2).
\label{eq_weyl}
}
Note the spinorial outer product $\la^\a \lat^\ad$ and the Lorentz contraction $\sigma^{\a\ad}_\mu p^\mu$ both yield
\ben
p^{\a\ad} = \mqty( p_0 - p_3 & -p_1 + \ii p_2 \\ -p_1 - \ii p_2 & p_0 + p_3).
\label{eq_bispinorupper}
\een
Writing our spinors in column form with conventions $\aket{\la} \equiv \la^\a \equiv \smqty(a \\ b)$ and $\sket{\la} \equiv \lat_\ad \equiv \smqty(\tilde a \\ \tilde b)$,
we move from $p^{\a\ad} \equiv \la^\a \lat^\ad$ to $p_{\ad\a} \equiv \lat_\ad \la_\a$ by lowering indices via contraction with the Levi--Civita symbol $\vepsi_{\a \b} = \vepsi_{\ad \bd} \equiv \smqty(0 & -1 \\ 1 & 0)$.
In fact, all invariants are obtained by the Levi--Civita symbol contractions because all $\SU(2)$ irreducible representations can be constructed from symmetric tensors with lower indices \cite{Georgi:1999wka}.
Proceeding accordingly, we find the representation $p_{\ad\a} \equiv \sketbraa{\la}$ by outer product $\lat_\ad \la_\a$ or by Lorentz contraction $\bar{\sigma}_{\ad \a}^\mu p_\mu$ to be
\ben
p_{\ad \a} = \mqty( p_0 + p_3 & p_1 - \ii p_2 \\ p_1 + \ii p_2 & p_0 - p_3).
\label{eq_bispinorlower}
\een
Note that the matrices of \cref{eq_bispinorupper,eq_bispinorlower} are singular and, thus, represent massless particles $p^2 = m^2 = 0$.
However, they are not quite negatives of each other because the energy $p_0$ \emph{does not} pick up a minus sign.
On the other hand, calculating $p_{\ad\a}$ using Lorentz index gymnastics
\ben
 p_{\ad\a} = p_\mu \bar{\sigma}_{\ad \a}^\mu = p_\mu ( g^{\mu \nu} \bar{\sigma}_{\nu \ad \a}) = p_\mu (-g^{\mu\nu} \sigma_\nu^{\a \ad}) = - p^{\a\ad},
\label{eq_gym_lorentz}
\een
as well as those for spinor indices
\ben
p_{\ad\a} = \left( -\id \cdot \vepsi ^2 \right) p_{\bd\b} = - \vepsi^{\a\b} \vepsi^{\ad\bd} p_{\bd\b} = - p^{\a\ad}
\label{eq_gym_spinor}
\een
both imply that $p_0$ \emph{does} pick up a minus sign.
The energy must remain positive, of course, so it is understood that it is remapped $E \equiv p_0 \mapsto -p_0$ during the index gymnastics of \cref{eq_gym_lorentz,eq_gym_spinor}.
See \cref{app_conventions} for more examples with detailed notation and conventions.

\subsection{Little group bases \label{sec_mhs_2}}

Let us turn our attention to on-shell massive particles of spin $s$ with energy $E^2 = p^2 + m^2$ in natural units ($c \equiv 1$). 
The little group is now $\SU(2)$, so a $(2s+1)$-dimensional irreducible representation as a tensor with $2s$ symmetric little group indices $(I_1 \cdots I_{2s})$ is guaranteed.

The simplest non-trivial case is massive spin $\tfrac12$, which requires only one $\SU(2)$ index.
Following \cite{Arkani-Hamed:2021a}, the massive helicity spinors are written as 
\ealin{
\aket{\bm\la} &\equiv \la^\a_I = \la^\a \zan_I + \eta^\a \zap_I, \nonumber \\
\sket{\bm\la} &\equiv \lat_{\ad I} = \lat_\ad \zap_I + \etat_\ad \zan_I.
\label{eq_mhsdef}
}
Writing the energy--momentum relation in factored form $m = \sqrt{E+p_z}\sqrt{E-p_z}$, both the energy and the spin in the arbitrary frame $p^\mu$ can be encoded as Weyl spinors
\ealin{
\la^\a &\equiv \frac{1}{\sqrt[4] 2} \sqrt{E + p_z} \, \zap[\a], 
	& \eta^\a \equiv \frac{1}{\sqrt[4] 2} \sqrt{E - p_z} \, \zan[\a],
\label{eq_helicity} \\
\lat_\ad &\equiv \frac{1}{\sqrt[4] 2} \sqrt{E + p_z} \, \zatn_\ad, 
	& \etat_\ad \equiv \frac{1}{\sqrt[4] 2} \sqrt{E - p_z} \, \zatp_\ad. 
\label{eq_helicitytilde}
}
In the high-energy limit, when $p_z \rightarrow E$, we see that $\la$ and $\lat$ grow into the ultraviolent as $\sqrt{2E}$ while $\eta$ and $\etat$ vanish into the infrared.
This factorization will feature prominently and is crucial for the coalescence of massive particle representations.

In addition to the usual left- and right-handed spinor bases $\{ \zap[\a], \zan[\a] \}$ and $\{ \zatn_\ad, \zatp_\ad \}$, we now also require a \emph{little group basis} $\{ \zan_I, \zap_I\}$ so that $\la^\a_I$ and $\lat_{\ad I}$ inherit the properties of $\la^\a$ and $\lat_\ad$ but with non-zero mass.
In frame $p^\mu$, recall that $\zap[\a] = \smqty( c \\ s )$ and $\zan[\a] = \smqty( -s^* \\ c )$ have been rotated by $c \equiv \cos \tfrac{\theta}{2}$ and $s \equiv e^{\ii \phi} \sin \tfrac{\theta}{2}$, and the right basis is identified with the conjugate of the left basis $\zatn_\ad \equiv (\zap[\a])^* = \smqty( c \\ s^*)$ and $\zatp_\ad \equiv (\zan[\a])^* = \smqty( -s \\ c)$.
A third independent basis for the little group can be defined without loss of generality as co-spinors of $\{ \zap[\a], \zan[\a] \}$ in the reference frame
where $\theta = \phi = 0$:
\ealin{
\zap_\a = \vepsi_{\a\b} \zap[\b] = \mqty(-s & c) &\longrightarrow \zap_I \equiv \mqty(0 & 1), \\
\zan_\a = \vepsi_{\a\b} \zan[\b] = \mqty(-c & -s^*) &\longrightarrow \zan_I \equiv \mqty(-1 & 0).
}
All three bases describe their respective spinorial spaces on the same footing. 
They all have the same orthonormality conditions $\abraket{\zeta^\pm \zeta^\mp} = \pm 1$ and $\abraket{\zeta^\pm \zeta^\pm} = 0$.
For example, $\zap_\a \zan[\a] = \zap_I \zan[I] = \abraket{\zap \zan} = 1$ and $\zatp_\ad \zatn[\ad] = \abraket{\zatp \zatn} = 1$.

Massive particles of general spin $s$ are represented as a $\smqty[2s \\ 2s]$-tensor, constructed as the product of $2s$ copies of $\la^\a_I$ and $\lat^\ad_I$ in total. 
We could express this as
\ben
\bm \la^{2s} \equiv \la^{\a_1\cdots}_{(I_1 \cdots} \lat^{\cdots \ad_{2s}}_{\cdots I_{2s})},
\label{eq_higherspin}
\een
where symmetrized indices (enclosed in round brackets) run through all copies of $\la$ and $\lat$.
Thankfully, the heavily indexed notation on the right hand side is unnecessary in practice because no ambiguity remains provided that each massive spin $s$ particle appears in each term of the amplitude exactly $2s$ times.
How these spinors are contracted will depend on the asymptotic state of the external particles.
We shall focus on the spin $\tfrac12$ processes, but many higher spin amplitudes appear in the literature.

The bi-spinor momentum of the massive particle is
\ealin{
\bm p^{\a\ad} &\equiv \aketbras{\bm \la} =  \la^\a_I \lat^{\ad I} \nonumber \\
\bm p_{\ad\a} &\equiv \sketbraa{\bm \la} = \lat_{\ad I} \la_\a^I
\label{eq_massivemomentum}
}
for any index $I \in \{I_1,\ldots, I_{2s} \}$.
Note, the momentum and spinor helicity variables corresponding to massive particles will always appear in boldface to distinguish them from massless particles.
Like $\a$ and $\ad$, the little group indices are raised/lowered by the Levi--Civita symbol $\vepsi_{IJ} \equiv \smqty(0 & -1 \\ 1 & 0)$, and the corresponding matrices must now be full-rank to span the entire symmetry group $\SU(2) \otimes \SU(2) = \SLtwoC$, the universal cover of the proper Lorentz group $\SO(1,3)$, such that $p^2 = \bm p_\mu \bm p^\mu = -\tfrac12 \bm p^{\a\ad} \bm p_{\ad\a} = m^2$.
Let us now check that this requirement holds.
Since $\bm p^{\a\ad} \bm p_{\ad\a} = \aketbras{\bm \la} \cdot \sketbraa{\bm \la} 
	= \abraket{\bm{\la \la}} \sbraket{ \bm{\la \la}}\,$, expand the angle brackets using the little group basis
\eali{
\abraket{\bm{\la \la}} 
	&= \Big( \abra{\la} \zan[I] + \abra{\eta} \zap[I] \Big) 
		\cdot \Big( \aket{\la} \zan_I + \aket{\eta} \zap_I \Big) \\
	&\hspace{-3mm}= \abraket{\la \la} \zan_I \zan[I] + \abraket{\eta \eta} \zap_I \zap[I] 
		+ \abraket{\la \eta} \zap_I \zan[I] + \abraket{\eta \la} \zan_I \zap[I] \\
	&\hspace{-3mm}= \abraket{\la \eta} - \abraket{\eta \la} = 2 \abraket{\la \eta}.
}
Applying the definitions in \cref{eq_helicity,eq_helicitytilde} yield
\ealin{
2 \abraket{\la \eta}
	&= 2 \left( \frac{1}{\sqrt[4]{2}} \sqrt{E + p_z} \zap_\a \right) 
		\cdot \left( \frac{1}{\sqrt[4]{2}} \sqrt{E - p_z} \zan[\a] \right) \nonumber \\
	&= \frac{2}{\sqrt{2}} \sqrt{E^2 - p_z^2} \abraket{\zap \zan} \nonumber \\
	&= \sqrt 2 m.
\label{eq_anglemass}
}
Similarly, for the square brackets, we obtain
\ealin{
\sbraket{\bm{\la \la}}
	&= \Big( \sbra{\la} \zap[I] + \sbra{\eta} \zan[I] \Big) 
		\cdot \Big( \sket{\la} \zap_I + \sket{\eta} \zan_I \Big) \nonumber \\
	&= \frac{2}{\sqrt 2} \sqrt{E^2 - p_z^2} \abraket{\zatn \zatp}  \nonumber \\
	&= - \sqrt 2 m,
\label{eq_squaremass}
}
showing indeed that $\bm p^{\a\ad} \bm p_{\ad\a} = -2 m^2$.

We shall soon generalize the kinematics to complex-valued momentum, where the little group weight $z$ is no longer restricted to a pure phase $e^{\ii \phi}$ and where the universal cover becomes the product group $\SLtwoC_\Lcal \otimes \SLtwoC_\Rcal$.
Formulating helicity spinors for complex momenta has remarkable consequences for our work ahead: i) scattering amplitudes become analytic (holomorphic) functions whose structure can be studied using the powerful results of residue theory, ii) the three-point scattering amplitude no longer vanishes and can thus serve as the fundamental building block for all amplitudes, and iii) the bi-spinor representation of particles, their interactions, and even spacetime itself can all be encoded onto the projective geometry of twistor theory.

\subsection{Projective geometry of massive helicity spinors \label{sec_mhs_3}}

Before moving on to dynamics, we conclude this section by showing how helicity spinors encode mass and spin.
Let us then try to see why their distinct properties might confer a simpler reformulation of scattering theory.

From the get-go, helicity spinors transform correctly under the little group and thus possess the right structure and degrees of freedom to represent particles.
The complex number $z$ encodes little group transformations and appropriately cancels itself out in the bi-spinor representation of momentum in \cref{eq_bispinorupper,eq_bispinorlower}.
Thus, when constructed from helicity spinors, scattering amplitudes are Lorentz invariant objects upon arrival, as they should be.
This is, of course, not the case in QFT.
Amplitudes written in terms of $4$-momenta require the so-called polarization ``vectors'' (though they are not Lorentz covariant!) to introduce redundancies between vector components in order to eliminate the superfluous degrees of freedom of the vector and tensor fields.
Unavoidable in field theory, this artifice is responsible for the familiar complications of having to sum terms obtained from all possible Feynman diagrams.
Cross sections lead to further overhead when expanding the square of the magnitude of this sum.

The spinor helicity story is consistent for particles of any mass and spin.
Mass acquisition categorically changes particle kinematics by providing additional spin degrees of freedom and by peeling the worldline off of the lightcone surface and into its timelike interior. 
For massive spin $\tfrac12$, both changes can be understood as consequences of \cref{eq_mhsdef}, which shows how a massive spinor can be represented as a linear combination of massless ones.
There is no difference for general spin $s$.
The massive spinor will appear in each term of the amplitude $2s$ times to represent a completely symmetric $\SU(2)$ tensor with $2s$ little group indices, as in \cref{eq_higherspin}.

Lastly, new insights into the fundamental laws of scattering and spacetime may follow from a spinor-helicity reformulation of particle interactions.
The complex bispinors of \cref{eq_bispinorupper,eq_bispinorlower} have a split-signature $(2,2)$, e.g., $\det p_{\ad\a} = p_0^2 - p_1^2 + p_2^2 - p_3^2$, and their bi-spinor representation transforms under $\SLtwoC_\Lcal \otimes \SLtwoC_\Rcal$ as $p^{\a\ad} \rightarrow L^\a_\b \la^\b \lat^\bd R^\ad_\bd$.
In contrast to the Lorentz signature $(1,3)$ with $4$-vectors $p^\mu$ transforming under $\SO(1,3)$, this splits the particle representation into two factors $\la^\a \lat^\ad$, and each helicity spinor can thus be thought of as a kind of square root of $p^\mu$.
That the twistors of Penrose \cite{Penrose:1967} have similar split-signatures, transformations, and geometric properties \cite{Penrose:1986}, is what leads us in \cref{sec_twistor} to explore the possibility of subsuming helicity spinors into twistor theory, where spacetime can also emerge from null, lightlike, and timelike twistors in an abstract projective geometry.
For now, we shall merely keep the above perspective in mind as we re-examine the analytical structure of scattering amplitudes and cross sections next.

\section{Scattering amplitudes \label{sec_amp}}

With the kinematic structure of the spinor helicity variables in place, we now turn to dynamics.
Since bispinor momenta in \cref{eq_bispinorupper,eq_bispinorlower,eq_massivemomentum} are always constrained by the energy-momentum relation, 
on-shell diagrams become the natural choice for encoding scattering amplitudes written in this formalism.
The transformation properties, singularity structure, and crossing symmetries of the simplest on-shell subdiagrams will be examined to develop the analytical tools needed for our work ahead.
To demonstrate and validate the formalism, we end by deriving amplitudes for Bhabha annihilation and Compton scattering, both elementary and ubiquitous in nature.

\subsection{On-shell diagrams are multi-channel amplitudes \label{sec_amp_1}}

Formulating scattering amplitudes using helicity spinors in a complexified spacetime eliminates the need for virtual particles, since the fundamental three-particle interactions underlying all scattering processes do not generically vanish for complex momenta.
The benefits that follow from this first change are remarkable.
The three-particle interaction, or \emph{coupling} for brevity, is fully constrained by Poincar\'e symmetry; it is the non-perturbative object containing everything we need to know about the particles and how they interact.
And so, constructing an on-shell diagram by gluing together couplings, first, implies that all internal lines are on-shell and, second, that the resulting amplitude is guaranteed to be a unitary and well-defined analytic function of the asymptotic particle states.

For a unitary theory, the spectral decomposition of quantum fields guarantees that each pole of the $S$-matrix corresponds to a particular internal line going on shell.
All internal lines in the spinor helicity formalism are on-shell; therefore, a tree-like on-shell diagram is singular and fully constrained by momentum- and energy-conservation.
In other words, a tree-like on-shell diagram represents a factorization channel, not an amplitude.
The on-shell diagram corresponding to a scattering amplitude, i.e., to the analytic function of \emph{generic} asymptotic particle states, requires at least one degree of freedom, which can be conferred by adding an internal bridge to any of the tree-like factorization channel diagrams.
In this way, the Feynman diagrams for the usual $s$, $t$, and $u$ channels are simultaneously embedded into a \emph{single} on-shell diagram representing the entire scattering process \cite{Arkani-Hamed:2016}.
To illustrate this, consider Compton scattering $e^- \g \rightarrow \g e^-$.

\begin{widetext}
\begin{fmffile}{compton}
\begin{equation}
\begin{gathered}
\begin{fmfgraph*}(100, 80)
        \fmfleft{i1,i2}
        \fmfright{o1,o2}
        
        \fmf{fermion, label=$p_1$, label.side=left}{i1,v1}
        \fmf{photon, label=$p_2$}{i2,v1}
        \fmf{fermion, label=$p_4$, label.side=left}{v2,o1}
        \fmf{photon, label=$p_3$}{v2,o2}
        \fmf{fermion}{v1,v2}

\end{fmfgraph*}
\end{gathered}
\quad
+
\quad
\begin{gathered}
\begin{fmfgraph*}(100, 80)
        \fmfleft{i1,i2}
        \fmfright{o1,o2}
        
        \fmf{fermion, label=$p_1$, label.side=left}{i1,v1}
        \fmf{phantom}{v1,i2}
        \fmf{fermion, label=$p_4$, label.side=left}{v2,o1}
        \fmf{phantom}{v2,o2}        
        \fmf{photon, rubout, tension=0}{i2,v2}
        \fmf{photon, tension=0}{v1,o2}
        \fmf{fermion}{v1,v2}
        
        \fmfv{label=$p_2$,label.angle=158,label.dist=20mm}{v2}
        \fmfv{label=$p_3$,label.angle=22,label.dist=20mm}{v1}
            
\end{fmfgraph*}
\end{gathered}
\quad\quad \longleftrightarrow \quad\quad
\begin{gathered}
\begin{fmfgraph*}(100, 80)
	\fmfpen{thick}
        \fmfleft{i1,i2}
        \fmfright{o1,o2}

        \fmf{fermion, tension=1.5}{i1,v1}
        \fmf{fermion}{v1,v2,v3,v4}
        \fmf{fermion, tension=1.5}{v4,o1}
        \fmf{photon, tension=1.5}{i2,v2}
        \fmf{photon, tension=1.5}{v3,o2}
        \fmf{photon}{v1,v4}

        \fmflabel{$\bm{1}^{\sfrac{1}{2}}$}{i1}
        \fmflabel{$2^{-1}$}{i2}
        \fmflabel{$\bm{4}^{\sfrac{1}{2}}$}{o1}
        \fmflabel{$3^{+1}$}{o2}        

        \fmfv{decor.shape=circle, decor.filled=empty, decor.size=4mm}{v1}
        \fmfv{decor.shape=circle, decor.filled=full, decor.size=4mm}{v2}
        \fmfv{decor.shape=circle, decor.filled=empty, decor.size=4mm}{v3}
        \fmfv{decor.shape=circle, decor.filled=full, decor.size=4mm}{v4}

\end{fmfgraph*}
\end{gathered}
\label{fig_compton}
\end{equation}
\end{fmffile}
\end{widetext}

The left-hand side shows the two Feynman diagrams, representing the $s$ and $u$ channels that prescribe how to formulate the amplitude in QFT.
On the right-hand side is the equivalent on-shell diagram drawn with thicker lines and colored vertices to distinguish it from Feynman diagrams.
Perhaps the most important difference between these formalisms is a moral one.
This single on-shell diagram represents the entirety of the scattering process at leading order or ``tree level.''
Indeed, the amplitude derived from the on-shell diagram is a monolithic and physical object possessing the correct transformation properties and nothing else.
In contrast, while the terms coming from individual Feynman diagrams do yield the correct amplitude \emph{as a sum}, they are unfortunately meaningless \emph{on their own} due to their lack of gauge invariance.
A secondary difference is that the on-shell diagram looks more loop-like than tree-like, a consequence of the bridge that forms the internal square in the diagram.
The bridge introduces the needed degrees of freedom that turn the internal (on-shell) lines into variables, of course, still constrained by the external states or ``scattering data.''
The couplings of the amplitude must have opposing helicity, and this information is provided by the diagram's colored vertices.
White corresponds to a mostly positive helicity coupling where all $\la$ variables are parallel, and black corresponds to a mostly negative helicity one with parallel $\lat$ variables.
Inspection of the diagram reveals the embedded factorization channels.
For example, we obtain the $s$ channel diagram by letting the internal line through the bottom edge of the box vanish, thus also eliminating two lower vertices.
The left and right edges of the box become lines $\bm1$ and $\bm4$, respectively. 
The $u$ channel diagram is harder to see.
Let the right edge of the box vanish and thus also the two right-side vertices.
The bottom (massless) edge of the box takes up the role of line $3$, while the top (massive) edge of the box becomes line $\bm4$.
(To recover the typical $u$ channel diagram shape, imagine dragging the remaining black vertex downward to be horizontal with the remaining white vertex and adjust the new external legs.)
An equivalent $u$ channel can be obtained by instead deleting the left edge of the box.
There are no other valid factorizations.
If we try to delete the top edge of the box, then we cannot recover the two external photons; therefore, this diagram is excluded.
Finally, note that the external particles of the on-shell diagram are not labeled by their momentum but rather by their mass and spin.
Massive particles appear in boldface with a superscript indicating their spin, while massless ones appear in regular type with a helicity superscript.
For example, $\bm{1}^{\sfrac{1}{2}}$ is a massive spin $\tfrac12$ particle, and $2^{-1}$ is a massless spin $1$ particle with negative helicity.

Unitary demands the consistent factorization of scattering amplitudes written as analytic functions of on-shell particles.
The above point makes no distinction about mass and thus applies to all particles.
Finding a generalization of the BCFW recursion algorithm \cite{Britto:2005} to efficiently calculate amplitudes for all masses and spins is an active area research \cite{Wu:2022,Christensen:2022nja}.
Nevertheless, the analytic structure of the amplitude, especially its residues, will depend on the particle spectrum in interesting ways, as we shall see next.

\subsection{Amplitude transformation and factorization \label{sec_amp_2}}

Scattering amplitudes $\Mcal \equiv \mel{f}{\Mcal}{i}$ are Lorentz invariant objects that transform under the little group.
With Feynman rules, this property is hidden behind the contraction of each quantum field and its polarization vector, e.g., $\epsilon^\mu A_\mu$.
In contrast, when constructing amplitudes with helicity spinors, the little group is made manifest by the presence of little group indices $I$ for massive particles and helicity indices $h$ for massless ones.

In the massless case, the little group is $\mathrm{GL}(1)$ and is represented by the complex weight $z$ in \cref{eq_weyl}, since rescalings $\la^\a \rightarrow z \la^\a$ and $\lat^\ad \rightarrow z^{-1} \lat^\ad$ leave the momentum $\la^\a \lat^\ad$ unchanged.
As mentioned above, Poincar\'e symmetry fully determines the massless coupling, and thus any massless scattering amplitude can be constructed by gluing together couplings.
Whatever the amplitude, it must scale by $W^h = z^{-2h}$ under a little group transformation of each external particle with helicity $h$ \cite{Elvang:2015}.
Famously, the Parke--Taylor formula
\ben
\Mcal(1^+ 2^+ \cdots j^- \cdots k^- \cdots n^+) 
= \frac{\abraket{jk}^4}{\abraket{12} \abraket{23} \cdots \abraket{n1}}
\label{eq_parketaylor}
\een
owes its elegance to this fact, which is clearly seen by counting the powers of any of the $n$ gluons \cite{Schwartz:2013}.
The negative helicity momenta $j$ and $k$ each have power $2$, while the remaining postivity helicity momenta each have power $-2$.
Therefore, any scattering amplitude involving a massless external particle with momentum $\la^\a \lat^\ad$ and helicity $h$ transforms under that particle's little group as 
\ben
\Mcal(\la^\a, \lat^\ad) \longrightarrow z^{-2h} \Mcal(\la^\a, \lat^\ad).
\label{eq_transformhelicity}
\een
In the massive case, the little group is $\SU(2)$, and an irreducible representation for spin $s$ can always be constructed as a tensor $W^{(I_1\cdots I_{2s})}$ with totally symmetric indices.
Therefore, any amplitude involving a massive external particle of momentum $\la^\a \lat^\ad$ and spin $s$ transforms under that particle's little group as
\ben
\Mcal(\la^\a, \lat^\ad) \longrightarrow W^{(I_1\cdots I_{2s})} \Mcal(\la^\a, \lat^\ad).
\label{eq_transformlittlegroup}
\een
We pause to note the simplicity that $W^{(I_1\cdots I_{2s})}$ offers compared to the standard notation in quantum mechanics \cite{Cohen-Tannoudji:1991b}.
The above suggests that there is no need to choose an arbitrary axis $z$ and diagonalize the spin angular momentum $S_z$ to form the complete set of commuting observables $S^2 \ket{s,m} = s(s+1) \ket{s,m}$ and $S_z \ket{s,m} = m \ket{s,m}$.
Instead, the electron's spin state can be simply described by a tensor $W^I$ without indicating a privileged direction.
For example, the quantum electrodynamics (QED) coupling in the on-shell diagram of \cref{fig_compton} is $\Mcal^h_{I J}$.
This coupling is manifestly Lorentz-invariant and transforms most generally under the little group as $\Mcal(p_1, p_2, p_3) \longrightarrow z^{-2h} W_I W_J \Mcal(p_1, p_2, p_3)$.

With this notation in place, we see how unitarity guarantees that a four-particle amplitude must factorize into two couplings, usually denoted by left $\Lcal$ and right $\Rcal$, along a massive internal line by Levi--Civita contraction 
\ben
\Mcal \longrightarrow \frac{ \Lcal^h_{I J} \, \vepsi^{JK} \,
	\Rcal^{-h}_{K L} } 
	{\xi-m^2} \qq{[massive internal line]}
	\label{eq_factorizemassive}
\een
or along a massless line by summing over the internal helicity states
\ben
\Mcal \longrightarrow \sum_{\pm h} \frac{ \Lcal^h_{I J } \,\,
	\Rcal^{-h}_{K L} } 
	{\xi} \qq{[massless internal line]}
	\label{eq_factorizemassless}
\een
with $\xi$ representing the pole where $\Mcal(\xi)$ becomes singular with residue
\ealin{
R_\xi \equiv \Res(\Mcal, m^2) 
	&= \lim_{\xi \rightarrow m^2} \left( \xi - m^2 \right) \Mcal(\xi) \nonumber \\
	&= \begin{cases}
        \Lcal^h_{I J} \, \vepsi^{JK} \, \Rcal^{-h}_{K L}  & \text{for } m > 0, \vspace{1mm} \\
        \sum_{\pm h} \Lcal^h_{I J } \,\, \Rcal^{-h}_{K L}  & \text{for } m = 0.
        \end{cases}
}
The cluster decomposition principle requires that all possible factorizations occur in this way \cite{Weinberg:1995}.

\subsection{Analytical structure of couplings \label{sec_amp_3}}

Let us now calculate amplitudes by factorizing them along internal lines according to \cref{eq_factorizemassive,eq_factorizemassless}.
The massless coupling is singular, and the residue for one factorization channel will have a pole from \emph{another} channel \cite{Benincasa:2008}.
Its amplitude is a non-perturbative quantity, completely fixed by helicity constraints
\ben
\Mcal^{h_1 h_2 h_3} =
\begin{cases}
g \abraket{12}^{h_3 - h_1 - h_2} \abraket{23}^{h_1 - h_2 - h_3} \abraket{31}^{h_2 - h_3 - h_1} \vspace{1mm} \\
\tilde{g} \sbraket{12}^{h_1 + h_2 - h_3} \sbraket{23}^{h_2 + h_3 - h_1} \sbraket{31}^{h_3 + h_1 - h_2} 
\end{cases}
\label{eq_massless3}
\een
with dimensionless coupling constants $g$ and $\tilde{g}$ and where the case with angle/square brackets applies when the sign of $h_1 + h_2 + h_3$ is negative/positive.
Referring back to \cref{fig_compton}, black vertices correspond to the coupling with mostly negative helicity in terms of angle brackets $\abraket{\cdot}$ and white vertices to the mostly positive helicity coupling in terms of square brackets $\sbraket{\cdot}$.
To see the singularity structure of a massless scattering amplitude, let us calculate the $s$-channel residue of the massless spin 1 amplitude $1^- 2^+ 3^- 4^+$ according to \cref{eq_massless3}.
We obtain
\ealin{
R_s &= \Lcal^{-+-} \Rcal^{+-+}
	= \left(g \frac{\abraket{P1}^3}{\abraket{12} \abraket{2P}} \right) \left(\tilde{g} \frac{\sbraket{4P}^3}{\sbraket{34} \sbraket{P3}} \right) \nonumber \\
	&= g \tilde{g} \frac{\abraket{1P}^3 \sbraket{P4}^3}
	{\abraket{12} \sbraket{34} \abraket{2P} \sbraket{P3}}
	= g \tilde{g} \frac{\abraket{13}^2 \sbraket{24}^2}{u},
\label{eq_massless3nonlocal}
}
where we used $P^\mu = p_1^\mu + p_2^\mu = p_3^\mu + p_4^\mu$ in the last step.
We find that the residue of the $s$-channel itself has a $u$-channel pole. 
Therefore, the amplitude factorized in this way has in fact \emph{two} poles: $su$.
This is a violation of the cluster decomposition principle, which guarantees that a \emph{local} theory cannot have more than a single pole or branch cut.
Although spinor helicity amplitudes are manifestly unitary, we see that they cannot always be formulated in a local way.
This is the first of multiple indications that the spinor helicity formalism exposes a fundamental tension between quantum mechanics and spacetime.
The $u$-channel pole arises because the massless couplings $\Lcal^{-+-}$ and $\Rcal^{+-+}$ are themselves singular when either of the brackets in the denominator vanish.
Indeed, any scattering amplitude that can be factorized with a massless coupling will be non-local.

Despite its singularity structure, the massless coupling is remarkably simple.
Recall \cref{sec_mhs_3}, where we noted that complex momenta transform covariantly under the product group $\SLtwoC_\Lcal \otimes \SLtwoC_\Rcal$, with a left subgroup $\Lcal$ spanned by an undotted basis $\{\la_\a\}$ and a right subgroup $\Rcal$ spanned by a dotted basis $\{\lat_\ad\}$.
Yet \cref{eq_massless3} prescribes a single tensor structure for massless couplings that always transforms \emph{either} entirely within the $\Lcal$ subgroup if $\sum_i h_i <0$ \emph{or} entirely within the $\Rcal$ subgroup if $\sum_i h_i >0$.
Massive couplings, on the other hand, are more complicated because they generically transform in \emph{both} $\Lcal$ and $\Rcal$, and they always do so with multiple independent tensor structures.
This will constrain the structure of the massive couplings and the configuration of scattering amplitudes constructed from them.
In contrast to QFT, we can appreciate that the spinor helicity formalism becomes simpler at higher energies.
When taken to high energy, a massive scattering amplitude breaks up into a simpler massless amplitude for each admissible \emph{spin configuration} of the external particles' spin and helicity values.
Coalescence is the reverse of this process, where ultrarelativistic configurations to form a single massive amplitude at low energies.

Now we shall see that \emph{massive} couplings are also singular in certain cases when couplings consist of two massive particles, each with a set of little group indices $\{I_1, \ldots, I_{2s}\}$, and one massless particle with helicity variable $h$.
Importantly, QED is one such case with $s=\tfrac12$.
The issue is more delicate than the all-massless case. 
There is no analogue of \cref{eq_massless3} because helicity is not conserved for massive particles.
Taking a constructive approach, we expose the crux of the matter---a degeneracy in spanning the Lorentz cover $\SLtwoC$---by expressing any massive amplitude as a contraction of tensors: one transforming under the little group and the other transforming explicitly under the Lorentz group.
For example, the QED coupling, having little group indices $I$ and $J$, and helicity $h$, can be written as the contraction
\ben
\Mcal_{IJ}^h = \frac{1}{\sqrt2 m} \la_{1I}^\a \la_{2J}^\b M_{\a\b}^h,
\label{eq_qed3}
\een
with the $\sqrt2 m$ appearing for dimensional consistency according to \cref{eq_anglemass,eq_squaremass}.
We identify the first piece $W_{IJ}^{\a\b} = \la_{1I}^\a \la_{2J}^\b$ as a natural choice for a rank-$2$ tensor transforming under the $\SU(2)$ little group of the \emph{massive} helicity spinors.
The remaining piece $M_{\a\b}^h$ is the coupling with little group indices stripped off, thus transforming only under the Lorentz group with indices $\a$ and $\b$.
We can construct $M_{\a\b}^h$ using two distinct tensors $(\,\cdot\,)_{\a\b}$ because irreducible representations of the Lorentz group are determined by the direct sum of algebras $\so(1,3) = \su(2) \oplus \su(2)$ for half-integer spins $s_1$ and $s_2$, and there are exactly $N = s_1 + s_2 - \abs{s_1 - s_2} + 1$ such representations \cite{Schwartz:2013}.
For example, QED has $s_1=s_2=\tfrac12$ and thus $N=2$ distinct couplings: a \emph{minimal} one representing momentum and helicity constraints, and an \emph{auxiliary} one between the electron's magnetic dipole and the electromagnetic field, as we shall see.

The aforementioned degeneracy of $M_{\a\b}^h$ in spanning $\SLtwoC$ occurs when the two massive particles have equal mass. 
This is clearly a prevalent issue, since fundamental interactions involving elementary particles are precisely of this nature.
Let us find two basis spinors $u_\a$ and $v_\a$ to construct $M_{\a\b}^h$.
We can simply use the helicity spinor from the massless particle to define $u_\a \equiv \la_{3\a} = \abra{3}$.
The second basis spinor will require the massive particles.
Simply choosing $\bm p_1^\mu$, for now, we can write
\ben
v_\a \equiv \frac{1}{\sqrt2 m} \lat_3^\ad \bm p_{1\ad\a}  = \frac{1}{\sqrt2 m} \sbra{3} \sketbraa{\bm1}\,.
\label{eq_vbasis1}
\een
These basis spinors point in distinct directions provided that \ben
v_\a u^\a = \frac{1}{\sqrt2 m} \sbraket{3\bm 1} \abraket{\bm 1 3}
\een
does not vanish.
In the case of equal masses $m_1=m_2$, conservation of momentum $\bm p_1^\mu + p_3^\mu = \bm p_2^\mu$ implies that $2 \bm p_1 \vdot p_3 = \abraket{\bm 1 3} \sbraket{3 \bm 1} = m_2 - m_1 = 0$.
Therefore, the basis $\{u,v\}$ is degenerate for equal masses and
no longer spans $\SLtwoC$.
This difficulty is an indication that something interesting is happening.
A crucial observation by Arkani-Hamed \emph{et al.}~is that $u_\a$ and $v_\a$ are parallel in Lorentz space but not in little group space \cite{Arkani-Hamed:2021a}.
According to \cref{eq_weyl}, $u$ scales under the little group as $\la_3 \rightarrow z \la_3$, while $v$ scales as $\lat_3 \rightarrow \tfrac{1}{z} \lat_3$. 
Therefore, we introduce the proportionality factor $x$, such that 
\ben
v_\a = x u_\a.
\label{eq_xprop}
\een
Notice that $x$ must scale like $\lat$ under the little group as $x \rightarrow \tfrac{1}{z} x$ for consistency
\ben
v_\a = x u_\a \longrightarrow \frac{1}{z} \left(\frac{1}{z} x \right) (z u_\a ) = \frac{1}{z} x u_\a = \frac{1}{z} v_\a.
\label{eq_xweight}
\een

Bringing everything together, the general stripped amplitude $M^h_{\{\a\} \{\b\}}$ with indices $\{\a\} \equiv \a_1, \ldots, \a_{s_1}$ and $\{\b\} \equiv \b_1, \ldots, \b_{s_2}$ can be constructed as the sum of $N$ distinct tensors
\ben
M^h_{\{\a\} \{\b\}} 
	= \hspace{-2mm} \sum_{i=\abs{s_1 - s_2}}^{s_1 + s_2} 
		\hspace{-2mm} g\, (\sqrt2 m)^{s_1+s_2-i} x^{h+i} 
		\left( u^{2i} \vepsi^{s_1 + s_2 - i} \right)_{\{\a\} \{\b\}}
\label{eq_construct3ampundot}
\een
where the $x$ factor carries photon helicity $h$ and all possible combinations of $u_\a$, $v_\a$, and $\vepsi_{\a\b}$ provide exactly $2s_1 + 2s_2$ Lorentz indices.
The $x$ factor clearly plays a central role in this generative formula for interactions involving massive particles.
In addition to its little-group scaling and its helicity weight, the $x$ factor is singular.
This can be seen by contracting $v_\a = x u_\a$ in \cref{eq_vbasis1} with a generic reference spinor $\chi^\a$ and then letting $\chi^\a \rightarrow \la_3^\a$:
\ben
\quad x = \frac{v_\a \chi^\a}{u_\a \chi^\a} 
		= \frac{1}{\sqrt2 m} \frac{\sbraket{ 3 \bm1} \abraket{ \bm 1 \chi}}{\abraket{3 \chi}}
		= \frac{1}{\sqrt2 m} \frac{\sbraketa{3}{\bm p_1}{\chi}}{\abraket{3 \chi}}.
\label{eq_xfactorsimple}
\een
Similar to the case for all-massless couplings in \cref{eq_massless3nonlocal}, we see that the factorization residue for massive couplings also possess an  \emph{additional} pole, thus revealing its non-local nature.
Finally, note that our choice of basis spinor $v_\a$ in \cref{eq_vbasis1} could have been written instead with $\bm p_2$.
Therefore, we need both momenta to avoid the trivial result of forward scattering, where $\bm p_1^\mu = \bm p_2^\mu$.
Defining $v_\a$ as the average of incoming and outgoing momenta $\tfrac12 (\bm p_{1\ad\a} + \bm p_2^{\a\ad})$ or via \cref{eq_gym_lorentz} as $\tfrac12 (\bm p_{1\ad\a} - \bm p_{2\ad\a})$ leads to the formal definition of the $x$ factor
\ealin{
x	&= \frac{1}{2\sqrt2 m} \left( \frac{\sbraket{3 \bm1} \abraket{\bm1 \chi}}{\abraket{3 \chi}} + \frac{\abraket{\phi \bm2} \sbraket{\bm 2 3}}{\abraket{\phi 3}} \right) \nonumber \\
	&= \frac{1}{2\sqrt2 m} \frac{\sbraketa{3}{(\bm p_1 - \bm p_2)}{\chi}}{\abraket{3 \chi}}.
\label{eq_xfactor}
}
In the last equality, we put $\phi = \chi$ for simplicity. 
The minus sign between the momenta of the fermions is crucial for spin statistics under $\bm 1 \leftrightarrow \bm 2$ particle exchange.
The $x$ factor will feature prominently in our work to study the singularities of amplitudes with massive particles.

Returning to \cref{eq_construct3ampundot}, note that the first term in the series is the minimal coupling with the lowest-dimensional operator $(\cdot)_{\{\a\} \{\b\}}$ constructed from the smallest possible number of basis spinors.
In the case of equal spin $s_1=s_2$, the minimal coupling operator consists of $s_1+s_2$ copies of the Levi--Civita symbol and is devoid of basis spinors.
The remaining terms are the auxiliary couplings with high-dimensional operators and thus coupling constants with mass dimension $g_i \equiv g \, (\sqrt2 m)^{-i}$.
The auxiliary couplings represent multipole moment interactions with copies of the basis spinors.
For example, consider the QED interaction of electron lines $\bm 1$ and $\bm 2$ and a photon line $3$ with positive helicity $h=+1$.
Using \cref{eq_construct3ampundot}, the stripped amplitude is
\ben
M^{+1}_{\a\b} = g\,\sqrt2 m x \left( \vepsi_{\a\b} + \frac{x}{\sqrt2 m} \left( u u \right)_{\a\b} \right).
\label{eq_qed3stripped+}
\een
This formula prescribes how the external lines must contract in the minimal and auxiliary couplings.
Notice that the $x$ factor in the minimal coupling $x \vepsi_{\a\b}$ provides the correct $h=+1$ weight for the positive helicity photon.
Dressing the stripped amplitude with the massive external lines, according to \cref{eq_qed3}, yields the full amplitude
\ben
\Mcal^{+1}_{IJ} = g \left( x \abraket{\bm 1 \bm 2} + x^2 \frac{\abraket{\bm 1 3}  \abraket{3 \bm 2}}{\sqrt2 m} \right).
\label{eq_qed3full+}
\een
We see that the minimal coupling simply contracts the external electron lines using the Levi--Civita symbol $\vepsi_{\a\b}$, while the auxiliary coupling represents the interaction between the electron magnetic dipole moment (lines $\bm1$ and $\bm2$) and electromagnetic field (line $3$).
Indeed, the spinor matrix $(uu)_{\a\b}$ corresponds to the familiar additional term $\tfrac{e}{2} F_{\mu \nu} \sigma^{\mu\nu}$ appearing in Dirac's equations of motion when an electron is coupled to a photon \cite{Schwartz:2013}, as well as to the Bohr magneton term $\tfrac{e}{2m} \vec B \vdot \vec \sigma$ appearing in the non-relativistic Schr\"odinger--Pauli equation \cite{Cohen-Tannoudji:1991a}.

Our final point about the analytical structure of massive couplings has to do with their representation in the complexified momentum space $\SLtwoC_\Lcal \otimes \SLtwoC_\Rcal$.
Note that the calculations in this section could have been performed using a right-handed basis with \emph{dotted} Lorentz indices $\{\tilde u_\ad, \tilde v_\bd\}$ instead of the undotted basis.
The procedure is parallel to that above, where we first define right-handed basis spinors $\tilde u_\ad \equiv \lat_{3\ad}$ and $\tilde v_\ad$ analogously to \cref{eq_vbasis1} with the difference being that the $x$ factor must be negated and inverted in order to provide the correct little-group weight, a consequence of the conjugation rules that we shall explore in the next section.
Therefore, instead of \cref{eq_xprop}, we would have $\tilde v_\ad = - \frac{1}{x} \tilde u_\ad$
and would arrive at an expansion of $N$ coupling terms analogous to those in \cref{eq_construct3ampundot} but now using $\lat$.
Indeed, we now see that massive couplings do generically transform in \emph{both} $\Lcal$ and $\Rcal$ subgroups.
However, these expansions must be redundant, since there can only be $N$ independent tensor structures, not $2N$, determined entirely by the spin of the massive particles $s_1$ and $s_2$ as discussed in the paragraph following \cref{eq_qed3}.
Indeed, we will show in the next section that these expansions are not independent but rather conjugates of each other.

As it often turns out in physics, one of the bases will be more practical than the others for solving a given problem. 
Although we can make this determination by appealing to prior knowledge about the QED coupling \cite{Arkani-Hamed:2021a,Guevara:2019,Schwartz:2013,Arkani-Hamed:2021b}, we shall approach this here by examining how these couplings coalesce from the high-energy limit (HE).
Consider the coupling for a positive helicity photon given by \cref{eq_qed3full+}.
In the dotted basis, we have
\ben
\Mcal_{IJ}^{+1} = -\tilde g x  \left( \sbraket{\bm1 \bm2} - \frac{ \sbraket{\bm1 3} \sbraket{3 \bm 2} }{\sqrt2 m x} \right).
\label{eq_qed3full+dot}
\een
Although the minimal coupling $-\tilde g x \sbraket{\bm1 \bm2}$ seems fine, it actually coalesces from a \emph{negative} helicity photon, while $g x \abraket{\bm1 \bm2}$ coalesces from a positive helicity one.
To see this, note that the electrons in either coupling must have opposite spin, as one enters the vertex and the other exits, so that angular momentum is conserved.
Then in the high-energy limit, either coupling $\Mcal^h_{IJ}$ has exactly two spin configurations $(\bm{1}^{\sfrac{1}{2}},\, \bm{2}^{\sfrac{1}{2}}, 3^h) \hel (1^{\pm\sfrac{1}{2}},\,2^{\mp\sfrac{1}{2}}, 3^h)$.
The positive helicity coupling $\Mcal^{+1}_{IJ}$ at high energy thus satisfies the mostly positive helicity case of the massless coupling in \cref{eq_massless3}, and $\Mcal^{-1}_{IJ}$ satisfies the mostly negative helicity case.
Averaging over both configurations for each coupling yields their high-energy limits:

\begin{widetext}
\begin{fmffile}{qed3HE+}
\begin{equation}
\begin{gathered}
\begin{fmfgraph*}(60,60)
	\fmfpen{thick}
        \fmfleft{i1}
        \fmfright{o1,o2}

        \fmf{fermion, tension=1}{i1,v1,o2}
        \fmf{photon, tension=1}{v1,o1}

        \fmflabel{$\bm{1}^{\sfrac{1}{2}}$}{i1}
        \fmflabel{$3^{+1}$}{o1}        
        \fmflabel{$\bm{2}^{\sfrac{1}{2}}$}{o2}

        \fmfv{decor.shape=circle, decor.filled=empty, decor.size=4mm}{v1}

\end{fmfgraph*}
\end{gathered}
\quad = g x \abraket{\bm1\bm2} 
\hel
\left.
\begin{cases}
\gt \frac{\sbraket{31}^2}{\sbraket{12}}
	&\qq*{for $(1^{+\sfrac{1}{2}},2^{-\sfrac{1}{2}},3^{+1})$} \vspace{1mm} \\
\gt \frac{\sbraket{23}^2}{\sbraket{12}}
	&\qq*{for $(1^{-\sfrac{1}{2}},2^{+\sfrac{1}{2}},3^{+1})$}
\end{cases}
\right\}
\longrightarrow
\gt \frac{\sbraket{13}^2 + \sbraket{23}^2}{2 \sbraket{12}},
\label{fig_qed3HE+}
\end{equation}
\end{fmffile}
\vspace{4mm}

\begin{fmffile}{qed3HE-}
\begin{equation}
\begin{gathered}
\begin{fmfgraph*}(60,60)
	\fmfpen{thick}
        \fmfleft{i1}
        \fmfright{o1,o2}

        \fmf{fermion, tension=1}{i1,v1,o2}
        \fmf{photon, tension=1}{v1,o1}

        \fmflabel{$\bm{1}^{\sfrac{1}{2}}$}{i1}
        \fmflabel{$3^{-1}$}{o1}        
        \fmflabel{$\bm{2}^{\sfrac{1}{2}}$}{o2}

        \fmfv{decor.shape=circle, decor.filled=full, decor.size=4mm}{v1}

\end{fmfgraph*}
\end{gathered}
\quad = - \frac{\gt}{x} \sbraket{\bm{1}\bm{2}} 
\hel
\left.
\begin{cases}
g \frac{\abraket{23}^2}{\abraket{12}}
	&\qq*{for $(1^{+\sfrac{1}{2}},2^{-\sfrac{1}{2}},3^{-1})$} \vspace{1mm} \\ 
g \frac{\abraket{31}^2}{\abraket{12}}
	&\qq*{for $(1^{-\sfrac{1}{2}},2^{+\sfrac{1}{2}},3^{-1})$}
\end{cases}
\right\}
\longrightarrow
g \frac{\abraket{13}^2 + \abraket{23}^2}{2 \abraket{12}}.
\label{fig_qed3HE-}
\end{equation}
\end{fmffile}
\end{widetext}

Therefore, $g x \abraket{\bm{1} \bm{2}}$ is the natural choice for the minimal coupling with a positive helicity photon and $-\tilde g x \sbraket{\bm{1} \bm{2}}$ is natural for the negative helicity photon.
A similar argument can be made for the auxiliary couplings.
We write the full QED couplings, manifestly transforming in both $\Lcal$ and $\Rcal$ subgroups each with exactly $N=2$ independent terms as follows:
\ealin{
\Mcal_{IJ}^{+1} &= g x \abraket{\bm{1} \bm{2}} + g x^2 \frac{ \abraket{\bm{1} 3} \abraket{3 \bm{2}} }{\sqrt{2} m} \nonumber \\
\Mcal_{IJ}^{-1} &= - \frac{\tilde g}{x} \sbraket{\bm{1} \bm{2}} + \tilde g \frac{ \sbraket{\bm{1} 3} \sbraket{3 \bm{2}} }{\sqrt{2} m}
\label {eq_qed3full}
}
These couplings are the building blocks for scattering amplitudes.
To calculate cross sections, we require their conjugation and crossing symmetry relations, derived next.

\subsection{Conjugation and crossing symmetry \label{sec_amp_4}}

Scattering amplitudes constructed by gluing the spinor helicity couplings shown above will possess many of the same properties as those derived using QFT.
We can still interpret $\Mcal$ as an $S$-matrix element $\mel{f}{\Mcal}{i}$ whose asymptotic states $\ket{p_a,\sigma_a}$ are encoded by the external helicity spinors with little group indices $I_a$ and helicities $h_a$.
It is unitary because $\Mcal$ is factorizable, its poles correspond to on-shell particles, and the bi-spinor representation of the phase space $\SLtwoC_\Lcal \otimes \SLtwoC_\Rcal$ is complete.
It is manifestly Lorentz invariant, since the Lorentz indices $\a, \ad, \b, \bd, \ldots$ do not appear in the full couplings presented in \cref{sec_amp_3} and, thus, cannot appear in any scattering amplitude constructed from them.

On the other hand, we have seen important differences with QFT.
The spinor helicity formalism is a particle theory not a field theory, and its phase space is the complex-valued bi-spinor representation of momentum, not the real-valued $4$-vector representation.
Consequently, Lagrangian densities, virtual particles, and gauge redundancies associated with $4$-vector particle embedding are absent.
In a unitary and Lorentz-invariant field theory, the $CPT$ theorem guarantees the invariance of Lagrangian density terms.
Furthermore, it implies that the scattering amplitude for any given process is the same as that for its inverse process, where particles are replaced with antiparticles with reversed spin \cite{Weinberg:1995}.
Despite the absent field theory, we shall expect our unitary and Lorentz-invariant spinor helicity amplitudes to inherit $CPT$ invariance.
In deriving the conjugation rules for helicity spinors below, it will be necessary to combine the Hermitian adjoint $(\cdot)^\dagger$ and the crossing relation $p^\mu \rightarrow -p^\mu$ in order to produce the net action of sending particles to antiparticles with spin (or helicity) and handedness reversed.
These rules are succinctly referred to in the literature as \emph{crossing symmetry} \cite{Dixon:2014,Elvang:2015}, where all particles are taken to be incoming, and thus all outgoing particles are assigned the negative of their physical $4$-momentum.

Here, we shall continue using physical momenta (and watching our signs) in order to better appreciate how crossing symmetry works and also to compare our amplitudes and cross sections with standard results in the literature.
The cross section $|\Mcal|^2 = \Mcal^\dagger \Mcal$ requires the Hermitian adjoint.
This usually involves the replacement of kets $\ket{\phi}$ by bras $\bra{\phi}$ and vice versa \cite{Cohen-Tannoudji:1991a}, but our bi-spinor representation has \emph{two} ket types: a left-handed $\aket{\bm \la}$ and a right-handed $\sket{\bm \la}$.
How should we proceed?
Recall that Dirac $4$-spinor fields $\psi(x)$  transform in the reducible $\left( \frac12, 0 \right) \oplus \left( 0, \frac12 \right)$ representation of $\SLtwoC$. 
In that case, the appropriate conjugation $\bar \psi(x) \equiv \psi(x)^\dagger \g^0$ swaps the left- and right-handed components to account for the anti-Hermitian boost generators of the Lorentz group.
Thus, Dirac conjugation replaces an incoming left-handed fermion $\psi(x)$ by an outgoing right-handed one $\bar \psi(x)$ with momentum and spin unchanged \cite{Schwartz:2013}.

Proceeding accordingly for a massless helicity spinor, we might simply expect that $\overline{\aket \la} = \sbra \la$ and similarly that $\overline{\sket \la} = \abra \la$, so handedness changes but momentum does not
\ben
\overline{\aket \la \sbra \la} = \overline{\sbra \la} \overline{\aket \la} = \aket \la \sbra \la.
\label{eq_momentumadjoint}
\een
This is indeed the adjoint operation on helicity (Weyl) spinors seen in the literature, which we shall refer to as Weyl conjugation.
Note that we can move from $\aket\la \propto \zap[\a]$ to $\sbra\la \propto \zatn[\ad]$ using Hermitian conjugation and the $\vepsi$ symbol (see basis spinors in \cref{app_conventions}).
However, an overall minus sign reverses the $4$-momentum, but this can be prevented with the crossing relation $p^\mu \rightarrow -p^\mu$, yielding the correct result
\ealin{
\overline{\aket \la} 
	&= \overline{\la^\a}
		= \frac{1}{\sqrt[4] 2} \sqrt{-E - p_z} \, \vepsi_{\a\b} \left(\zap[\b] \right)^\dagger \nonumber \\
	&= \frac{1}{\sqrt[4] 2} \sqrt{-E - p_z} \, \mqty(-s^* & c) \nonumber \\
	&= \frac{1}{\sqrt[4] 2} \sqrt{E + p_z} \, \mqty(s^* & -c) \nonumber \\
	&= \frac{1}{\sqrt[4] 2} \sqrt{E + p_z} \, \zatn[\ad] \nonumber \\
	&= \lat^\ad 
		= \sbra \la.
\label{eq_adjointmassless}
}
Thus, Weyl conjugation reverses the spin by swapping the basis-spinor sign and the handedness by toggling the tilde/dot notation: $\za^{\pm \a} \leftrightarrow \zat^{\mp \ad}$.
The little group basis $\{\za^{\pm}_I\}$ has the same transformation properties as the spinor bases $\{\za^{\pm \a}\}$ and $\{\zat^{\pm \ad}\}$, which means that Weyl conjugation reverses spin and raises/lowers the index: $\za^\pm_I \leftrightarrow \za^{\mp I}$.
Taken together, we obtain the conjugation rules for massive helicity spinors
\ealin{
\overline{\aket{\bm \la}}
	&= \overline{\la^\a_I} 
		= \overline{\aket{\la} \za^-_I} + \overline{\aket{\eta} \za^+_I}
		= \sbra \la \za^{+I} + \sbra \eta \za^{-I}
		= \lat^{\ad I}
		= \sbra{\bm \la}, \\	
\overline{\sket{\bm \la}}
	&= \overline{\lat_{\ad I}} 
		= \overline{\sket{\la} \za^+_I} + \overline{\sket{\eta} \za^-_I}
		= \abra \la \za^{-I} + \abra \eta \za^{+I}
		= \la_\a^I
		= \abra{\bm \la}.
\label{eq_adjointmassive}
}
These rules are a straightforward generalization of the massless ones in \cref{eq_adjointmassless}.
Indeed, the momentum for massive particles is unchanged under conjugation, and we can now appreciate how the adjoint of $\abraket{\bm{1} 2}\sbraket{3 \bm{4}}\,$ will take the simple and suggestive form $\overline{\abraket{\bm{1} 2}\sbraket{3 \bm{4}}\,} = \abraket{\bm{4} 3}\sbraket{2 \bm{1}}\,$, where the particles have reversed roles in the scattering process.

The last ingredient of $\Mcal$ to conjugate is the $x$ factor.
We can proceed by rearranging \cref{eq_xfactorsimple}
\ben
x = \frac{1}{\sqrt2 m} \frac{\sbraketa{3}{\bm p_1}{\chi}}{\abraket{3 \chi}}
\Longrightarrow \sqrt2 m \abraket{3 \chi} x = \sbraketa{3}{\bm p_1}{\chi},
\een
and conjugating both sides of the equation to find
\ealin{
& \overline{ \sqrt2 m \abraket{3 \chi} x } = \overline{ \sbraketa{3}{\bm p_1}{\chi} } \nonumber \\
&\Rightarrow \sqrt2 m \sbraket{\chi 3} \overline x = \sbraketa{\chi}{\bm p_1}{3} \nonumber \\
&\Rightarrow \overline x = \frac{1}{\sqrt2m} \frac{\sbraketa{\chi}{\bm p_1}{3}}{\sbraket{\chi 3}}.
\label{eq_xfactoradjoint}
}
This last quantity can be directly related to $x$ by expanding the spinors and multiplying numerator and denominator by $-\sqrt2 m$
\ealin{
\overline x 
	&= \frac{1}{\sqrt2m} \frac{\chit^\ad \bm p_{1\ad\a} \la_3^\a}{\chit^\ad \lat_{3\ad}}
		= \frac{-\sqrt2 m}{-2 m^2} \frac{\bm p_{1\ad\a} \la_3^\a}{\lat_{3\ad}} \nonumber \\
	&= -\frac{\sqrt2 m}{\bm p_1^{\a\ad} } \frac{ \la_3^\a}{\lat_{3\ad}} 
		= -\sqrt2 m \frac{\chi_\a}{\chi_\a} \frac{ \la_3^\a}{ \bm p_1^{\a\ad} \lat_{3\ad}}  \nonumber \\
	&= -\sqrt2 m \frac{\abraket{\chi 3}}{\abrakets{\chi}{\bm p_1}{3}}
		= -\sqrt2 m \frac{\abraket{3 \chi}}{\sbraketa{3}{\bm p_1}{\chi}} 
		= - \frac{1}{x}. 
	\label{eq_xfactorreciprocal}
}
We see that the right-handed basis $\{ \tilde u_\ad, \tilde v_\ad \}$ discussed in the paragraph following \cref{eq_qed3full+} of \cref{sec_amp_3} is simply the adjoint of the left-handed basis, where $\overline{v_\a} = \overline{x u_\a} \Rightarrow \tilde v_\ad = - \frac{1}{x} \tilde u_\ad$.
With everything now in place, let us draw the on-shell diagrams for Bhabha annihilation and Compton scattering and construct their amplitudes by gluing together couplings, subject to momentum conservation and helicity constraints.

\subsection{Bhabha and Compton amplitudes \label{sec_amp_5}}

Our work ahead will focus on Bhabha annihilation and Compton scattering. 
These processes provide a natural starting point as they are relatively simple yet embody the most prevalent motifs encountered in particle physics.
We calculate their minimal coupling amplitudes in QED here using the spinor helicity formalism.

We begin with Bhabha annihilation $e^- e^+ \rightarrow \mu^+ \mu^-$.
This is the simplest scattering process in QED at leading order.
Ironically, already at 1-loop in QFT, this particular amplitude reveals an important theoretical issue in particle physics, that of infrared divergences.
(More will follow on this in \cref{sec_mass}.)
For now, at leading order, the only Feynman diagram is the $s$ channel, where a photon is exchanged.
Therefore, the on-shell diagram should only factorize along a massless internal line.
Consider the corresponding diagrams shown below.

\begin{widetext}
\begin{fmffile}{bhabha}
\begin{equation}
\begin{gathered}
\begin{fmfgraph*}(100,80)
        \fmfleft{i1,i2}
        \fmfright{o1,o2}
        
        \fmf{fermion, label=$p_1$, label.side=right}{i1,v1}
        \fmf{fermion, label=$p_2$}{v1,i2}
        \fmf{fermion, label=$p_3$}{o2,v2}
        \fmf{fermion, label=$p_4$}{v2,o1}
        \fmf{photon}{v1,v2}

\end{fmfgraph*}
\end{gathered}
\quad\quad \longleftrightarrow \quad\quad
\begin{gathered}
\begin{fmfgraph*}(100,80)
	\fmfpen{thick}
        \fmfleft{i1,i2}
        \fmfright{o1,o2}

        \fmf{fermion, tension=1.5}{i1,v1}
        \fmf{fermion}{v1,v2}
        \fmf{fermion, tension=1.5}{v2,i2}
        \fmf{fermion, tension=1.5}{o2,v4}
        \fmf{fermion}{v4,v3}
        \fmf{fermion, tension=1.5}{v3,o1}
        \fmf{photon}{v1,v3}
        \fmf{photon}{v2,v4}

        \fmflabel{$\bm{1}^{\sfrac{1}{2}}$}{i1}
        \fmflabel{$\bm{2}^{\sfrac{1}{2}}$}{i2}
        \fmflabel{$\bm{4}^{\sfrac{1}{2}}$}{o1}
        \fmflabel{$\bm{3}^{\sfrac{1}{2}}$}{o2}        

        \fmfv{decor.shape=circle, decor.filled=empty, decor.size=4mm}{v1}
        \fmfv{decor.shape=circle, decor.filled=full, decor.size=4mm}{v2}
        \fmfv{decor.shape=circle, decor.filled=full, decor.size=4mm}{v3}
        \fmfv{decor.shape=circle, decor.filled=empty, decor.size=4mm}{v4}

\end{fmfgraph*}
\end{gathered}
\label{fig_bhabha}
\end{equation}
\end{fmffile}
\end{widetext}

The on-shell diagram looks similar to Compton scattering in \cref{fig_compton}, except that its lines $\bm2$ and $\bm3$ are now fermions (a positron and an anti-muon, respectively) and that the top edge of the box is now a photon.
Yet, these differences completely change the behavior of the process.
Like the Feynman diagram, we see a continuous fermionic path between lines $\bm1$ to $\bm2$ as well as between lines $\bm3$ to $\bm4$, with both paths exchanging force carriers.
Indeed, there are only two consistent factorizations here: those obtained by the deletion of either the top or bottom edges in the box, where a single photon line is exchanged.
Note that the two other factorizations arising from the deletion of the left or right edges of the box must vanish because they allow two massless lines to escape, in conflict with the external all-massive data.
The only difference between the two valid factorizations is the internal photon's helicity. 
Summing over both cases according to \cref{eq_factorizemassless} thus yields the amplitude
\ealin{
\Mcal(\bm{1}, \bm{2}, \bm{3}, \bm{4})
	&= \sum_{h=\pm1} \frac{ \Lcal^h_{I J } \; \Rcal^{-h}_{K L} } {s} \nonumber \\
	&= \frac{ \Lcal^{+1}_{I J } \; \Rcal^{-1}_{K L}}{s} 
		+ \frac{ \Lcal^{-1}_{I J } \; \Rcal^{+1}_{K L}}{s}.
}
The minimal couplings in \cref{eq_qed3full} provide the leading order contribution
\ealin{
\Mcal(\bm{1}, \bm{2}, \bm{3}, \bm{4})
	&= \bigg( g x_{12} \abraket{\bm1 \bm2} \bigg) \frac{1}{s} \bigg( -\tilde g \frac{1}{x_{34}} \sbraket{\bm3 \bm4} \bigg) \nonumber \\
		&\quad+ \bigg( -\tilde g \frac{1}{x_{12}} \sbraket{\bm1 \bm2} \bigg) \frac{1}{s} \bigg( g x_{34} \abraket{\bm3 \bm4} \bigg) \nonumber \\
	&= -\frac{g\gt}{s} \left( \frac{x_{12}}{x_{34}} \abraket{\bm1 \bm2} \sbraket{\bm3 \bm4} + \frac{x_{34}}{x_{12}} \abraket{\bm3 \bm4} \sbraket{\bm1 \bm2} \right) \nonumber \\
	&= \frac{2e^2}{s} \left( \frac{x_{12}}{x_{34}} \abraket{\bm1 \bm2} \sbraket{\bm3 \bm4} + \frac{x_{34}}{x_{12}} \abraket{\bm3 \bm4} \sbraket{\bm1 \bm2} \right)
\label{eq_minampbhabha}
}
after setting the coupling constants to $g \equiv -\sqrt{2}e$ and $\gt \equiv -g$ for QED.
This expression is the scattering amplitude for \emph{all} possible asymptotic states of the four spin-$\tfrac12$ particles.
Let us pause to examine some remarkable features.
First, its manifest symmetry, with one term being the conjugate of the other, justify the simplicity of this scattering process and its corresponding diagram.
Nevertheless, the term $\abraket{\bm1 \bm2} \sbraket{\bm3 \bm4}\,$ is not \emph{too} simple, as it prescribes how all the Lorentz and little group indices of these massive helicity spinors are being contracted. 
Thus are encoded the rich dynamics of the interaction.
For example, only four of the $2^4$ distinct spin configurations conserve angular momentum; the rest are inadmissible and must have vanishing amplitudes.
The four configurations are those where the two external lines on both sides have opposite spin.
In that case, each bracket leads to a generically non-zero factor like $\sbraketa{\bm1}{\bar\sigma_\mu}{\bm2}$ and $\abrakets{\bm1}{\sigma_\mu}{\bm2}$.
All other configurations have at least one side with equal spins in both lines, and the amplitude vanishes due to the appearance of at least one factor like $\abraketa{\bm1}{\sigma_\mu}{\bm2}=0$ or $\sbrakets{\bm1}{\bar\sigma_\mu}{\bm2}=0$.
Finally, let us examine the quotient of $x$ factors.
We saw in \cref{sec_amp_3} that $x$ can encode the presence of additional poles and thus indicate the non-locality of the amplitude.
For Bhabha \emph{annihilation}, there cannot be any other singularities because the incoming electron \emph{must} annihilate the incoming positron to produce an outgoing muon--antimuon pair.
The amplitude can only have one pole; therefore, the process must be local.
Using \cref{eq_xfactorsimple,eq_xfactoradjoint,eq_xfactorreciprocal} with the internal massless line $k^\mu$ playing the role of basis spinor for \emph{both} couplings, we have
\ben
x_{12} = \frac{1}{\sqrt2 m} \frac{ \sbraketa{k}{\bm p_1}{\chi} }{ \abraket{k \chi}}
\qqtext{and}
\frac{1}{x_{34}} = \frac{-1}{\sqrt2 m} \frac{ \sbraketa{\phi}{\bm p_3}{k} }{ \sbraket{\phi k}},
\een
where $\chi$ and $\phi$ are the reference spinors.
The quotient is then
\ealin{
\frac{x_{12}}{x_{34}} 
	&= -\frac{1}{2m^2}
		\frac{\sbraketa{\phi}{\bm p_3}{k} \sbraketa{k}{\bm p_1}{\chi}}
		{\sbraket{\phi k} \abraket{k \chi}} \nonumber \\
	&= -\frac{1}{2m^2}
		\frac{\sbraketa{\phi}{\bm p_3 \vdot k \vdot \bm p_1}{\chi}}
		{\sbraketa{\phi}{k}{\chi}} 
	= -\frac{\bm p_1 \vdot \bm p_3}{2 m^2},
\label{eq_xquotientbhabha}
}
where, in the last step, we lowered the $k$ indices and raised the $\bm p_1$ indices to make simplifying cancellations.
But something is wrong here because the reciprocal $x_{34}/x_{12}$ leads to the same result and thus implies $\bm p_1 \vdot \bm p_3 = \pm 2 m^2 = \const$, which contradicts the effect of scattering angle.
Indeed, by eschewing the formal $x$ factor defined in \cref{eq_xfactor}, we omitted the degree of freedom \emph{between} $\bm p_1$ and $\bm p_2$, namely their scattering angle!
Repeating the above calculation with the relevant scattering information by formalizing, say, $x_{12}$, leads to the slightly more descriptive amplitude for Bhabha annihilation
\ben
\Mcal(\bm{1}, \bm{2}, \bm{3}, \bm{4})
	= - \frac{e^2}{2} \frac{(\bm p_1 - \bm p_2) \vdot \bm p_3}{s m^2} \bigg( \abraket{\bm1 \bm2} \sbraket{\bm3 \bm4} + \abraket{\bm3 \bm4} \sbraket{\bm1 \bm2}\, \bigg).
\label{eq_minampbhabha_scatter}
\een
The amplitude is manifestly local with its single pole in the $s$ channel.
That no additional poles appear in the calculation hinges on the fact that for Bhabha annihilation both couplings share a common basis spinor.
Finally, the amplitude vanishes at forward scattering $\bm p_1^\mu = \bm p_2^\mu$, when the incident particles have parallel trajectories and annihilation does not take place.

Next, let us consider Compton scattering $e^- \g \rightarrow \g e^-$, which has two Feynman diagrams corresponding to the $s$ and $u$ channels shown in \cref{fig_compton}.
As above, we can again factorize along the $s$ channel, but this time the internal line is the massive $\bm k^\mu$.
Using \cref{eq_factorizemassive}, at minimal coupling we have
\ealin{
\Mcal(\bm{1}, 2^-, 3^+, \bm{4})
	&= \frac{ \Lcal^{-1}_{I J} \vepsi^{J K} \Rcal^{+1}_{K L} } {s - m^2} \nonumber \\
	&= \bigg(-\gt \frac{1}{x_{12}} \sbraket{\bm k \bm 1} \bigg) \frac{1}{s - m^2} 
		\bigg(g x_{34} \abraket{\bm 4 \bm k} \bigg) \nonumber \\
	&= - \frac{g \gt}{s - m^2} \frac{x_{34}}{x_{12}} \abraket{\bm 4 \bm k} \sbraket{\bm k \bm 1}.
\label{eq_minampcompton_temp}
}
Unlike Bhabha annihilation, we expect another pole to appear in the $x$ factors of the residue. 
This time the external photon lines $2$ and $3$ provide \emph{distinct} basis spinors for the left and right sides, respectively.
Along with the external electron lines $\bm1$ and $\bm4$, the quotient becomes
\ben
\frac{x_{34}}{x_{12}} 
	= -\frac{1}{2 m^2} 
		\frac{\sbraketa{\phi}{\bm p_1}{2} \sbraketa{3}{\bm p_4}{\chi}}{\sbraket{\phi 2} \abraket{3 \chi}}.
\label{eq_xquotientcompton_temp}
\een
Now, by choosing the reference spinor of each side to be the basis spinor of the other side, i.e., let $\sbra \chi \rightarrow \sbra 3$ and  $\abra \phi \rightarrow \abra 2$, the denominator reveals a second pole: $\sbraket{3 2} \abraket{3 2} = -2 p_2 \vdot p_3 = t$.
Although there is no $t$ channel for Compton scattering, we did factorize along the $s$ channel so that $s-m^2=0$, and thus $s + t + u = 2m^2$ implies $t = -(u-m^2)$.
Therefore, we can write $-\sbraket{3 2} \abraket{2 3} = u - m^2$, and our residue is indeed singular along the $u$ channel, as expected.
Further simplification is possible by noting that conservation of momentum requires that $\sum_{j=1}^4 \abraket{i j} \sbraket{j k} = 0$ for any null momenta $i$ and $k$. 
Therefore, $\sbraket{3 \bm1} \abraket{\bm1 2}\, + \sbraket{3 \bm4} \abraket{\bm4 2} = 0$, which implies $\sbraketa{3}{\bm p_1}{2} = -\sbraketa{3}{\bm p_4}{2}$ and that each side of this last equation must also be equal to $\tfrac12 \sbraketa{3}{(\bm p_1 - \bm p_4)}{2}$.
Substituting this last expression into \cref{eq_xquotientcompton_temp} leads to the more descriptive quotient
\ben
\frac{x_{34}}{x_{12}} 
	= - \frac{1}{8 m^2}
		\frac{\sbraketa{3}{(\bm p_1 - \bm p_4)}{2}^2}{u-m^2},
\label{eq_xquotientcompton}
\een
which forces the amplitude to vanish at forward scattering $\bm p_1^\mu = \bm p_4^\mu$, and where the $u$ channel is made manifest.
Note that the square in the numerator acts on the entire bracket.
Combining this with \cref{eq_minampcompton_temp} and writing the internal line as $\bm k^\mu = \bm p_1^\mu + p_2^\mu$ yields the amplitude at leading order
\ealin{
\Mcal&(\bm{1}, 2^-, 3^+, \bm{4}) \nonumber \\
	&=\frac{g \gt}{(s - m^2)} 
		\frac{\sbraketa{3}{(\bm p_1 - \bm p_4)}{2}^2}{8 m^2 (u - m^2)} \abrakets{\bm 4}{(\bm p_1 + p_2)}{ \bm 1} \nonumber \\
	&= g \gt \frac{\sbraketa{3}{(\bm p_1 - \bm p_4)}{2}}{(s - m^2)(u - m^2)}
		\frac{\sbraketa{3}{(\bm p_1 - \bm p_4)}{2} \abrakets{\bm 4}{(\bm p_1 +  p_2)}{ \bm 1}}{8 m^2} \nonumber \\
	&= -\frac{e^2}{4} \frac{\sbraketa{3}{(\bm p_1 - \bm p_4)}{2}}{(s - m^2)(u - m^2)}
		\bigg( \abraket{\bm1 2} \sbraket{\bm4 3} + \abraket{\bm4 2} \sbraket{\bm1 3}\, \bigg)
\label{eq_minampcompton}
}
after replacing the coupling constants with $g=-\sqrt2 e$ and $\gt=-g$, and after simplifying the last factor using \cref{eq_anglemass,eq_squaremass} as well as $\abraket{\bm1 2} \sbraket{2 \bm1} = \abraket{\bm4 2} \sbraket{2 \bm4} = 0$ in the last step.
Upon visual inspection, the structure of this amplitude says a great deal about the scattering process.
The external photons appear as $\aket 2$ and $\sket 3$, indicating that line $2$ has negative helicity and line $3$ has positive helicity, as labeled in the diagram.
Indeed, the amplitude with those helicities reversed $\Mcal(\bm{1}, 2^+, 3^-, \bm{4})$ looks exactly like \cref{eq_minampcompton} after exchanging the symbols $2 \leftrightarrow 3$.
Analogously, the handedness of the scattered electron also changes from left $\abra{\bm1}$ to right $\sbra{\bm4}$ in the first term and vice versa in the second.
Moreover, the two terms transparently encode the factorization channels: the first term is the $s$ channel, coupling lines $\bm1$ with $2$ and lines $3$ with $\bm4$, while the second term is the $u$ channel, coupling lines $\bm1$ with $3$ and $2$ with $\bm4$.

Remarkably thus does the spinor helicity formalism allow us to calculate scattering amplitudes.
We obtain a single object that manifestly transforms within a representation of the Poincar\'e group determined by the helicity and little group indices of its asymptotic states according to \cref{eq_transformhelicity,eq_transformlittlegroup}, respectively.
That its analytical structure inherently prioritizes mass and spin over $4$-momentum, and that it encodes multiple factorization channels all at once, suggests that the primacy of locality and even spacetime ought to be re-evaluated.
Lest we get ahead of ourselves, it is appropriate to first compare these results with their well-established QFT counterparts.
We thus turn to the task of calculating their associated cross sections, whence a direct comparison may be made.
Proceeding accordingly, we shall restrict our work in this paper to the leading-order contributions obtained from minimal coupling.

\section{Cross sections \label{sec_cross}}

In this section, we lay out the challenges, methods, and results for calculating cross sections in the spinor helicity formalism.
We first show that massive spinor helicity amplitudes are tensor objects and represent all spin configurations of the scattering process at once.
Second, we developed two complementary methods to expand the scattering amplitude over its spin configurations to calculate unpolarized cross sections.
Finally, we calculate cross sections for Bhabha annihilation and Compton scattering from the previous section.
Taken together, we aim to demonstrate how to use massive spinors effectively and simultaneously provide non-trivial checks by comparing our results with well-known theoretical data from QFT.

\subsection{Square of the amplitude and unpolarized scattering \label{sec_cross_1}}

Equipped with the methods and results laid out in \cref{sec_mhs,sec_amp}, we can now tackle the problem of calculating the rate of any scattering process from its unpolarized cross section
\ben
\sigma = \frac{1}{N} \sum_{\mathrm{spin}} | \Mcal |^2.
\label{eq_crosssection}
\een
We must average over all $N$ distinct spin configurations in order to compare with experiment.
Given $\Mcal$, we can easily calculate its conjugate $\Mcal^\dagger$, but the task of expanding the square of the (modulus of the) amplitude $| \Mcal |^2 = \Mcal^\dagger \Mcal$ is delicate because $\Mcal$ in the spinor helicity formalism is an explicit function of the external helicity states but \emph{not} of the external spin states.
All possible configurations are implicitly represented in the tensor structure of the scattering amplitude.
Therefore, to calculate the unpolarized cross section, we must extract the configurations, square them separately, and add them up.
We shall learn how to do this in two ways in the following sections, but first it is instructive to review how it is done in QFT using Dirac algebra \cite{Schwartz:2013}.

Consider the Bhabha annihilation process shown in \cref{fig_bhabha}.
Focusing only on the left hand side of the process, we have the QED interaction
$- \ii e \bar v(p_2) \g^\mu u(p_1)$.
Its contribution to the cross section is proportional to
\ben
\sum_{s'} \sum_s \left[ \bar u^{s'}_b (p_1) \g^\mu_{b c} v^s_c (p_2) \, \bar v^s_d (p_2) \g^\nu_{d a} u^{s'}_a (p_1) \right],
\label{eq_diracleft_temp}
\een
where Latin letters $a,\,b,\,c,\,d$ are the Dirac $4$-spinor indices, and we sum over all spin states $s$ and $s'$.
As Dirac taught us, we sum over spins using the outer products
\ealin{
\sum_s u^s_a (p) \bar u^s_b (p) &= \big( \ps + m \id \big)_{a b} \nonumber \\
\sum_s v^s_a (p) \bar v^s_b (p) &= \big( \ps - m \id \big)_{a b}
}
and trace over the Dirac indices, reducing \cref{eq_diracleft_temp} to an expression that is covariant only with the Lorentz group
\ealin{
\big( \slashed{p}_1 + m \id \big)_{a b} &\g^\mu_{b c} \big( \slashed{p}_2 - m \id \big)_{c d} \g^\nu_{d a} \nonumber \\
	&= \Tr \left[ \big( \slashed{p}_1 + m \big) \g^\mu \big( \slashed{p}_2 - m \big) \g^\nu \right] \nonumber \\
	&= 4 \left( p_1^\mu p_2^\nu + p_2^\mu p_1^\nu - p_1^\rho p_2^\rho g^{\mu\nu} \right) - 4m^2 g^{\mu\nu}.
\label{eq_diracleft}
}
Thus are the non-trivial rules for coupling fermions beautifully encoded in QFT: as the Dirac algebra over spinors equipped with multiplication defined by the gamma matrix $\g_{a b}^\mu$.
The incoming momenta $p_1^\mu$ and $p_2^\mu$ on the left side of the process get paired-up in \cref{eq_diracleft} in such a way that, when contracted with the outgoing momenta on the right side of the process, we obtain a Lorentz invariant quantity where the left- and right-handed chiralities do not mix.
Namely, the $(p_1 \vdot p_2) (p_3 \vdot p_4)$ term cancels out to yield the unpolarized Bhabha cross section at leading order
\ealin{
\sigma_\mathrm{B} = &\frac{8 e^4}{s^2} \Big[ (p_1 \vdot p_3) (p_2 \vdot p_4) 
		+ (p_1 \vdot p_4) (p_2 \vdot p_3) \nonumber \\
		&+ m^2 ( p_1 \vdot p_2 +  p_3 \vdot p_4 ) + 2m^4 \Big].
\label{eq_bhabhacrosssection}
}

Although this is by now a familiar calculation, the expansion of $| \Mcal |^2$ is clearly a rather delicate process in QFT.
What are the massive spinor helicity analogues to Dirac's $\g$-matrix and the spinor outer product and trace operations?
The short answer is coalescence.
Massive spinor helicity amplitudes and cross sections find their analogous coupling rules in the high-energy limit, which is fully determined by the massless coupling \cref{eq_massless3}.
In the next section, we shall formulate a method using the high-energy limit to calculate massless cross sections and a \emph{quasi}-limit to calculate massive cross sections like \cref{eq_bhabhacrosssection}.
The constructive method that was presented for calculating scattering amplitudes will also be extended to cross sections, to serve as a building block similar to \cref{eq_diracleft} for directly calculating massive cross sections.

Before moving on, we take the opportunity to make contact with QFT and show explicitly how the Dirac algebra manifests itself in the spinor helicity formalism.
Consider the QED interaction between two electrons and a positive helicity photon.
Summing over both spin configurations, we have
\ealin{
\sum_s \bar u(p_2)^{-s} \slashed{\epsilon}_3^+ u(p_1)^s
	&= \abrakets{\bm u_2}{\epsilon_{3\mu}^+ \g^\mu}{\bm u_1}
		+ \sbraketa{\bm u_2}{\epsilon_{3\mu}^+ \g^\mu}{\bm u_1} \nonumber \\
	&= \abrakets{\bm2}{\epsilon_{3\mu}^+ \sigma^\mu}{\bm1}
		+ \sbraketa{\bm2}{\epsilon_{3\mu}^+ \bar \sigma^\mu}{\bm1},
}
where in the last step, we contracted all the Dirac indices using the standard gamma matrix $\g_{ab}^\mu$ and
\ealin{
	\aket{\bm u_p} = \mqty(\la^\a_I \\ 0), &\quad \sket{\bm u_p} = \mqty(0 \\ \lat_{\ad I}), \nonumber \\
	\abra{\bm u_p} = \mqty(\la_\a^I & 0), &\quad \sbra{\bm u_p} = \mqty(0 & \lat^{\ad I}). 
}
Converting from a Lorentz $4$-vector index $\mu$ to bi-spinor indices $\a \ad$ with the identities 
\ealin{
\epsilon_p^{+\nu} &= \frac12 \bar \sigma_{\ad\a}^\nu \epsilon_p^{+\a\ad}, \nonumber \\
\bar \sigma_{\ad\a}^\nu &= \vepsi_{\a\b} \vepsi_{\ad\bd} \sigma^{\nu \b \bd}, \nonumber \\
g_{\mu\nu} \sigma^{\mu\a\ad} \sigma^{\nu\b\bd} &= 2 \vepsi^{\a\b} \vepsi^{\ad\bd},
}
and writing out the polarizations using the spinor helicity notation \cite{Schwartz:2013}
\ben
\big[\epsilon_p^{+}(\chi)\big]^{\a\ad} = \sqrt2 \frac{\aket \chi \sbra p}{\abraket{\chi p}}
	\qq{and}
	\big[\epsilon_p^{+}(\chi)\big]_{\ad\a} = \sqrt2 \frac{\sket p \abra \chi}{\abraket{\chi p}}
\label{eq_polidentities}
\een
leads us back to the positive helicity minimal coupling
\ealin{
\hspace{-3mm}\abrakets{\bm2}{\epsilon_{3\mu}^+ \sigma^\mu}{\bm1}
		&+ \sbraketa{\bm2}{\epsilon_{3\mu}^+ \bar \sigma^\mu}{\bm1} \nonumber \\
	&\hspace{-3mm}= \abrakets{\bm2}{\epsilon_3^+}{\bm1} + \sbraketa{\bm2}{\epsilon_3^+}{\bm1} \nonumber\\
	&\hspace{-3mm}= \la_{2\a}^I \epsilon_3^{+\a\ad} \lat_{1\ad I} + \lat_2^{\ad I} \epsilon_{3\ad\a}^+ \la_{1 I}^\a \nonumber \\
	&\hspace{-3mm}= \sqrt2 \frac{\abraket{\bm2 \chi} \sbraket{3 \bm1}}{\abraket{\chi 3}} + \sqrt2 \frac{\sbraket{\bm2 3} \abraket{\chi \bm1}}{\abraket{\chi 3}} \nonumber \\
	&\hspace{-3mm}= \sqrt2 \frac{\abraket{\chi \bm1}\sbraket{\bm1 3} - \abraket{\chi \bm2}\sbraket{\bm2 3}}{\sqrt2 m \abraket{\chi 3}} \abraket{\bm1 \bm2} \nonumber\\
	&\hspace{-3mm}= - 2\sqrt2\, x \abraket{\bm1 \bm2}.
\label{eq_qftminangle}
}

Taken together, we see that the Dirac algebra in QFT is encoded in the massive couplings of \cref{eq_qed3full}.
Direct links from Dirac to Weyl $(\g^\mu_{ab} \rightarrow \sigma^{\mu\a\ad})$ and from Lorentz $4$-vectors to bi-spinors $(g_{\mu\nu} \rightarrow \vepsi^{\a\b}\vepsi^{\ad\bd})$ are evident in this calculation and suggest that the physics described by the QED interaction $\sum_s \bar u(p_2)^{-s} \slashed{\epsilon}_3^+ u(p_1)^s$ has a simpler formulation as $x \abraket{\bm1 \bm2}$.

\subsection{High-energy limit and quasi-limit expansions \label{sec_cross_2}}

We showed in \cref{fig_qed3HE+} how $x \abraket{\bm1 \bm2}$ coalesces from couplings at high energy with a positive helicity photon, i.e., where only $\sket 3$ appears and not $\aket 3$.
Our approach was to use conservation of angular momentum to determine the admissible spin configurations and use them to calculate the high-energy limit directly from the massless coupling in \cref{eq_massless3}.
Here, we discuss how to formally take such limits by returning to our definition of massive helicity spinors in \cref{eq_mhsdef,eq_helicity,eq_helicitytilde}, where each spin state is represented by a massless helicity spinor paired with a little group basis spinor $\za^{\pm}_I$.
We can determine the high-energy limit of any \emph{particular} spin state of a massive helicity spinor by keeping the term with the correct basis spinor and by taking the mass to zero via $p_z \rightarrow E$.
For example, consider $\abra{\bm \la} = \la_\a^I = \la_\a \za^{-I} + \eta_\a \za^{+I}$ and
$\sbra{\bm \la} = \lat^{\ad I} = \lat^\ad \za^{+I} + \etat^\ad \za^{-I}$.
The negative spin state $s=-\tfrac12$ of $\abra{\bm \la}$ is $\la_\a \za^{-I}$, and there are no issues letting $p_z \rightarrow E$ for $\la_\a$.
So, the high-energy limit of $\abra{\bm\la}$, when insisting on spin state $-\tfrac12$ is simply $\abra \la$, and we write
\ben
\lim_\mathrm{-HE} \abra{\bm \la} = \abra \la
\qq{or equivalently}
\abra{\bm\la} \hel[-] \abra\la.
\een
On the other hand, the positive spin state of $\abra{\bm\la}$ is $\eta_\a \za^{+I}$, but $\abra\eta$ will vanish as $p_z \rightarrow E$.
It is $\sbra{\bm \la}$ that has the non-vanishing positive spin $\lat^\ad \za^{+I}$ component at high energy, so we can exchange angle and square brackets using $\bm p_{\ad\a} \bm p^{\a\ad} = \sbraket{\bm\la \bm\la} \abraket{\bm\la \bm\la} = -2m^2$:
\ben
\lim_\mathrm{+HE} \abra{\bm \la} 
	= \lim_\mathrm{+HE} \frac{\sbraket{\bm\la \bm\la}}{-\sqrt2 m} \abra{\bm\la}
	= \lim_\mathrm{+HE} \sbra{\bm\la} \frac{\bm p_{\ad\a} \bm p^{\a\ad}}{-2 m^2}
	= \sbra \la.
\een
Therefore, the high-energy limit for a single massive helicity spinor is simple:
replace its boldface type with $\la$ for $s=-\tfrac12$ or $\lat$ for $s=+\tfrac12$ as follows
\begin{equation}
\aket{\bm\la},\, \sket{\bm\la} \hel[]
\begin{cases}
\aket{\la} & \text{for $s = -\tfrac12$} \\
\sket{\la} & \text{for $s = +\tfrac12$,}
\end{cases}
\label{eq_spinorhel}
\end{equation}
and similarly for the bras $\abra{\bm\la}$ and $\sbra{\bm\la}$.
For a given spin configuration, each massive particle will be assigned a particular spin state $s$.
In the case of a spin-$\tfrac12$ particle, the amplitude can only have one spinor appear for that particle in each term of the amplitude to indicate its spin as either $s=-\tfrac12$ with an angle bracket $\la$ or $+\tfrac12$ with a square bracket $\lat$.
For higher-spin particles, each term of the amplitude will have exactly $2s$ spinors to collectively represent the particular spin state $s$ as products of spinors.
For example, a spin-$1$ particle must have two spinors to collectively represent $s=-1$ as $\la \la$, $s=0$ as $\tfrac12 (\la \lat + \lat \la)$, and $s=+1$ as $\lat \lat$.
We see again that a particle's spin determines exactly how many spinors for that particle must appear in each term of any scattering amplitude.
The admissible spin configurations are thus determined by the amplitude and by its relation to the massless coupling in the high-energy limit.
We shall see that the high-energy limit is an effective tool for calculating cross sections because it allows us to naturally expand the sum in \cref{eq_crosssection} by breaking up the scattering amplitude into its spin configurations.

In order to apply this method to massive amplitudes, we shall take a small step \emph{backward} from the high-energy limit to restore mass to the helicity spinors that represent the interacting massive particles while retaining the expansion over spin configurations.
Of course, this backward step would be invalid for a \emph{free} particle with spin $s>0$, due to the categorical change in degrees of freedom, going from $2$ to $2s+1$ as it becomes massive.
However, the degrees of freedom of \emph{interacting} particles are constrained by angular momentum conservation and are accounted for when their massive helicity spinors are contracted along the little group basis $\{ \zan_I, \zap_I\}$.
We therefore define the high-energy \emph{quasi}-limit (qHE) as the result of the following procedure: first, take the high-energy limit to expand the scattering amplitude into its spin configurations, and then restore the boldface type of the corresponding massive helicity spinors.
The quasi-limit enables us to work with the massive amplitude just before it breaks up into its massless components.

We close by using these methods to reproduce the result of \cref{fig_qed3HE+}.
Expanding the $x$ factor with \cref{eq_xfactor} leads to
\ealin{
\Mcal^{+1}_{IJ} = g x \abraket{\bm1 \bm2}
	&= g \frac{\sbraketa{3}{(\bm p_1 - \bm p_2)}{\chi}}{2\sqrt2 m \abraket{3 \chi}} \abraket{\bm1 \bm2} \nonumber \\
	&= -\frac{g}{2} \frac{ \sbraket{3 \bm1} \abraket{\bm2 \chi} + \sbraket{3 \bm2} \abraket{\bm1 \chi} }{\abraket{3 \chi}}.
}
Taking the high-energy limit leads to
\ben
\Mcal^{+1}_{IJ} \hel[]
\begin{cases}
-g \frac{\sbraket{3 1} \abraket{2 \chi}}{\abraket{3 \chi}}
	&\qq*{for $(1^{+\sfrac{1}{2}},2^{-\sfrac{1}{2}}, 3^{+1})$,} \vspace{1mm} \\
+g \frac{\sbraket{3 2} \abraket{1 \phi}}{\abraket{3 \phi}}
	&\qq*{for $(1^{-\sfrac{1}{2}},2^{+\sfrac{1}{2}}, 3^{+1})$.}
\end{cases}
\een
The average of these two configurations should give the high-energy limit from \cref{fig_qed3HE+}, but we cannot make progress with massless spinors.
Thus, we take the quasi-limit of each configuration, choose references $\chi = \bm1$, $\phi=\bm2$, and average over configurations
\ealin{
\Mcal^{+1}_{IJ} &\hel[q] 
\left.
\begin{cases}
-g \frac{\sbraket{3 \bm1} \abraket{\bm 2 \bm1}}{\abraket{3 \bm1}}
	&\qq*{for $(1^{+\sfrac{1}{2}},2^{-\sfrac{1}{2}}, 3^{+1})$} \vspace{1mm} \\
+g \frac{\sbraket{3 \bm2} \abraket{\bm1 \bm2}}{\abraket{3 \bm2}}
	&\qq*{for $(1^{-\sfrac{1}{2}},2^{+\sfrac{1}{2}}, 3^{+1})$}
\end{cases}
\right. \nonumber \\
&\longrightarrow
-\frac{g}{2} \left( \frac{\sbraket{3 \bm1} \abraket{\bm 2 \bm1}}{\abraket{3 \bm1}} - \frac{\sbraket{3 \bm2} \abraket{\bm1 \bm2}}{\abraket{3 \bm2}} \right).
}
We can manipulate the brackets with $\abraket{\bm1 \bm2}\sbraket{\bm2 \bm1} = 2m^2$ and $\abraket{\bm1 3}\sbraket{3 \bm1} = 2 \bm p_1 \vdot p_3 = -2m^2$ or $\abraket{\bm2 3}\sbraket{3 \bm2} = 2 \bm p_2 \vdot p_3 = 2m^2$, where the minus sign between $\bm p_1 \vdot p_3$ and $\bm p_2 \vdot p_3$ is again
due to incoming line $\bm1$ and outgoing line $\bm2$.
We indeed recover the high-energy limit with $\gt = -g$:
\eali{
\Mcal^{+1}_{IJ} &\hel[q] -\frac{g}{2} \left( \frac{\sbraket{3 \bm1} \abraket{\bm 2 \bm1}}{\abraket{3 \bm1}} - \frac{\sbraket{3 \bm2} \abraket{\bm1 \bm2}}{\abraket{3 \bm2}} \right) \\
	&\hspace{3mm}= -\frac{g}{2} \left( \frac{\sbraket{3 \bm1}^2}{\sbraket{\bm1 \bm2}} \frac{2m^2}{2 \bm p_1 \vdot p_3} + \frac{\sbraket{3 \bm2}^2}{\sbraket{\bm1 \bm2}} \frac{2m^2}{2 \bm p_2 \vdot p_3} \right) \\
	&\hel \gt \frac{\sbraket{13}^2 + \sbraket{23}^2}{2 \sbraket{12}}.
}
In the following sections, we shall see how these limits can be used to calculate massive cross sections.

\subsection{Gluing partial cross sections \label{sec_cross_3}}

The modularity of spinor helicity amplitudes suggests an alternative method for calculating cross sections by gluing together smaller pieces, obtained from couplings, that we shall call \emph{partial cross sections}.
It should be analogous to \cref{eq_diracleft}, possessing bi-spinor Lorentz indices and nothing else.
We shall require a generalized form of the square of the couplings in \cref{eq_qed3}, where the Lorentz indices are symmetrized and where the little group indices of the massive spinors are fully contracted, such that only external momenta $\bm p_{\ad\a}$ remain.
The stripped coupling and its conjugate will provide the ``glue'' for adhesion
\ben
\left| M^h_{\a\b} \right|^2 = \left( M^h_{\a\b} \right) \left( M^h_{\a\b} \right)^\dagger
	= \left( M^h_{\a\b} \right) \left( M^{-h\ad\bd} \right).
\een
Minimal coupling in QED leads to a single term: $2m^2 g^2 \vepsi_{\a\b} \vepsi^{\ad\bd}$.
If we include the auxiliary coupling from \cref{eq_qed3full}, then we obtain additional terms that couple the electrons with the photon helicity spinors $\la^\a$ and $\lat_\ad$
\ealin{
\left| M^h_{\a\b} \right|^2 = &g^2 \left[ 2 m^2 \vepsi_{\a\b} \vepsi^{\ad\bd}
	- \frac{\sqrt2 m}{x} \vepsi_{\a\b} \big( \lat \lat \big)^{\ad\bd} \right. \nonumber \\
	&\left. + \sqrt2 m x \big( \la \la \big)_{\a\b} \vepsi^{\ad\bd}
	- \big( \la \la \big)_{\a\b} \big( \lat \lat \big)^{\ad\bd} \right].
}
The electromagnetic field strength $F^{\mu\nu}$ in the spinor helicity formalism is $F_{\a\b}^{\ad\bd} = \vepsi_{\a\b} \lat^\ad \lat^\bd + \la_\a \la_\b \vepsi^{\ad\bd}$, and we can see its direct role in the auxiliary terms.
Now, let us define the partial cross section as the square of the coupling
\ben
\sigma_{\a\b}^{\ad\bd} \equiv \left| \frac{\la_{1} \la_{2}}{\sqrt{2} m} M_{(\a\b)}^h \right|^2
	= \frac{\la_1 \la_2}{\sqrt{2} m} \left[ M^2 \right]_{(\a\b)}^{(\ad\bd)} \frac{\lat_2 \lat_1}{\sqrt{2} m}.
\een
Here, the Lorentz indices from the stripped amplitude are symmetrized
\ben
\left[ M^2 \right]_{(\a\b)}^{(\ad\bd)} = 2 m^2 g^2 \vepsi_{(\a\b)} \vepsi^{(\ad\bd)}
= m^2g^2 \left[ \vepsi_{\a\b} \vepsi^{\ad\bd} + \vepsi_{\b\a} \vepsi^{\bd\ad} \right].
\een
Also, the spinor indices for massive particles $a$ and $b$ are suppressed and will be contracted in all three possible ways: intra-particle to yield the original momenta $\lat_{a \ad I} \la_{a \a}^I = \bm p_{a \ad \a}$, inter-particle to yield the inner products $\bm p_{a \dot\delta \delta} \bm p_b^{\delta \dot\delta} = \bm p_a \vdot \bm p_b$, and to yield mass terms $\tfrac{1}{\sqrt2 m} \lat_a^{\dot\delta I} \bm p_{a \dot\delta \delta} \la_{aI}^\delta = 2m^2$.
Putting everything together and averaging over these $2 \times 3 = 6$ possibilities yields the QED partial cross section at minimal coupling
\ealin{
\sigma_{\a\b}^{\ad\bd}
	&= \frac{g^2}{6} \left[ \bm p_{1\ad\a} \bm p_{2\bd\b} 
		+ \bm p_{2\ad\a} \bm p_{1\bd\b} \right. \nonumber \\ 
		&\hspace{3mm}\left.- \big( \vepsi_{\a\b} \vepsi^{\ad\bd} 
		+ \vepsi_{\b\a} \vepsi^{\bd\ad} \big) \big( \bm p_1 \vdot \bm p_2 + 2m^2 \big) \right] \nonumber \\
	&= \frac{g^2}{3} \left[ \bm p_{1(\a}^{(\ad} \bm p_{2 \b)}^{\bd)} 
	- \vepsi_{\a\b} \vepsi^{\ad\bd} \big( \bm p_1 \vdot \bm p_2 + 2m^2 \big) \right].
\label{eq_partialsectionmin}
}
With the relations $\bm p_{a \ad\a} \leftrightarrow p_a^\mu$ and $\vepsi_{\a\b} \vepsi^{\ad\bd} \leftrightarrow g^{\mu\nu}$, we see that the partial cross section is analogous to the left-hand side of the cross section $p_1^{(\mu} p_2^{\nu)} - g^{\mu\nu} \big( p_1^\rho p_2^\rho + 4m^2 \big)$ that was calculated in \cref{eq_diracleft}.
Consequently, the cross section for Bhabha annihilation at minimal coupling is the gluing of two copies of \cref{eq_partialsectionmin}
\ben
\sigma_\mathrm{B} = \sigma_{12\a\b}^{\ad\bd} \frac{1}{s^2} \sigma_{34\ad\bd}^{\a\b}.
\een 
The partial cross section for particles of general spin $s$ is derived in \cref{app_partialsection} using the above combinatorial arguments.
In the general case, there are $4s+1$ ways to distribute the massive spinor indices, and we find
\begin{widetext}
\ben
\sigma_{\a_i \b_i}^{\ad_i \bd_i}  = \frac{g^2}{4s+1} \sum_{n=0}^{2s} \frac{(-1)^{n+1}}{2}
	\big( \bm p_1 \bm p_2\big)^{2s-n\, (\ad_i \bd_i}_{\quad\quad(\a_i \b_i} \big( \vepsi \tilde{\vepsi} \big)^{n\, \ad_j \bd_j)}_{\;\;\a_j \b_j)} \left[ (\bm p_1 \vdot \bm p_2)^n + m^{2n} \right].
\label{eq_partialsectiongen}
\een
\end{widetext}
In order to contract everything, we have symmetrized indices with labels $1\le i < j \le s$, as well as $2s-n$ copies of the momenta $\bm p_1$ and $\bm p_2$, and finally $n$ copies of the Levi--Civita symbols.
This approach for calculating cross sections bypasses scattering amplitudes altogether and may be suitable for computational methods.
We shall not pursue this method further here, however, and will instead return to our study of scattering amplitudes and their high-energy behavior.

\subsection{Bhabha cross section \label{sec_cross_4}}

Let us now calculate the cross section for Bhabha annihilation of \cref{fig_bhabha}.
The scattering amplitude in \cref{eq_minampbhabha} written in terms of the quotient $x_{12}/x_{34}$ has a simple conjugate due to the property $\overline x = - 1/x$.
We obtain
\ealin{
\Mcal^\dagger 
	&= \frac{2e^2}{s} \left( \frac{\overline x_{12}}{\overline x_{34}} \overline{\abraket{\bm1 \bm2} \sbraket{\bm3 \bm4}\,} 
	+ \frac{\overline x_{34}}{\overline x_{12}} \overline{\abraket{\bm3 \bm4} \sbraket{\bm1 \bm2}\,} \right) \nonumber \\
	&= \frac{2e^2}{s} \left( \frac{x_{34}}{x_{12}} \abraket{\bm4 \bm3} \sbraket{\bm2 \bm1}\, 
	+ \frac{x_{12}}{x_{34}} \abraket{\bm2 \bm1} \sbraket{\bm4 \bm3}\, \right).
}
Na\"\i vely expanding the square of the amplitude yields
\ealin{
|\Mcal|^2 &= \frac{4 e^4}{s^2} \bigg\{ 
	2s^2 + \abraket{\bm1 \bm2}^2 \sbraket{\bm3 \bm4}^2 \bigg[ \left( \frac{x_{12}}{x_{34}} \right)^2 + \left( \frac{x_{34}}{x_{12}} \right)^2 \bigg] \bigg\}.
\label{eq_bhabhasectionnaive}
}
Interesting, perhaps, in its own right but bearing little resemblance to the known formula in \cref{eq_bhabhacrosssection}, which is written entirely in terms of the external momenta of the process.
The problem is that the na\"\i ve approach bypasses the expansion over spin configurations and the subsequent contractions with the conjugate amplitude.
Let us instead consider the high-energy limit of $\Mcal$ in \cref{eq_minampbhabha}.
The massive spinors in the first term represent four possible spin configurations
\ben
\frac{x_{12}}{x_{34}} \abraket{\bm1 \bm2} \sbraket{\bm3 \bm4} \hel[]
\begin{cases}
	\abraket{\eta_1 2} \sbraket{3 \eta_4} \rightarrow \sbraket{13}\abraket{24} 
		&\qq*{for $1^+2^-3^+4^-$} \\
	\abraket{\eta_1 2} \sbraket{\eta_3 4} \rightarrow \sbraket{14}\abraket{32} 
		&\qq*{for $1^+2^-3^-4^+$} \\
	\abraket{1 \eta_2} \sbraket{3 \eta_4} \rightarrow \abraket{14}\sbraket{32} 
		&\qq*{for $1^-2^+3^+4^-$} \\
	\abraket{1 \eta_2} \sbraket{\eta_3 4} \rightarrow \abraket{13}\sbraket{24} 
		&\qq*{for $1^-2^+3^-4^+$}
\end{cases}
\een
noting that $x \hel 1$ because its numerator and denominator are both proportional to $m$, as seen in \cref{eq_xquotientbhabha}.
Averaging over these configurations as well as those corresponding to the second term in \cref{eq_minampbhabha} yields the high-energy limit of the full amplitude
\ben
\Mcal \hel \frac{e^2}{s} \Big( \underset{a}{\abraket{14}\sbraket{32}\,} 
	+ \underset{b}{\abraket{13}\sbraket{24}\,}
	+ \underset{\bar b}{\abraket{24}\sbraket{13}\,} + \underset{\bar a}{\abraket{32}\sbraket{14}\,} \Big),
\een
labeling the terms $a$, $b$, $\bar b$, and $\bar a$.
The high-energy limit of $\Mcal^\dagger$ is simply $\tfrac{e^2}{s} (\bar a + \bar b + b + a)$.
Sixteen terms are obtained by squaring the amplitude, but $a^2 = b^2 = \bar a^2 = \bar b^2 = 0$, since $\abraket{ii} = \sbraket{ii} = 0$ for lightlike $p_i^\mu$, $i \in \{1,2,3,4 \}$.
Furthermore, $ab = \bar a b = a \bar b = \overline{ab} = 0$, since $\abraket{ij}^2 = \sbraket{ij}^2 = 0$ for all $i \neq j$.
What remains is $2(a \bar a + b \bar b)$, leading to the correct result at high energy
\ealin{
\sigma_\mathrm{B} &\hel \frac{2e^4}{s^2} \Big( \abraket{14}\sbraket{41} \abraket{23} \sbraket{32} + \abraket{13}\sbraket{31} \abraket{24} \sbraket{42} \Big) \nonumber \\
	&= \frac{2e^4}{s^2} \Big( t^2 + u^2 \Big).
\label{eq_bhabhacrossmassless}
}
Following the above expansion, we take the quasi-limit, restoring mass to all spinors.
Now all $16$ terms survive
\ealin{
\sigma_\mathrm{B} \hel[q] &\frac{2e^4}{s^2} \Big( \bm a^2 + \bar{\bm a}^2 + \bm b^2 + \bar{\bm b}^2 \nonumber \\
&+ 2 \big( \bm{ab} + \bm{ a \bar b} + \bar{\bm a} \bm b + \overline{\bm{ab}} + \bm a \bar{\bm a} + \bm b \bar{\bm b} \big) \Big),
}
where boldfaced labels refer to the massive spinors they represent.
Each of the four square-terms contributes $4m^4$.
For example, $\bm a^2 =  \abraket{\bm1 \bm4}^2 \sbraket{\bm3 \bm2}^2 = \abraket{\bm1 \bm1} \abraket{\bm4 \bm4} \sbraket{\bm3 \bm3} \sbraket{\bm2 \bm2}\, = 4m^2$.
The $\bm a \bar{\bm a}$ and $\bm b \bar{\bm b}$ terms lead to the $t$- and $u$-channel invariants, namely,
\ealin{
\bm a \bar{\bm a} &= \abraket{\bm1 \bm4} \sbraket{\bm4 \bm1} \abraket{\bm2 \bm3} \sbraket{\bm3 \bm2}\, = \left(t - 2m^2 \right)^2, \\
\bm b \bar{\bm b} &= \abraket{\bm1 \bm3} \sbraket{\bm3 \bm1} \abraket{\bm2 \bm4} \sbraket{\bm4 \bm2}\, = \left(u - 2m^2 \right)^2.
}
The cross terms provide the $s$ invariant
\ealin{
\bm{ab} 
&= \abraket{\bm1 \bm4} \sbraket{\bm3 \bm2} \abraket{\bm3 \bm1} \sbraket{\bm4 \bm2}\, 
= - \abrakets{\bm1}{\bm p_3}{\bm2} \abrakets{\bm1}{\bm p_4}{\bm2} \nonumber \\
&= - \frac12 \abraket{\bm{11}} \sbraket{\bm{22}}\, \bm p_3 \vdot \bm p_4 
= m^2 \big( s - 2m^2 \big),
}
and similarly for $\bm{ a \bar b}$, $\bar{\bm a} \bm b$, and $\overline{\bm{ab}}$.
Taken together, we find
\ealin{
\sigma_\mathrm{B} &\hel[q] \frac{2e^4}{s^2} \Big[ 8m^4 + 4m^2 \big( s - 2m^2 \big) +  \big( t - 2m^2\big)^2 + \big( u - 2m^2\big)^2 \Big] \nonumber \\
&= \frac{2e^4}{s^2} \Big[ 8m^4 + 8m^2 s - 4m^2 \big( s + t + u \big) + t^2 + u^2 \Big] \nonumber \\
&= \frac{2e^4}{s^2} \Big[ t^2 + u^2 + 8m^2 s - 8m^4 \Big],
\label{eq_bhabhacrossmassive}
}
which is the correct result with the Mandelstam invariants $s + t + u = 4m^2$.

\subsection{Compton cross section \label{sec_cross_5}}

Now we calculate the cross section for Compton scattering shown in \cref{fig_compton}.
The scattering amplitude in \cref{eq_minampcompton} was obtained by contracting left and right minimal couplings given by \cref{eq_minampcompton_temp}
\ben
\Lcal^{-1} = - \frac{\gt}{x_{12}} \sbraket{\bm k \bm1} 
\qq{and} 
\Rcal^{+1} = g x_{34} \abraket{\bm4 \bm k}. 
\een
Applying the result in \cref{eq_qftminangle} as well as its negative helicity counterpart, we have
\ealin{
x_{34} \abraket{\bm4 \bm k} 
	&= \frac12 \left( \abrakets{\bm4}{\epsilon^+_{3\mu} \sigma^\mu}{\bm k}\, + \sbraketa{\bm4}{\epsilon^+_{3\mu} \sigma^\mu}{\bm k}\, \right), \\
\frac{1}{x_{12}} \sbraket{\bm1 \bm k} 
	&= \frac12 \left( \Big[ \bm1 \Big| \epsilon^{-*}_{2\nu} \sigma^\nu \Big| \bm k \Big> + \Big< \bm1 \Big| \epsilon^{-*}_{2\nu} \sigma^\nu \Big| \bm k \Big] \right).
}
Combining both pieces, summing over polarizations $\ii g_{\mu\nu} = \sum_{i = \pm1} \epsilon^i_\mu \epsilon^{i*}_\nu$, and taking the high-energy limit
\ealin{
\frac{x_{34}}{x_{12}} \abraket{\bm4 \bm k} \sbraket{\bm1 \bm k} 
	&= \frac12 \Big( \abrakets{\bm4}{\sigma^\mu}{\bm k}\, g_{\mu\nu} \sbraketa{\bm1}{\sigma^\nu}{\bm k} \nonumber \\
	&\hspace{3mm}+  \sbraketa{\bm4}{\sigma^\mu}{\bm k}\, g_{\mu\nu} \abrakets{\bm1}{\sigma^\nu}{\bm k}\, \Big) \nonumber \\
	&\hel \abraket{4k}\sbraket{1k}\, + \sbraket{4k}\abraket{1k},
}
where the identity in \cref{eq_polidentities} for $g_{\mu\nu} \sigma^\mu \sigma^\nu$ was used in the last step.

It is interesting that the high energy electrons have opposite spin in both configurations.
Since the photons are external with fixed helicities $2^-$ and $3^+$, the first term corresponds to the $s$ channel, thus coupling lines $1$ with $2$ and $3$ with $4$.
The second term corresponds to the $u$ channel, swapping lines $2 \leftrightarrow 3$.
Writing the scattering amplitude at high energy as
\ben
\Mcal = \Mcal_s + \Mcal_u \hel g\gt \left( \frac{\abraket{4k} \sbraket{1k}}{s} + \frac{\abraket{1k} \sbraket{4k}}{u} \right),
\een
we immediately see that $\Mcal_s \Mcal_u^\dagger \hel 0$ and $\Mcal_s^\dagger \Mcal_u \hel 0$ due to the contraction of identical massless spinors.
Therefore, the high-energy limit does not have cross terms with $1/su$, and the \emph{unpolarized} cross section reduces to the simple form known in the literature
\ealin{
\sigma_\mathrm{C} 
	&\hel \frac{1}{2} \sum_{h = \pm1} \left( \left| \Mcal^h_s \right|^2 + \left| \Mcal^h_u \right|^2 \right) \nonumber \\
	& = \frac{g^2 \gt^2}{2} \left( \frac{1}{s^2} + \frac{1}{u^2} \right) \abraket{4k}\sbraket{1k}\abraket{1k}\sbraket{4k} \nonumber \\
	&= 2e^4 \left( \frac{1}{s^2} + \frac{1}{u^2} \right) \abrakets{4}{(p_1 + p_2)}{1} \abrakets{1}{(p_2 + p_4)}{4} \nonumber \\
	&= 2e^4 \left( \frac{1}{s^2} + \frac{1}{u^2} \right) \abraket{42}\sbraket{21}\abraket{12}\sbraket{24} \nonumber \\
	&= 2e^4 \left( \frac{u}{s} + \frac{s}{u} \right)
\label{eq_comptoncrossmassless}
}
with eight spin configurations in total, four per external photon helicity $2^\mp$ and $3^\pm$.

To get the massive cross section, we take the quasi-limit, where the cross terms now survive
\ben
| \Mcal |^2 \hel[q] \left| \Mcal_s \right|^2 + \left| \Mcal_u \right|^2 + \Mcal_s \Mcal_u^\dagger + \Mcal_s^\dagger \Mcal_u.
\een
Starting with the first term, we have
\ben
\left| \Mcal_s \right|^2 = -g^2 \gt^2 \frac{ \abraket{ \bm{4k} } \sbraket{ \bm{1k} } \abraket{ \bm{k1} } \sbraket{ \bm{k4}} }{\big( s-m^2 \big)^2},
\een
and the $s$ channel momentum $\bm k^\mu = \bm p_1^\mu + p_2^\mu$ leads to the numerator
\ealin{
\abraket{ \bm{4k} } &\sbraket{ \bm{1k} } \abraket{ \bm{k1} } \sbraket{ \bm{k4}} \nonumber \\
	&= \Big( \abraket{\bm{44}} \sbraket{\bm{41}}\, + \abraket{\bm{42}} \sbraket{\bm{21}}\,  \Big) \Big( \abraket{\bm{12}} \sbraket{\bm{24}}\, + \abraket{\bm{14}} \sbraket{\bm{44}}\, \Big) \nonumber \\
	&= (s-m^2)(u-m^2) - 2m^2 (t - 2m^2) + 4m^2 \bm p_1 \vdot p_2 \nonumber \\
	&= (s-m^2)(u-m^2) - 4m^2 (s-m^2) + 4 m^4
}
using $t=0$ and $\bm p_1 \vdot p_2 = s - m^2$.
Similarly, the second term with $u$ channel momentum $\bm k^\mu = p_2^\mu + \bm p_4^\mu$ has the same numerator after swapping $\bm p_1 \leftrightarrow \bm p_4$ and $s \leftrightarrow u$:
\ben
\abraket{ \bm{1k} } \sbraket{ \bm{4k} } \abraket{ \bm{k4} } \sbraket{ \bm{k1}}
= (s-m^2)(u-m^2) - 4m^2 (u-m^2) + 4 m^4.
\een
The cross terms use both $s$- and $u$-channel momenta and have the form
\ben
\Mcal_s \Mcal_u^\dagger = g^2 \gt^2 \frac{ \abrakets{\bm4}{(\bm p_1 + p_2)}{\bm1} }{s-m^2} \left( \frac{ \sbraketa{\bm4}{(p_2 + \bm p_4)}{\bm1} }{u-m^2} \right)^\dagger
\een
with numerator
\ealin{
\abrakets{\bm4}{(\bm p_1+p_2)}{\bm1} &\abrakets{\bm1}{(p_2+\bm p_4)}{\bm4} \nonumber \\
	&\hspace{-15mm}= \big( \bm p_1^2 + \bm p_1 \vdot p_2 \big) \big( \bm p_4^2 + p_2 \vdot \bm p_4\big) \nonumber \\
	&\hspace{-15mm}= 4m^4 - 2m^2 \big( \bm p_1 \vdot p_2 + p_2 \vdot \bm p_4 \big) + \bm p_1 \vdot p_2^2 \vdot \bm p_4 \nonumber \\
	&\hspace{-15mm}= 4m^4 - 2m^2 \big( s + u - 2m^2 \big) \nonumber \\
	&\hspace{-15mm}= 4m^4.
}
Note that lightlike $p_2^2 = 0$ and that the second term in the penultimate line vanishes due to the Mandelstam invariants $s + t + u = 2m^2$.
This result implies $\Mcal_s^\dagger \Mcal_u = \big( \Mcal_s \Mcal_u^\dagger \big)^\dagger = 4m^4$.

Gathering everything leads to
\ealin{
\sigma_\mathrm{C} 
	&\hel[q] \frac{g^2 \gt^2}{2} \Bigg[ \frac{(s-m^2) (u-m^2) - 4m^2 (u-m^2) + 4m^4}{\big( s-m^2 \big)^2} \nonumber \\
	&\hspace{1cm}+ \frac{(s-m^2) (u-m^2) - 4m^2 (s-m^2) + 4m^4}{\big( u-m^2 \big)^2}  \nonumber \\
	&\hspace{1cm}+ \frac{8m^4}{(s-m^2)(u-m^2)} \Bigg] \nonumber \\
	&\hel 2e^4 \left( \frac{u}{s} + \frac{s}{u} \right),
}
consistent with \cref{eq_comptoncrossmassless} when $m \rightarrow 0$.
The quasi-limit simplifies to the well-known form in the literature, upon using the Mandelstam invariants $u-m^2 = - (s-m^2)$:
\begin{widetext}
\ealin{
\sigma_\mathrm{C} 
	&\hel[q] 2e^4 \Bigg[ \frac{u-m^2}{s-m^2} + \frac{s-m^2}{u-m^2} - 4m^2 \left( \frac{u-m^2}{\big( s-m^2\big)^2} + \frac{s-m^2}{\big( u-m^2\big)^2} \right) + 4m^4 \left( \frac{1}{\big( s-m^2\big)^2} + \frac{1}{\big( u-m^2\big)^2} \right) + \frac{8m^4}{(s-m^2)(u-m^2)} \Bigg] \nonumber \\
	&\hspace{3mm}= 2e^4 \Bigg[ \frac{u-m^2}{s-m^2} + \frac{s-m^2}{u-m^2} + 4m^2 \left( \frac{1}{s-m^2} + \frac{1}{u-m^2} \right) + 4m^4 \left( \frac{1}{s-m^2} + \frac{1}{u-m^2} \right)^2 \Bigg].
\label{eq_comptoncrossmassive}
}
\end{widetext}

Performing this calculation in QFT is lengthier and more convoluted due to the redundancy built into the Dirac spinor embeddings in $4$-vector fields.
It is striking that spinor helicity amplitudes like \cref{eq_minampbhabha,eq_minampcompton_temp} expand to reproduce the massive cross sections of \cref{eq_bhabhacrossmassive,eq_comptoncrossmassive}.
Expressed in terms of Mandelstam invariants, these cross sections are not tied to any particular formalism and thus provide a fair and direct way of comparing massive helicity spinors and quantum fields.

Now that we have performed several non-trivial checks of the massive spinor helicity formalism and have observed the computational benefits it offers, we turn to the examination of new insights it may reveal about the nature of scattering and spacetime.

\section{Mass acquisition \label{sec_mass}}

We have extended the spinor helicity formalism by including representations for massive particles and have calculated the scattering amplitude and cross section of well known processes.
It seems clear at this point that although the formulation and methods used here differ in significant ways from QFT, the standard results in the literature can nevertheless be reproduced with relative ease.
Our departure from a manifestly local theory is, perhaps, the most important difference with QFT.
Locality is a cornerstone of the Standard Model---upon which the central ideas of spontaneous symmetry breaking and effective field theory rest.
Consequently, let us proceed by examining the coalescence of massless helicity amplitudes and its perspective on how a particle can acquire mass as well as a localized trajectory through spacetime.

\subsection{Coalescence and the Higgs mechanism \label{sec_mass_1}}

The idea for an on-shell analog for the Higgs mechanism traces back to Conde and Marzolla, who suggested that the helicity spinors representing the reference null vector for a massive particle become the polarization vectors associated with a massless particle at high energy \cite{Conde:2016vxs}.
Arkani-Hamed \emph{et al.}~took this further by showing that the massive amplitude is the common low-energy limit of all possible external spin configurations of the corresponding massless amplitude \cite{Arkani-Hamed:2021a}.
Their introductory example shows the essence of the mechanism.
We can view a spin-$1$ particle with mass $m$ interacting with a massive Higgs scalar particle with mass $\sqrt{2} m$ as the coalescence of the two possible spin configurations between one massless spin-$1$ particle and two massless scalars.
The process has the high-energy limit $(\bm{1}^1,\, \bm{2}^1, \bm 3^0) \hel (1^h,\, 2^0, 3^0)$.
The diagram below shows how two high-energy coupling coalesce into a single massive one at low energy

\begin{widetext}
\begin{fmffile}{higgs}
\fmfcmd{%
  style_def wiggly_arrow expr p =
  cdraw (wiggly p);
  shrink (1.4);
  cfill (arrow p);
  endshrink;
  enddef;}
\begin{equation}
\hspace{-35mm}
g \frac{\abraket{\bm1\bm2} \sbraket{\bm2\bm1}}{m} = \quad
\begin{gathered}
\begin{fmfgraph*}(100,80)
	\fmfpen{thick}
        \fmfleft{i1,i2}
        \fmfright{o1}

        \fmf{wiggly_arrow, tension=1}{i1,v1,i2}
        \fmf{scalar, tension=1}{v1,o1}

        \fmflabel{$\bm{1}^1$}{i1}
        \fmflabel{$\bm{2}^1$}{i2}
        \fmflabel{$\bm 3^0$}{o1}        

        \fmfv{decor.shape=circle, decor.filled=shaded, decor.size=4mm}{v1}

\end{fmfgraph*}
\end{gathered}
\quad\quad
\hel
\quad
\left\{
\quad \quad
\begin{gathered}
\begin{fmfgraph*}(60,60)
	\fmfpen{thick}
        \fmfleft{i1,i2}
        \fmfright{o1}

        \fmf{photon, tension=1}{i1,v1} 
        \fmf{dashes, tension=1}{i2,v1,o1}

        \fmflabel{$1^{+1}$}{i1}
        \fmflabel{$2^0$}{i2}
        \fmflabel{$3^0 \;\; = \gt \frac{\sbraket{12}\,\sbraket{31}}{\sbraket{23}}$}{o1} 

        \fmfv{decor.shape=circle, decor.filled=empty, decor.size=4mm}{v1}
\end{fmfgraph*} \\ \vspace{2cm} \\
\begin{fmfgraph*}(60,60)
	\fmfpen{thick}
        \fmfleft{i1,i2}
        \fmfright{o1}

        \fmf{photon, tension=1}{i1,v1} 
        \fmf{dashes, tension=1}{i2,v1,o1}

        \fmflabel{$1^{-1}$}{i1}
        \fmflabel{$2^0$}{i2}
        \fmflabel{$3^0 \;\; = g \frac{\abraket{12}\,\abraket{31}}{\abraket{23}}$}{o1} 

        \fmfv{decor.shape=circle, decor.filled=full, decor.size=4mm}{v1}
\end{fmfgraph*}
\end{gathered}
\right.
\label{fig_higgs}
\end{equation}
\end{fmffile}
\end{widetext}

Indeed, this merging of high-energy amplitudes is what we have been calling coalescence and is what was crucial for the calculation of unpolarized cross sections performed earlier in this paper.
The general idea is that the two high-energy spin configurations of line $1^h$ and line $2^0$ with a total of $2 + 1$ spin degrees of freedom coalesce at low-energy to represent a single massive spin $1$ particle on lines $\bm{1}^1$ and $\bm{2}^1$ with three spin degrees of freedom.
Thus, one can interpret the massless scalar line $2^0$ as the Goldstone boson $\pi$.
The remaining massive scalar line $\bm{3}^0$ plays a similar role to the Higgs boson $\sigma$ that gets decoupled from the massive spin $1$ particle $A_\mu$ in the nonlinear sigma model \cite{Schwartz:2013}.

This interesting reformulation of mass acquisition presents new opportunities and also raises important questions.
The full particle spectrum after spontaneous symmetry breaking in QFT is tricky to interpret in the Abelian case and nigh intractable in non-Abelian and beyond-Standard-Model cases.
This is due to the variety of kinetic mixing terms that arise when expanding the Lagrangian density around the different vacua associated with the broken symmetry.
For this reason, the decoupling limit that yields the nonlinear sigma model is a reasonable simplification which trades off information about $\sigma$ in order to better understand the relationship between $A_\mu$ and $\pi$.
Indeed, the CCWZ method (named after Callan, Coleman, Wess, and Zumino) of deducing $\pi$ couplings in effective field theory from unbroken symmetry constraints in the full theory are accompanied by nonlinear transformations of $\pi$ that complicate the nature between phenomena at the effective and the fundamental energy scales \cite{Coleman:1969,Callan:1969}.
In stark contrast, without any Lagrangian formulation to deal with, these issues seem to be absent in the spinor helicity formalism, whose interactions are always accessible and in principle reducible to a conjunction of known couplings.
Therefore, perhaps by analyzing mass acquisition via processes like \cref{fig_higgs} more can be learned about effective field theory in general and Higgs physics in particular.

\subsection{Worldline representation \label{sec_mass_2}}

The bi-spinor representation of complexified momentum provides us with a more general setting to formulate scattering amplitudes by gluing together couplings.
We learned that couplings and amplitudes constructed in this way are generally simpler than their QFT counterparts, especially when massless particles are involved, but this simplicity comes at the price of giving up locality.
Let us dig deeper---before any interactions take place---by thinking about the worldline of a free particle moving through complexified Minkowski space $\CMbb$.

\begin{figure}[t!]
\begin{center}
\includegraphics[width=0.35\textwidth]{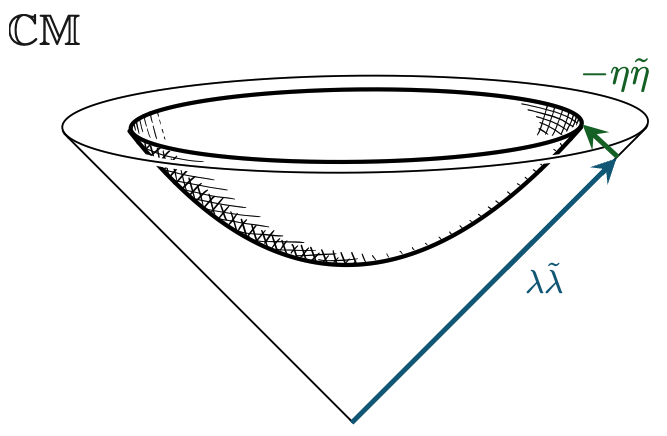}
\caption{Future lightcone in complexified Minkowski space $\CMbb$.
The null surface of the cone is parameterized by helicity spinors $\la$ and $\lat$.
Its corresponding hyperbolic surface (shaded) in the interior of the cone is mapped with two additional helicity spinors $\eta$ and $\etat$.}
\label{fig_lightcone}
\end{center}
\end{figure}

For a massless spin $s$ particle, we expect its linear momentum to be null; thus, the matrix $p^{\a\ad}$ will be singular with vanishing determinant $\det p^{\a\ad} = \det p_{\ad\a} = 0$ and zero mass $p^2 = m^2 = 0$.
The particle's worldline is therefore confined to lie on the null surface of the lightcone with null momentum given by the outer product $p^{\a\ad} = \la^\a \lat^\ad$ and little group $\U(1)$ parameterized by the number $z \in \Cbb$ in \cref{eq_weyl}.
Thinking of this lightcone surface at future time $t > 0$ as the celestial complex $2$-sphere, the required two complex degrees of freedom are provided by the Weyl spinors $\la^\a = \smqty(a \\ b)$ and $\lat_\ad = \smqty(\tilde a \\ \tilde b)$ constrained by $\det p^{\a\ad} = 0$.
A horizontal section of the future lightcone appears as a circle in \cref{fig_lightcone} instead of a sphere due to the suppression of one spatial dimension.
Thus, any point on the celestial sphere is identified by the helicity spinors $\la^\a$ and $\lat_\ad$, illustrated as a $4$-vector pointing to this circle.
Generally, we shall interpret lines in this and similar figures as $2$-surfaces of spacetime, and surfaces in such figures as volumes of spacetime.

The massive spin $s$ particle with a timelike momentum $p^2 = m^2 > 0$ and full-rank matrix $\det \bm p^{\a\ad} = m^2$ is defined by introducing two additional Weyl spinors $\eta^\a$ and $\etat_\ad$ according to \cref{eq_mhsdef}, which scale with mass $m$, so that $\bm p^{\a\ad} = \la^\a \lat^\ad - \eta^\a \etat^\ad \hel p^{\a\ad} = \la^\a \lat^\ad$.
In line with Conde and Marzolla's suggestion \cite{Conde:2016vxs}, we can think of $\la^\a \lat^\ad$ as the reference null vector directed toward a point on the celestial sphere that is parallel to $\bm p^{\a\ad}$.
In the high-energy limit $\eta^\a \etat^\ad$ vanishes with $m \rightarrow 0$, and $\la^\a \lat^\ad$ corresponds to null momentum $p^\mu$, which is perpendicular to the massless particle's polarization vectors $\epsilon_1^\mu$ and $\epsilon_2^\mu$.
At low energies, the survival of $\eta^\a \etat^\ad$ categorically increases the particle's degrees of freedom such that $p^\mu$ becomes timelike with $p^2 = m^2 > 0$, and the little group becomes $\SU(2)$ with $2s+1$ degrees of freedom.
The massive particle's worldline has thus been peeled off from the null surface of the lightcone. 
At each moment in time $t > 0$, it will intersect the corresponding hyperbolic $3$-surface inside the lightcone, as illustrated in \cref{fig_lightcone}.

From this kinematical perspective, the acquisition of mass indicates the excursion of the worldline inside the lightcone;
a worldline whose depth beneath the null surface is parameterized by the bi-spinor $\eta^\a \etat^\ad$, which scales with $m$ and is orthogonal to $\la^\a \lat^\ad$ as can be seen from \cref{eq_helicity,eq_helicitytilde}.
In other words, a massive particle is \emph{localized} within the lightcone's interior.
This point of view complements the dynamical aspects of the scattering amplitudes we studied earlier, which showed that amplitudes constructed entirely from massless couplings are manifestly non-local, those involving massive particles as well as residual $x$ factors still possess some non-local features, and only when all external particles become massive and all $x$ factors cancel does locality manifest itself.
This relationship between locality and mass is intriguing because it has always been obfuscated in QFT, where we imposed locality onto our quantum fields so that the breaking of their gauge symmetries could underlie the Higgs mechanism.
Further exploration of the interplay between mass and locality might enrich our understanding about particle physics and spacetime.
We pursue this in the next and final section by subsuming the spinor helicity formalism within twistor theory, whose intrinsically non-local projective geometry can account for both mass acquisition and the emergence of spacetime.

\section{Massive helicity spinors in twistor theory \label{sec_twistor}}

The spinor helicity formalism suggests that locality is not a fundamental property of the universe but one that emerges from the coalescence of particle interactions at low energy.
Despite their efficiency, helicity spinors lack a clear geometrical picture of the particles they represent.
Indeed, these complex-valued objects do not exist in our real spacetime, yet they do encode $4$-momentum, and they do transform correctly under the Poincar\'e group.
Sixty years ago, Roger Penrose proposed a new kind of algebra with a similar character, one which describes Minkowski space and its covariant Poincar\'e operations based on an abstract and by definition non-local geometry called projective twistor space \cite{Penrose:1967}.
Accordingly, we briefly review twistor theory and demonstrate how the ideas discussed in the previous sections---helicity spinors, complexified momentum, mass acquisition, and the emergence of a local spacetime---are all encapsulated within Penrose's theory.

\subsection{Twistors and the null rays of spacetime \label{sec_twistor_1}}

We begin by showing how null rays forming the lightcones of Minkowski space $\Mbb$ emerge from twistor theory.
The notation set by Penrose and Rindler \cite{Penrose:1986} is used here with slight modifications to suit our purpose.
A general introduction to the theory and its motivations is also given by Penrose \cite{Penrose:2004}.

The twistor $\Zsf^a$ is an object whose four components indexed by $a = 0, 1, 2, 3$ are coordinates into a complex-valued vector space called twistor space $\Tbb$.
By considering the three ratios of these coordinates, we obtain new (homogeneous) coordinates for a three-dimensional projection of $\Tbb$ called \emph{projective twistor space} $\PTbb$.
Its significance is that each \emph{point} in $\PTbb$ can be made to correspond to a \emph{ray} in spacetime; thus, a twistor variable in this projective space can represent the entire worldline of a particle or even a null ray of spacetime itself.
This is achieved in a remarkable way for our Lorentzian $4$-real-dimensional spacetime because in this particular case the celestial sphere is a conformal manifold whose homogeneous coordinates have the topology of the complex-valued Riemann sphere.
In this sense, Minkowski space $\Mbb$ is naturally generalized by twistor theory to a complex Minkowski space $\CMbb$, whose null rays and worldlines---and anything built from them---are analytic functions with a direct physical interpretation.

The key distinction between twistors representing worldlines and those representing null rays is that the former should encode spin (by ``twisting'') while the latter should not.
As we have seen throughout this paper, both the linear and angular momentum of a particle can be efficiently encoded as spinor helicity variables.
This fact is leveraged in twistor theory by recasting the four components of $\Zsf^a$ as two spinors $\omega^\a$ and $\lat_\ad$ respectively called the \emph{primary} and \emph{projection} parts of the twistor
\ben
\Zsf^a = \mqty(\omega^\a \\ \lat_\ad).
\een
We shall treat these parts exactly like helicity spinors, where $\lat_\ad$ is associated with linear momentum and $\omega^\a$ with angular momentum.
Keep in mind that all twistors in $\PTbb$ must correspond to \emph{some} direction and, therefore, $\lat_\ad \neq 0$.

\begin{figure}[t!]
\begin{center}
\includegraphics[width=0.45\textwidth]{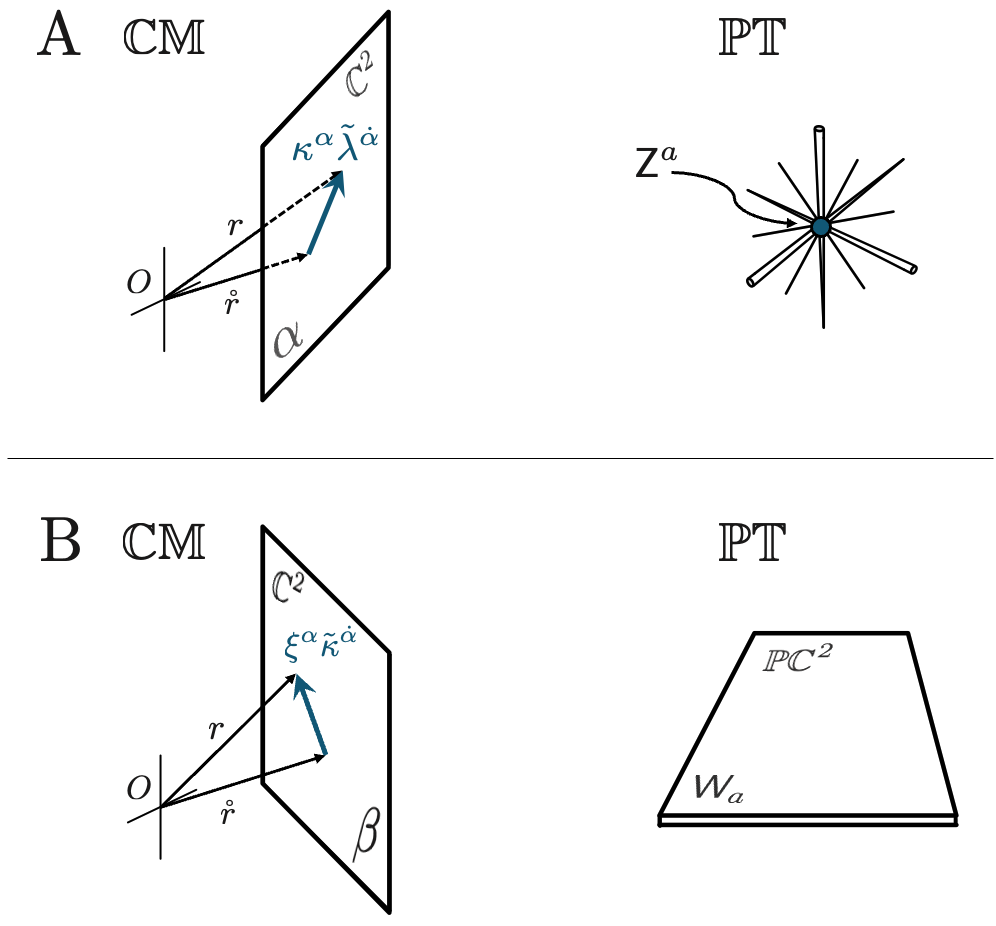}
\caption{Twistor $\alpha$-plane in $\CMbb$ and corresponding star of lines in $\PTbb$ (A). Dual twistor $\beta$-plane in $\CMbb$ and corresponding projective plane in $\PTbb$ (B).}
\label{fig_twistor}
\end{center}
\end{figure}

To establish a correspondence between a twistor in $\PTbb$ and a ray in Minkowski space, the spinor parts are constrained by the \emph{incidence relation}
\ben
\omega^\a = \ii r^{\a\ad} \lat_\ad,
\label{eq_incidence}
\een
which can be thought of as the locus of points $r^{\a\ad} \in \CMbb$ that satisfies this equation for \emph{fixed} $\omega^\a$ and $\lat_\ad$.
The particular locus of points where $\omega^\a$ vanishes is called the $\alpha$-plane, shifted from the origin by $\mathring{r}^{\a\ad}$ and given by
\ben
r^{\a\ad} = \mathring{r}^{\a\ad} + \kappa^\a \lat^\ad \qq{for all $\kappa^\a$.}
\een
That $\kappa^\a$ is arbitrary makes this a $2$-complex-dimensional surface in $\CMbb$.
In $\PTbb$, the $\alpha$-plane corresponds to the $2$-complex-parameter ``star of lines'' passing through the point $\Zsf^a$, as shown in part A of \cref{fig_twistor}.
We shall also require the complementary picture of this correspondence given by the \emph{dual} twistor $\Wsf_a$ with downstairs index $a$ and dual incidence relation
\ben
\Wsf_a = \mqty( \xi_\a & \pit^\ad ) \qq{such that} \pit^\ad = -\ii \xi_\a r^{\a\ad}.
\een
The locus of points where the spinor part $\pit^\ad$ of $\Wsf_a$ vanishes is called the $\beta$-plane, given by
\ben
r^{\a\ad} = \mathring{r}^{\a\ad} + \xi^\a \kappat^\ad \qq{for all $\kappat^\ad$.}
\een
The $\beta$-plane in $\CMbb$ corresponds to a $2$-complex-parameter ``projective plane'' in $\PTbb$, as shown in part B of \cref{fig_twistor}.
Thus, we obtain the full correspondence between twistor $\Zsf^a$ and dual twistor $\Wsf_a$: in $\CMbb$ the $\alpha$-plane is dual to the $\beta$-plane, and in $\PTbb$ the star of lines is dual to the projective plane.
It is from the \emph{incidence} of $\Wsf$ and $\Zsf$, given by $\Wsf_a \Zsf^a = 0$, which is surprisingly rich in structure, whence the geometry of spacetime and the worldline of elementary particles can emerge.
There is no privileged scale in projective geometry; instead, the vanishing contraction $\Wsf_a \Zsf^a = 0$ answers the question, `Where in $\PTbb$ do $\Zsf$ and $\Wsf$ intersect?'
Whether we are dealing with null rays or worldlines is determined by the spin $s$ associated with the twistors, which we can now define as follows
\ben
s = \frac12 \Zsf^a \Zsfbar_a = \frac12 \mqty( \omega^\a \\ \lat_\ad) \mqty( \la_\a & \omegat^\ad ) = \frac12 \left( \omega^\a \la_\a + \lat_\ad \omegat^\ad \right),
\label{eq_twistorspin}
\een
where $\Zsfbar_a$ is the dual twistor of $\Zsf^a$.

\begin{figure}[t!]
\begin{center}
\includegraphics[width=0.5\textwidth]{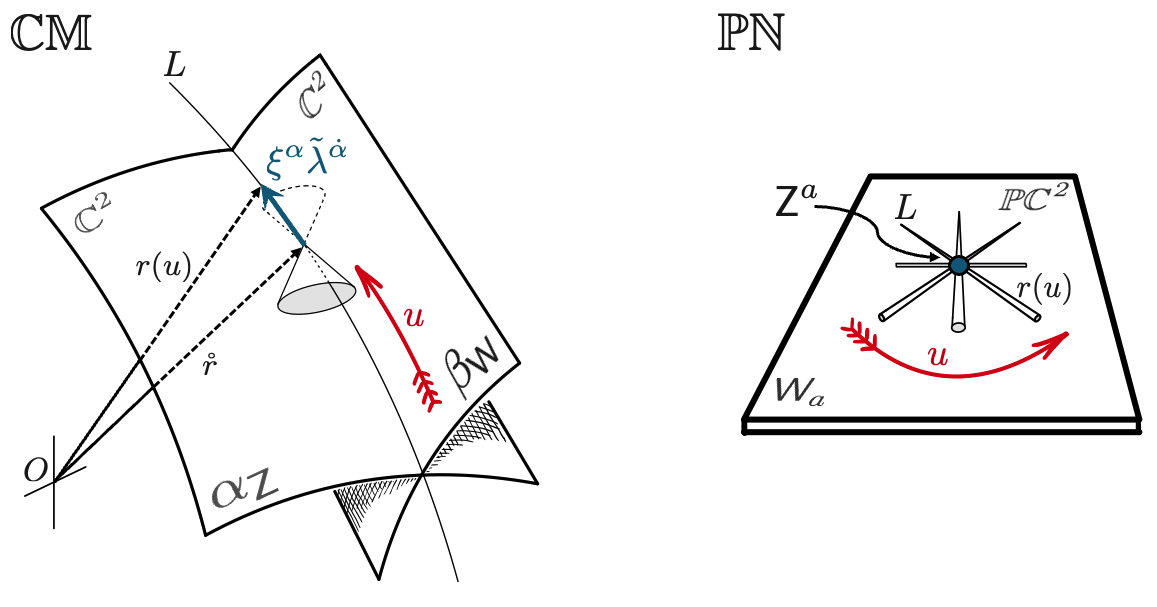}
\caption{A null geodesic is traced out by the incidence $\Wsf_a \Zsf^a = 0$, corresponding to a complex-valued null ray $L$ through $\CMbb$ and a plane pencil of lines (also $L$) in $\PNbb$ parameterized by $u \in \Cbb$.}
\label{fig_nullgeodesic}
\end{center}
\end{figure}

The projective twistor space is divided into three subsets based on $s$.
In this section, we focus only on twistors without spin ($s = 0$) called \emph{null twistors}, which sit in the null projective twistor space $\PNbb \subset \PTbb$.
The incidence of a null twistor $\Zsf^a$ with another null dual twistor $\Wsf_a$ corresponds to the intersection of the $\alpha_\Zsf$ and $\beta_\Wsf$ planes and thus traces out a generally complex-valued \emph{null geodesic} $L$ through $\CMbb$, given by
\ben
r^{\a\ad}(u) = \mathring{r}^{\a\ad} + u \xi^\a \lat^\ad \qq{for $u \in \Cbb$.}
\label{eq_nullgeodesic}
\een
In $\PTbb$, this corresponds to the intersection of the star of lines of $\Zsf^a$ and projective plane of $\Wsf_a$, which is a ``plane pencil of lines'' in $\PNbb$, as shown in \cref{fig_nullgeodesic}.
Since $L$ is parameterized by complex-valued $u$, it has the topology of the Riemann sphere $S^2$.
The null geodesic gets its name from the great circle on the Riemann sphere, where $u$ takes on real numbers.
However, $L$ must also contain some complex-valued points $r^{\a\ad}$ corresponding to values $u$ off of the great circle.
Only for the special case $\Wsf_a = \Zsfbar_a = \mqty( \la_\a & \omegat^\ad)$ does the locus $r^{\a\ad}$ only take on real values, thus describing the real null ray $\Zsf$
\ben
r^{\a\ad}(h) = \mathring{r}^{\a\ad} + h \la^\a \lat^\ad \qq{for $h \in \Rbb$.}
\label{eq_realnullray}
\een
And so we see that the set of null twistors in $\PNbb$ gives rise to all real null rays for all lightcones in real Minkowski space $\Mbb$.
It is in this way that spacetime emerges from twistor theory.

Before moving on to the non-null subsets of $\PTbb$, we now show that \cref{eq_realnullray} is the locus in $\Mbb$ of $\Zsf^a$ where $\omega^\a = 0$.
This calculation will be useful in the coming sections.
Still assuming that $\lat_\ad \neq 0$, consider the incidence relation 
\ben
\ii r^{\a\ad} \lat_\ad = \mathring{\omega}^\a,
\een
where we consider the angular momentum spinor at the origin $\mathring{\omega}^\a$, which is non-zero and orthogonal to $\la^\a$ so that  $\mathring{\omega}^\a \la_\a = 0$.
By contracting the incidence relation with $\mathring{\omegat}^\ad$, the solution of points can be obtained 
\ben
r^{\a\ad} = \left( \ii  \lat_\bd \mathring{\omegat}^\bd \right)^{-1} \mathring{\omega}^\a \mathring{\omegat}^\ad.
\label{eq_nullincidence}
\een
Note that the elements of the factor $\mathring{\omega}^\a \mathring{\omegat}^\ad$ are real because
\ben
\mathring{\omega}^\a \mathring{\omegat}^\ad 
	= \mathring{\omega}^\a \left( \mathring{\omega}^\a \right)^*
	= \left| \mathring{\omega}^\a \right| ^2 \in \Rbb.
\een
Also, since $\Zsf^a$ is a null twistor, \cref{eq_twistorspin} implies that
\ealin{
\Zsf^a \Zsfbar_a = 0 
	&\Rightarrow 
		\mathring{\omega}^\a \la_\a + \mathring{\omegat}^\ad \lat_\ad = 0
		\nonumber \\
	&\Rightarrow
		\mathring{\omega}^\a \la_\a = - \left(  \mathring{\omega}^\a \la_\a \right)^* \nonumber \\
	&\Rightarrow \mbox{$\mathring{\omega}^\a \la_\a$ is purely imaginary,} 
}
and thus $\ii \mathring{\omega}^\a \la_\a$ is real.
Therefore, we can conclude that all points $r^{\a\ad}$ satisfying \cref{eq_nullincidence} form a real locus in $\Mbb$.
Finally, to recover the parameterization of this locus in \cref{eq_realnullray}, note that one of these points must be the displacement of the ray from the origin $\mathring{r}^{\a\ad}$.
Consider the difference
\be
\begin{split}
\ii r^{\a\ad} \lat_\ad &= \mathring{\omega}^\a \nonumber \\
- \ii \mathring{r}^{\a\ad} \lat_\ad &= \mathring{\omega}^\a \nonumber \\
\hline
\left( r^{\a\ad} - \mathring{r}^{\a\ad} \right) \lat_\ad &= 0.
\end{split}
\ee
This means that the displacement along the null ray $r^{\a\ad} - \mathring{r}^{\a\ad}$ is real and also annihilates $\lat_\ad$.
We therefore identify it with a real-scaled version of $\la^\a$, and that is exactly what the term $h \la^\a \lat^\ad$ in \cref{eq_realnullray} represents.
Here, we considered the spinor part representing angular momentum evaluated at the origin $\mathring{\omega}^\a$.
It will be useful in the next section to note that the origin in $\CMbb$ is arbitrary, and $\omega^\a \rightarrow \omega^\a -\ii r^{\a\ad} \lat_\ad$, while $\lat_\ad$ remains fixed under a change in origin.
The opposite is true for dual twistors $\Wsf_a = \mqty( \xi_\a & \pit^\ad )$, where $\xi_\a$ is constant and $\pit^\ad \rightarrow \pit^\ad + \ii \xi_\a r^{\a\ad}$ \cite{Penrose:1986}.

\subsection{Lightlike null hyperplane \label{sec_twistor_2}}

The previous section showed that null twistors (with $s = 0$), give rise to the real null rays of spacetime, upon which $\omega^\a$ vanishes and no angular momentum is associated with the ray.
We also saw that null twistors only correspond to the null projective subspace $\PNbb \subset \PTbb$.
Now we consider the case where $\Zsf^a$ has spin $s \neq 0$.
From \cref{eq_twistorspin}, we have $\omega^\a \la_\a = s$, since $ \overline{\lat_\ad \omegat^\ad} = \omega^\a \la_\a$. 
This implies that $\omega^\a \neq 0$ for all real points in $\CMbb$.
Recall that $\omega^\a$ is the spinor field associated with angular momentum, and 
we shall now see that $\Zsf^a$ with $s \neq 0$ indeed represents a massless particle in $\Mbb$ with helicity.
For this reason, it is called a \emph{lightlike twistor}.

Twistors with $s > 0$ are said to have positive helicity and live in the subspace $\PTbb^+$, while those with $s < 0$ are their negative counterparts sitting in $\PTbb^-$.
Unlike the case of null twistors, the incidence relation in \cref{eq_incidence} for lightlike twistors can only be satisfied by a locus of complex points $r^{\a\ad} \in \CMbb$.
In other words, the $\a$-plane, upon which $\omega^\a$ vanishes, contains no real points.
The lightlike dual twistor $\Wsf_a = \mqty( \xi_\a & \pit^\ad )$ also has $s \neq 0$, and thus $\xi_\a \pi^\a = s$. 
The dual incidence relation $\pit^\ad = -\ii \xi_\a r^{\a\ad}$ is satisfied for complex $r^{\a\ad} \in \CMbb$; its $\b$-plane, upon which $\pit^\ad$ vanishes, also contains no real points.
Therefore, when considering the incidence of lightlike twistors $\Wsf_a \Zsf^a = 0$, the null geodesic $r^{\a\ad}(u)$ will be entirely complex valued.
The lightlike $\a_\Zsf$ and $\b_\Wsf$ planes do still meet where $\omega^\a = \pit^\ad = 0$; however, this occurs on a locus of points forming a complex null geodesic.
These twistors nevertheless have a reality structure that possesses angular momentum, is null, and is remarkably non-local.
This offers an explanation for the non-local properties we previously encountered for massless helicity spinors, and it also justifies why we qualify these non-null twistors as being lightlike.
The structure in $\Mbb$ is not localized as a single ray, as it was for null twistors, but instead consists of a special family of null rays called a \emph{Robinson congruence} \cite{Penrose:1967} that forms a null hyperplane $\Pi$ and fills $\Rbb^3$ entirely.
Snapshots of this congruence in $\Mbb$ are strikingly beautiful and can be described mathematically as a stereographic projection of \emph{Clifford parallels} on the three-sphere $S^3$. 
Renditions shown elsewhere illustrate a system of oriented circles twisting around each other and filling all space, thus encoding the angular momentum of the system \cite{Penrose:1986,Penrose:2004}.

\begin{figure}[t!]
\begin{center}
\includegraphics[width=0.5\textwidth]{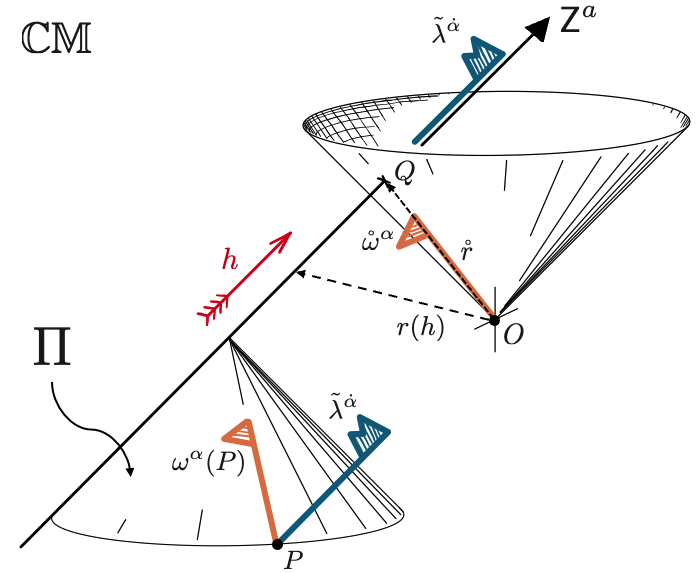}
\caption{Null hyperplane $\Pi$ of a lightlike twistor $\Zsf^a$ in complex Minkowski space $\CMbb$.
Both spinor parts $\omega^\a$ and $\lat_\ad$ are non-zero everywhere in $\CMbb$.
At any point $P$, $\omega^\a$ points along the lightcone whose vertex lies on the null ray given by $\Zsf^a$, thus encoding the non-zero angular momentum.
Point $Q$ on the null ray $\Zsf^a$ is closest to the origin $O$.
Adapted from fig.~6-2 of Penrose \& Rindler \cite{Penrose:1986}.}
\label{fig_hyperplane}
\end{center}
\end{figure}

To visualize this in $\CMbb$, note that the spinor part $\omega^\a$ of the lightlike twistor is non-zero and takes the null direction that intersects the single ray of the corresponding null twistor $\Zsf^a$ whose constant, linear momentum $p^{\a\ad} = \la^\a \lat^\ad$ matches that of the lightlike twistor.
This is illustrated in \cref{fig_hyperplane}.
Therefore, we can visualize the null hyperplane $\Pi$ of a lightlike twistor as \emph{the family of null directions for all lightcones} whose vertices lie on $\Zsf^a$.
It is in this sense that the lightlike twistor does not have a worldline and is therefore not localized.

Taken together, a lightlike particle is described by a twistor $\Zsf^a = \smqty( \omega^\a \\ \lat_\ad)$ up to a phase $e^{\ii \theta}$ for $\theta$ real.
Although not important for our purposes here, this phase arises from differences between the orientated plane elements defined by bi-spinors $\omega^\a \omega^\b$ and $\lat^\ad \lat^\bd$, as illustrated by the ``flags'' in \cref{fig_hyperplane}.
Whether it has a role to play in the physical theory remains unclear.
Its $4$-momentum $p^\mu \mapsto p^{\a\ad} \equiv \la^\a \lat^\ad$ and $6$-angular momentum relative to the origin are \cite{Penrose:1986} 
\ben
M^{\mu\nu} \mapsto M^{\a\ad\b\bd} \equiv \ii \omega^{(\a} \la^{\b)} \vepsi^{\ad\bd} - \ii \vepsi^{\a\b} \omegat^{(\ad} \lat^{\bd)}.
\label{eq_6angmom}
\een
Under a displacement $r^{\a\ad}$ of the origin, the linear momentum remains constant $p^{\a\ad} \rightarrow p^{\a\ad}$, and the angular momentum transforms according to
\ben
M^{\a\ad\b\bd} \rightarrow M^{\a\ad\b\bd} - 2 r^{[\a\ad} p^{\b\bd]}
\een
with square brackets representing antisymmetrization of the Lorentz indices.
We can derive a formula for $\Pi$ by attempting to find the locus $r^{\a\ad}$ satisfying the relativistic center-of-mass given by $p_\mu M^{\mu\nu} = 0$ \cite{Synge:1965}.
Without loss of generality, choose $\mathring{r}^\mu$ such that $p_\mu \mathring{r}^\mu = 0$ and $r^\mu(h) = \mathring{r}^\mu + h p^\mu$, for $h \in \Rbb$, as shown in \cref{fig_hyperplane}.
Now consider,
\ben
M^{\mu\nu}(\mathring{r}^\mu) p_\nu 
	= \mathring{M}^{\mu\nu} p_\nu - 2 \mathring{r}^{[\mu} p^{\nu]} p_\nu
	= \mathring{M}^{\mu\nu} p_\nu - \mathring{r}^\mu p^2 = 0.
\een
Solving for $\mathring{r}^\mu$ in the last equation, we obtain the real parameterization for the locus forming $\Pi$
\ben 
r^\mu(h) = \frac{\mathring{M}^{\mu\nu} p_\nu}{p_\sigma p^\sigma} + hp^\mu \qq{for $h \in \Rbb$.}
\een
If $p^\mu$ is lightlike, then $p^2 = 0$ and the relativistic center-of-mass $p_\mu M^{\mu\nu} = 0$ has no solution unless $\mathring{M}^{\mu\nu} = r^\mu p^\nu - p^\mu r^\nu$, which implies
\ben
\mathring{M}^{\mu\nu} p_\nu = r^\mu p^2 - p^\mu \left( r_\nu p^\nu \right) = - p^\mu \left( r_\nu p^\nu \right).
\een
Therefore, the relativistic center-of-mass exists only if $\mathring{M}^{\mu\nu} p_\nu$ is proportional to the linear momentum $p^\mu$ and $r_\nu p^\nu = k$ for constant $k$.
This last result follows from the fact that lightlike particles must have exactly two spin degrees of freedom and a constant spin vector
\ben
S_\mu = \star M_{\mu\nu} p^\nu =  \star \mathring{M}_{\mu\nu} p^\nu = \mathring{S}_\mu,
\een
where $\star$ indicates the Hodge dual $\star M_{\mu \nu} = \tfrac12 e_{\mu\nu\rho\sigma} M^{\rho \sigma}$.
This in turn implies
\ben
S_\mu = \star M_{\mu\nu} p^\nu = s p_\mu,
\een
where the spin $s$ is, of course, also constant and non-zero.
Therefore, we can conclude that the relativistic center-of-mass is satisfied on the locus of points $r^\mu$ where $r_\mu p^\mu = k$ is constant, which is indeed a 3-real-dimensional hyperplane $\Pi$ in $\CMbb$ corresponding to the Robinson congruence that fills $\Rbb^3$ in $\Mbb$.
It is not hard to see that $\Pi$ is the locus of points where $\omega^\a(r^{\a\ad})$ satisfies $\omega^\a \la_\a = s$, as was claimed at the beginning of this subsection, since the relativistic center-of-mass according to \cref{eq_6angmom} implies
\ealin{
p_{\mu} M^{\mu\nu} = 0 &\rightarrow p_{\ad\a} M^{\a\ad\b\bd} = 0 \nonumber \\
	&\Rightarrow \ii \la^\b \lat^\bd \left( -\omega^\a \lat_\a + \lat_\ad \omegat^\ad \right) = 0 \nonumber \\
	&\Rightarrow \omega^\a \la_\a = \lat_\ad \omegat^\ad.
}
Note that this is same constraint we found upon expanding $s = \tfrac12 \Zsf^a \Zsfbar_a$ in \cref{eq_twistorspin}.

\begin{figure}[h!]
\begin{center}
\includegraphics[width=0.5\textwidth]{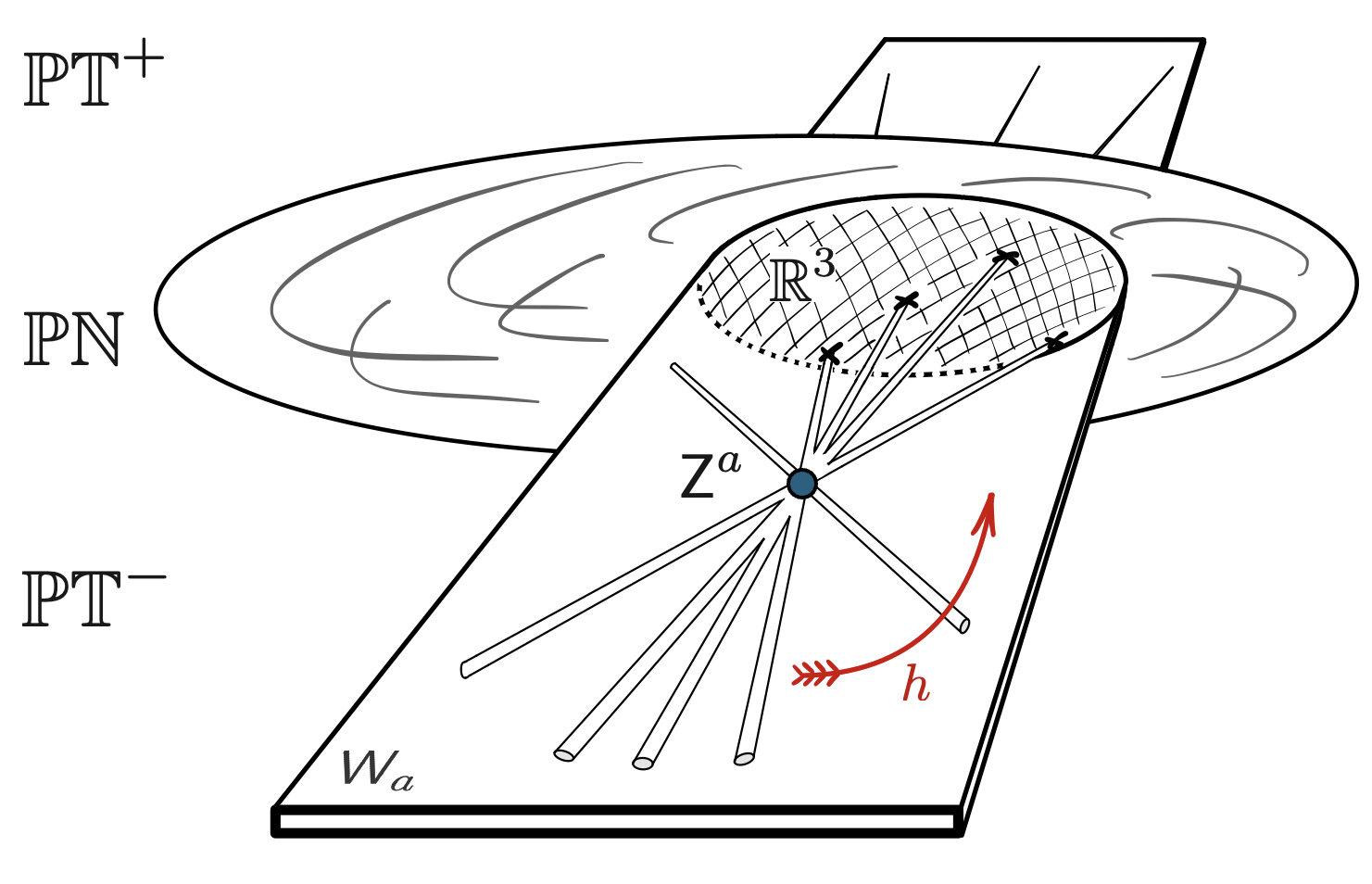}
\caption{Incidence of lightlike twistor $\Zsf^a$ and dual twistor $\Wsf_a$ in projective twistor space $\PTbb$.
The Robinson congruence is generated by those plane pencil lines intersecting $\PNbb$.
Adapted from fig.~33.14 of Penrose \cite{Penrose:2004}.}
\label{fig_lightlikeincidence}
\end{center}
\end{figure}

In projective twistor space, we can consider the corresponding incidence $\Wsf_a \Zsf^a = 0$, whose geometry is similar to that for null twistors, shown on the right hand side of \cref{fig_nullgeodesic}.
The main difference is that both $\Zsf^a$ and $\Wsf_a$ now extend beyond $\PNbb$.
For example, if $\Zsf^a$ has negative helicity, it will be located in $\PTbb^-$.
Some subset within its star of lines will, however, pass through $\PNbb$.
Since $\Wsf_a$ is dual to some lightlike twistor, its projective plane $\PCbb^2$ will also intersect with $\PNbb$.
Thus, the plane pencil of lines satisfying their incidence parameterizes a 3-real-dimensional region in $\PNbb$ whose points correspond to the Robinson congruence of rays in $\Mbb$, as shown in \cref{fig_lightlikeincidence}.

\subsection{Massive timelike worldline \label{sec_twistor_3}}

Earlier in this paper, we constructed massive helicity spinors and their scattering amplitudes by considering the linear combination of massless spinors $\la^\a$ and $\eta^\a$ and their duals, as in \cref{eq_mhsdef}.
This led to the description of a particle with timelike momentum $\bm p^{\a\ad} = \la^\a \lat^\ad - \eta^\a \etat^\ad$ and mass $\bm p^2 = m^2$.
We learned that massive helicity spinor couplings are non-local due to their $x$ factor, but their amplitudes can become local if all $x$ factors cancel.

In a similar fashion, we shall now see that a massive particle can be represented by the linear combination of lightlike twistors $\Xsf^a$ and $\Zsf^a$
\ben
\Ysf^a = \mqty(\psi^\a \\ \kappat_\ad) \equiv \beta \Xsf^a + \gamma \Zsf^a \qq{for $\beta,\,\gamma \in \Cbb$,}
\label{eq_massivetwistor}
\een
where
\ben
\Xsf^a = \mqty(\chi^\a \\ \etat_\ad) \qq{and} \Zsf^a = \mqty(\omega^\a \\ \lat_\ad)
\een
have helicity
\ben
s = \frac12 \Xsf^a \Xsfbar_a = \frac12 \Zsf^a \Zsfbar_a.
\een

Interestingly, since $\Ysf^a$ represents a massive particle with non-zero spin $s$, its worldline must also be non-local in $\Mbb$, since $\psi^\a \neq 0$ for all real points. 
In $\CMbb$, however, it turns out that $\Ysf^a$ \emph{does} have unique timelike worldline that runs along the incidence of the null hyperplanes $\Pi_\Xsf$ and $\Pi_\Zsf$. 
By definition, these hyperplanes intersect if and only if $\Xsf^a \Zsfbar_a  = \Zsf^a \Xsfbar_a = 0$.
Being two $3$-real-dimensional hyperplanes in $\CMbb$, their intersection is a $2$-real-dimensional surface or, equivalently, a $1$-complex-dimensional worldline.
Each point on the worldline must satisfy both incidence relations
\ealin{
\ii r^{\a\ad} \lat_\ad = \omega^a, \label{eq_inc1} \\
\ii r^{\a\ad} \etat_\ad = \chi^a. \label{eq_inc2}
}
Indeed, note that neither relation is satisfied on real points.
For example, 
\ben
\Zsf^a \Zsfbar_a = \omega^\a \la_\a + \lat_\ad \omegat^\ad = 2s 
	\Rightarrow \lat_\ad \omegat^\ad = 2s - \omega^\a \la_\a,
\een
and also
\ben
\omega^\a \la_\a = \ii r^{\a\ad} \lat_\ad \la_\a 
	\Rightarrow \lat_\ad \omegat^\ad = -\ii r^{\a\ad} \lat_\ad \la_a = 2s - \omega^\a \la_\a.
\een
Therefore,
\ben
\overline{\omega^\a \la_\a} = 2s - \omega^\a \la_\a \notin \Rbb \qq{for $s\neq 0$,}
\een
and furthermore
\ben
\overline{\ii r^{\a\ad} \lat_\ad \la_\a} = \overline{\omega^\a \la_\a} 
	= \lat_\ad \omegat^\ad = 2s - \ii r^{\a\ad} \lat_\ad \la_\a.
\een
This implies that the matrix $r^{\a\ad}$ is not Hermitian; therefore, $\Pi_\Xsf \cap \Pi_\Zsf$ must occur on a locus of complex points where $\chi^\a = \omega^\a = 0$.
Consequently, from \cref{eq_massivetwistor}, the primary part of $\Ysf^a$ also vanishes $\psi^\a = \beta \chi^\a + \gamma \omega^\a = 0$, and the above locus also defines the complex worldline of $\Ysf^a$.
Indeed, one can check by direct substitution using the incidence relations in \cref{eq_inc1,eq_inc2} that the complex locus is
\ben
r^{\a\ad} = \frac{\omega^\a \etat^\ad - \chi^\a \lat^\ad}{\ii \lat_\bd \etat^\bd}, \qq{where $\lat_\ad \; \slashed{\propto} \; \etat_\ad$.}
\een

So much for the worldline. 
The massive particle will also have non-zero spin
\ben
s_\Ysf = \frac12 \Ysf^a \Ysfbar_a
	= \frac{|\beta|^2}{2} \Xsf^a \Xsfbar_a + \frac{|\gamma|^2}{2} \Zsf^a \Zsfbar_a
	= \frac{|\beta|^2 + |\gamma|^2}{2} s
\een
and momentum
\ben
\kappat_\ad \kappa_\a = |\beta|^2 \etat_\ad \eta_\a + |\gamma|^2 \lat_\ad \la_\a
	+ \beta \gamma^* \etat_\ad \la_\a + \beta^* \gamma \lat_\ad \eta_\a
\een
for general $\beta,\,\gamma \in \Cbb$.
In the special case where $\gamma = \ii \beta$, we find a familiar result where the momenta simply add
\ealin{
p_{\Ysf \ad \a} \equiv \kappat_\ad \kappa_\a 
	&= |\beta|^2 \left( \etat_\ad \eta_\a + \lat_\ad \la_\a - \ii \etat_\ad \la_\a + \ii \lat_\ad \eta_\a \right) \nonumber \\
&= |\beta|^2 \left( p_{\Xsf\, \ad \a} + p_{\Zsf\, \ad \a} \right) \qq{[if $\etat_\ad \la_\a$ Hermitian]}
\label{eq_addmomenta}
}
This equation is identical to $\bm p_{\ad\a} = \lat_\ad \la_\a - \etat_\ad \eta_\a$ when $|\beta|^2 = 1$, with the exception of a minus sign indicating the opposite direction of $ p_{\Xsf\, \ad \a} = \etat_\ad \eta_\a$, which could be absorbed into the definition of $\Xsf_a$ in \cref{eq_massivetwistor}.
We thus recognize the massive helicity spinor as the projection part of what we shall call the \emph{timelike twistor}
\ben
\Ysf^a = \Xsf^a + \ii \Zsf^a = \mqty(\chi^\a + \ii \omega^\a \\ \etat_\ad + \ii \lat_\ad).
\een
The spin of $\Ysf^a$ is encoded by the \emph{total angular momentum twistor}, a bi-twistor with symmetrized indices
\ben
\Asf_{ab} = 2 \Xsfbar_{(a} \Isf_{b)c} \Xsf^c + 2 \Zsfbar_{(a} \Isf_{b)c} \Zsf^c,
\qq{with $\Isf_{ab} \equiv \mqty( 0 & 0 \\ 0 & \vepsi^{\ad \bd})$.}
\label{eq_addangularmomenta}
\een
This is the sum of the angular momenta from $\Xsf^a$ and $\Zsf^a$. 
For example,
\ben
\Asf_{\Zsf ab} = 2 \Zsfbar_{(a} \Isf_{b)c} \Zsf^c 
	= \mqty( 0 & \la_\a \lat^\bd \\ \lat^\ad \la_\b & -2 \omegat^{(\ad} \lat^{\bd)} ).
\label{eq_angulartwistor}
\een
The mass follows from the ``square'' of the angular momentum \cite{Penrose:1972}
\ealin{
-\frac12 \Asf_{ab} \Asfbar^{ab} 
	&= 2 \Big( \Xsfbar_{(a} \Isf_{b)c} \Xsf^c + \Zsfbar_{(a} \Isf_{b)c} \Zsf^c \Big) 
		\Big( \Xsf^{(a} \Isf^{b)c} \Xsfbar_c + \Zsf^{(a} \Isf^{b)c} \Zsfbar_c \Big) \nonumber \\
	&= 2 \left| \Xsf^a \Zsf^b \Isf_{ab} \right|^2 \nonumber \\
	&= 2 \left| \mqty(\chi^\a \\ \etat_\ad) \mqty(\omega^\b \\ \lat_\bd) \mqty( 0 & 0 \\ 0 & \vepsi^{\ad \bd}) \right|^2 \nonumber \\
	&= 2 \left| \etat_\ad \lat^\ad \right|^2.
}
Note the product of like-terms in the first line will vanish due to the fact that $\etat_\ad \etat^\ad = \lat_\ad \lat^\ad =0$; therefore, only the cross-terms appear in the second line.
We recognize this last expression as $p_\Ysf^2 = p_{\Ysf \ad \a} \,p_\Ysf^{\a\ad} = m^2$, since only cross-terms in $\big( \kappat_\ad \kappa_\a \big) \big( \kappa^\a \kappat^\ad \big)$ will be non-zero:
\ealin{
p_\Ysf^2 &= |\beta|^2 |\gamma|^2 \left( \etat_\ad \la_\a \lat^\ad \eta^\a 
	+ \lat_\ad \eta_\a \etat^\ad \la^\a \right) \nonumber \\
	&= 2 |\beta|^2 |\gamma|^2 \left| \etat_\ad \lat^\ad \right|^2 \nonumber \\
	&= m^2.
\label{eq_twistormass}
}
Therefore, linear combinations of lightlike twistors can give rise to mass $m^2 = p_\Ysf^2 = -\tfrac12 \Asf_{ab} \Asfbar^{ab}$.
The resemblance of this last equation to the spinor helicity version $m^2 = \bm p^2 = -\tfrac12 \bm p_{\ad\a} \bm p^{\a\ad}$ from \cref{sec_mhs_2} is clear and suggests, once again, that $\bm p_{\ad\a}$ is just a fragment of the more informative $\Asf_{ab}$ in \cref{eq_angulartwistor}.

Finally, knowing that the linear and angular momenta of $\Xsf^a$ and $\Zsf^a$ simply add up, as in \cref{eq_addmomenta,eq_addangularmomenta}, we recover two more familiar relations in Lorentz indices
\ealin{
p_\Ysf^\mu &= p_\Xsf^\mu + p_\Zsf^\mu \\
M_\Ysf^{\mu\nu}(r^\mu) &= M_\Xsf^{\mu\nu}(r^\mu) + M_\Zsf^{\mu\nu}(r^\mu).
}
It thus follows that the relativistic center-of-mass of $\Ysf^a$, being the locus of points $r^\mu$ satisfying $p_{\Ysf\mu} M_\Ysf^{\mu\nu} = 0$, takes up the unique complex timelike and localized worldline
\ben
r_\Ysf^\mu(h) = \frac{\mathring{M}_\Ysf^{\mu\nu} p_{\Ysf\nu}}{ p_\Ysf^2 } + h p_\Ysf^\mu \qq{for $h \in \Rbb$,}
\een
since the first term is now a well defined constant with $p_\Ysf^2 > 0$.

\section{Outlook \label{sec_outlook}}

The on-shell formulation of scattering amplitudes has been growing quickly in the recent years since it became possible to represent massive particles with helicity spinors.
Initially limited to gluonic scattering, this development has extended the reach of the spinor helicity formalism, in principle, to any process being studied in a collider experiment.
In addition to providing efficient tools to calculate amplitudes and cross sections, this formalism also offers a rare glimpse of an underlying particle theory that is distinct from the quantum field theory of the Standard Model.

Our contributions in this paper advance the field in three ways.
First, in light of the rapidly progressing literature, we filled in much technical detail on the notation and formulation of massive helicity spinors (\cref{sec_mhs}) in order to further examine the diagrammatic and analytical structure of scattering amplitudes constructed from them (\cref{sec_amp}).
We chose a pedagogical tack to invite new students to the on-shell program by frequently comparing the spinor helicity formalism to quantum electrodynamics.
Second, these details enabled us to develop new cross sectional methods, which we utilized to calculate Bhabha and Compton cross sections, comparing our results to known ones and confirmed their agreement at both high and low energies (\cref{sec_cross}).
Third, the frequent appearance of coalescence and non-locality during our study of massive scattering amplitudes prompted us to examine their relationship with mass acquisition.
In addition to the dynamical aspect presenting itself as a coalescence of interactions, we explored the kinematical aspect, which is closely related to the locality of particle worldlines (\cref{sec_mass}).
This last point led us to consider how helicity spinors might be naturally incorporated into twistor theory, where we found some correspondence to further pursue a deeper physical interpretation of the formalism as well as an explanation for its efficiency (\cref{sec_twistor}).

We hope that these contributions provide new perspectives for the various avenues currently being pursued in the literature.
On the practical side, the establishment of a comprehensive toolkit for calculating the rates of processes in collider experiments would be an asset for phenomenological work.
The auxiliary couplings will play an important role in precision physics, and their contributions to the cross section will surely be complicated by high-spin interactions.
Methodology for computing observables efficiently, perhaps, via quasi limits or partial cross sections, could be the first step in understanding how their higher-order contributions relate to Feynman's loop diagrams and how they may account for infrared divergences.
With such a refined on-shell toolkit, one that is bereft of limitations inherent to nonlinear sigma models, perhaps the finer details of the Higgs boson can be mapped out and tested.
Efforts along these lines are already underway within the Standard Model \cite{Christensen:2020}, and important benchmarks have been achieved for tree-level calculations of the electroweak sector \cite{Bachu:2020,Durieux:2020} and quantum chromodynamics \cite{Ballav:2022}.
Beyond the Standard Model, the on-shell counterpart of the supersymmetrized Higgs mechanism appears to be a straightforward extension of the coalescence of superamplitudes, yet constraints on the Higgs supermultiplet and its interactions at low energy will require further analysis before new observables can be identified \cite{Herderschee:2019}.
Regardless of the particular model, the advent of massive helicity spinors makes possible new on-shell analyses of effective field theory \cite{Balkin:2022}, Yang--Mills and Yukawa theories \cite{Bachu:2024}, and the deeper connection between spontaneously broken gauge theory and the unitary scattering of massive vector bosons \cite{Liu:2022}.
On this last point, the spinor helicity formalism's departure from a manifestly local framework raises important questions about what mass acquisition means for non-local representations of particles and their interactions.

It seems to us that twistor theory is a natural generalization of the spinor helicity formalism, rich enough to provide a unified representation of spacetime and particle content for all masses and spins.
The earlier foundational works by Arkani-Hamed and collaborators that considered massless spinors for the amplituhedron and positive Grassmannian frameworks were surely a guide \cite{Arkani-Hamed:2014,Arkani-Hamed:2016}.
Inroads probing the relationship between massive helicity spinor and twistor representations have since gotten underway.
Notably, Joon-Hwi Kim and collaborators have shown that the particle spectrum of massive helicity-spinors and quantized massive twistors are equivalent \cite{Kim:2021} and their high-energy limit may be seen as a unified representation of both spin and spacetime \cite{Kim:2025}.
Albonico, Geyer, and Mason, have also utilized massive helicity spinors to incorporate mass in the twistorial framework of string theory known as ambitwistor-strings \cite{Albonico:2024}.
These works, in addition to the already large corpus of literature presenting applications of massive spinors both within and beyond the Standard Model are a resource for new theoretical data to help narrow down the general theoretical framework outlined here.

\appendix
\section{Spinor helicity conventions and bases \label{app_conventions}}

Massless helicity spinors and Levi--Civita symbols:
\ealin{
\la^\a &= \aket \la =\mqty(a \\ b), \quad \lat^\ad = \sket \la = \mqty(\tilde a \\ \tilde b), \quad \vepsi_{\a\b} = \mqty(0 & -1 \\ 1 & 0)  \nonumber \\
\vepsi_{\a\b} &= - \vepsi^{\a\b} = - \vepsi^{\b\a} = \vepsi_{\ad\bd} = \vepsi_{IJ} \nonumber
}
Raising and lowering indices:
\ealin{
\la_\a &= \abra \la = \vepsi_{\a\b} \la^\b,  & \la_\a &= \la^\b \vepsi_{\a\b} 
	= \la^\b (-\vepsi_{\b\a}) = - \la^\b \vepsi_{\b\a} \nonumber \\
\lat^\ad &=  \sbra \la = \vepsi^{\ad\bd} \lat_\bd, & \lat^\ad &= \lat_\bd \vepsi^{\ad\bd} 
	= \lat_\bd (-\vepsi^{\bd\ad}) = - \lat_\bd \vepsi^{\bd\ad} \nonumber
}
Inner products are antisymmetric because of $\vepsi_{\a\b}$ contraction:
\ealin{
\abraket{12} &= \la_1^\b \la_2^\a \vepsi_{\a\b} 
	= - \la_1^\b \la_{2\b} = - \la_{2\b} \la_1^\b
	= -\abraket{21} \nonumber \\
\sbraket{12} &= \lat_{1\bd} \lat_{2\ad} \vepsi^{\ad\bd} 
	= - \lat_{1\bd} \lat_2^\bd = -\lat_2^\bd \lat_{1\bd}
	= -\sbraket{21} \nonumber
}
Note above that, unlike spinor \emph{fields}, the spinor helicity variables must commute: $\la_1 \la_2 = \la_2 \la_1$.
\be
\abraket{\la \chi} = - \abraket{\chi \la} \ \mbox{and}\  \sbraket{\la \chi} = - \sbraket{\chi \la} 
\ \Leftrightarrow\  \abraket{\la \la} = \sbraket{\la \la} = 0
\ee
Thus, outer products like $\la^\a \chi_\a$ and $\lat_\ad \chit^\ad$ are superfluous since
$\left. 1 \right> \left< 2 \right. = \abraket{12}$ and $\left. 1 \right] \left[ 2 \right. = \sbraket{12}$.
Only the ``mixed'' outer product is used as it defines momentum:
\be
p^{\a\ad} = \la^\a \lat^\ad = \aketbras{\la} = \mqty(a \\ b) \mqty(\tilde a & \tilde b)
	= \mqty(a \tilde a & a \tilde b \\ b \tilde a & b \tilde b).
\ee
Spinor bases for general frame $p^\mu$, rotated away from reference frame $k^\mu$ by angles $\theta$ and $\phi$:
\be
\za^{+\a} \equiv \mqty(c \\ s) \qq{and} \za^{-\a} \equiv \mqty(-s^* \\ c)
\ee
with $c \equiv \cos \tfrac{\theta}{2}$ and $s \equiv e^{\ii \phi} \sin \tfrac{\theta}{2}$ being diametrically opposed on $S^2$.
Remaining spinor bases are derived using contraction rules:
\ealin{
\za^+_\a &= \vepsi_{\a\b} \za^{+\b} = \mqty(-s & c), & \za^-_\a &= \vepsi_{\a\b} \za^{-\b} = \mqty(-c -s^*), \nonumber \\
\zat^-_\ad &\equiv \left( \za^{+\a} \right)^* = \mqty(c \\ s^*), & \zat^+_\ad &\equiv \left( \za^{-\a} \right)^* = \mqty(-s \\ c), \nonumber \\ 
\zat^{-\ad} &= \vepsi^{\ad\bd} \zat^-_\bd = \mqty(s^* & -c), & \zat^{+\ad} &= \vepsi^{\ad\bd} \zat^+_\bd = \mqty(c & s). \nonumber
}
Little group basis is identified with $\za^{+\a}$ and $\za^{-\a}$ for reference $k^\mu \Rightarrow \theta = \phi= 0 \Rightarrow c = 1, s = 0$:
\ealin{
\za^{+I} &= \mqty(1 \\ 0) \leftrightarrow \za^{(k)+\a}, & \za^{-I} &= \mqty(0 \\ 1) \leftrightarrow \za^{(k)-\a}, \nonumber \\
\za^+_I &= \mqty(0 & 1) \leftrightarrow \za^{(k)+}_{\quad\a}, & \za^-_I &= \mqty(-1 & 0) \leftrightarrow \za^{(k)-}_{\quad\a}. \nonumber
}
The standard $4$-point Mandelstam variables are
\eali{
s &= (p_1^\mu + p_2^\mu)^2 = (p_3^\mu + p_4^\mu)^2, \\
t &= (p_2^\mu - p_3^\mu)^2 = (p_1^\mu - p_4^\mu)^2, \\
u &= (p_1^\mu - p_3^\mu)^2 = (p_2^\mu - p_4^\mu)^2.\\
}

\section{Partial cross section for general spin \label{app_partialsection}}

For minimal coupling $(\bm{1}^s,\, \bm{2}^s, 3^0)$ with spin $s$, the partial cross section is the $\smqty[4s \\ 4s]$-rank tensor
\be
\sigma_{\{\a_1 \cdots \a_{2s}\} \{\b_1 \cdots \b_{2s}\}}^{\{\ad_1 \cdots \ad_{2s}\} \{\bd_1 \cdots \bd_{2s}\}}
\ee
or, concisely, $\sigma_{\a_i \b_i}^{\ad_i \bd_i}$ with $i \in \{1,2,\ldots, 2s\}$.
The case for $s=\tfrac12$ is shown in \cref{eq_partialsectionmin}.
Consider the case $s=1$.
There are $4$ undotted and $4$ dotted indices that will distribute over the variables $\la_1 \la_1 \la_2 \la_2$.
Concisely,
\be
\sigma_{\a_i \b_i}^{\ad_i \bd_i} = \left| \frac{(\la_1)^2_{I_i} (\la_2)^2_{J_i} }{m^2} M_{\a_i \b_i}^{+1} \right|^2
\ee
with symmetrized indices appearing below for
\ealin{
\left| M \right|^{2\, (\ad_1 \ad_2 \bd_1 \bd_2)}_{(\a_1 \a_2 \b_1 \b_2)} 
	&= g^2 m^2 \vepsi_{(\a_1 \a_2} \vepsi^{(\ad_1 \ad_2} \vepsi_{\b_1 \b_2)} \vepsi^{\bd_1 \bd_2)} \nonumber \\
	&\hspace{-1cm} = \frac{g^2 m^2}{4!} \sum_{\pi \in S_4} \vepsi_{\pi_1 \pi_2} \vepsi^{\dot \pi_1 \dot \pi_2} \vepsi_{\pi_3 \pi_4} \vepsi^{\dot \pi_3 \dot \pi_4},
}
where $S_4$ is the symmetric group of $4$ elements $1234$, and $\pi$ is a permutation of these elements.
With still more conciseness, the pattern becomes more apparent
\be
\left| M \right|^{2\, (\ad_i \bd_i)}_{(\a_i \b_i)} = g^2 m^2 \left( \vepsi \tilde \vepsi \right)_{(\a_i \b_i)}^{(\ad_i \bd_i)}.
\ee
Note that the round brackets $\left( \vepsi \tilde \vepsi \right)$ are \emph{not} symmetrization indices.
This leads to
\ealin{
\sigma_{\a_i \b_i}^{\ad_i \bd_i} &= \frac{g^2}{5} \Bigg[ 
	\bm p_{1\,(\a_1}^{(\ad_1} \bm p_{1\,\a_2}^{\ad_2} \bm p_{2\,\b_1}^{\bd_1} \bm p_{2\,\b_2)}^{\bd_2)} \Bigg. \nonumber \\
	&\Bigg. \quad\quad - \frac12 \bm p_{1\,(\a_1}^{(\ad_1} \bm p_{1\,\a_2}^{\ad_2} \vepsi_{\b_1 \b_2)} \vepsi^{\bd_1 \bd_2)} \left( \bm p_{12} + m^2 \right) \Bigg. \nonumber \\
	&\Bigg. \quad\quad + \frac12 \vepsi_{(\a_1 \a_2} \vepsi^{(\ad_1 \ad_2} \vepsi_{\b_1 \b_2)} \vepsi^{\bd_1 \bd_2)} \left(\bm p_{12}^2 + m^4 \right) \Bigg], \nonumber
}
where $\bm p_{12} \equiv \bm p_1 \vdot \bm p_2$.
We see that there are generally $5$ ways to distribute the indices, yielding terms: $\bm p_a^{\a\ad}$, $\bm p_{ab}$, $m^2$, $\bm p_{ab}^2$, and $m^4$.
Also, a minus sign appears for each $\bm p_{12} + m^2$ contraction.

Altogether, for arbitrary spin $s$, there are $4s+1$ ways to distribute the indices, so the general formula claimed in the main text in \cref{eq_partialsectiongen} is indeed
\begin{widetext}
\ben
\sigma_{\a_i \b_i}^{\ad_i \bd_i}  = \frac{g^2}{4s+1} \sum_{n=0}^{2s} \frac{(-1)^{n+1}}{2}
	\big( \bm p_1 \bm p_2\big)^{2s-n\, (\ad_i \bd_i}_{\quad\quad(\a_i \b_i} \big( \vepsi \tilde{\vepsi} \big)^{n\, \ad_j \bd_j)}_{\;\;\a_j \b_j)} \left[ \bm p_{12}^n + m^{2n} \right]
\een
for fully symmetrized indices with labels $1\le i < j \le s$, as well as $2s-n$ copies of the momenta $\bm p_1$ and $\bm p_2$, and $n$ copies of the Levi--Civita symbols.
\end{widetext}

\begin{acknowledgments}
We thank Nima Arkani-Hamed for stimulating discussions that motivated the project as well as James Halverson and Tomasz Taylor for helpful feedback about the results.
We also thank two anonymous reviewers for their suggestions and for spotting an error in a previous version of the manuscript.
This work was supported in part by a Faculty Professional Development Award at Northeastern University.
\end{acknowledgments}

\bibliography{gomezbibliography}

\begin{thebibliography}{39}%
\makeatletter
\providecommand \@ifxundefined [1]{%
 \@ifx{#1\undefined}
}%
\providecommand \@ifnum [1]{%
 \ifnum #1\expandafter \@firstoftwo
 \else \expandafter \@secondoftwo
 \fi
}%
\providecommand \@ifx [1]{%
 \ifx #1\expandafter \@firstoftwo
 \else \expandafter \@secondoftwo
 \fi
}%
\providecommand \natexlab [1]{#1}%
\providecommand \enquote  [1]{``#1''}%
\providecommand \bibnamefont  [1]{#1}%
\providecommand \bibfnamefont [1]{#1}%
\providecommand \citenamefont [1]{#1}%
\providecommand \href@noop [0]{\@secondoftwo}%
\providecommand \href [0]{\begingroup \@sanitize@url \@href}%
\providecommand \@href[1]{\@@startlink{#1}\@@href}%
\providecommand \@@href[1]{\endgroup#1\@@endlink}%
\providecommand \@sanitize@url [0]{\catcode `\\12\catcode `\$12\catcode
  `\&12\catcode `\#12\catcode `\^12\catcode `\_12\catcode `\%12\relax}%
\providecommand \@@startlink[1]{}%
\providecommand \@@endlink[0]{}%
\providecommand \url  [0]{\begingroup\@sanitize@url \@url }%
\providecommand \@url [1]{\endgroup\@href {#1}{\urlprefix }}%
\providecommand \urlprefix  [0]{URL }%
\providecommand \Eprint [0]{\href }%
\providecommand \doibase [0]{https://doi.org/}%
\providecommand \selectlanguage [0]{\@gobble}%
\providecommand \bibinfo  [0]{\@secondoftwo}%
\providecommand \bibfield  [0]{\@secondoftwo}%
\providecommand \translation [1]{[#1]}%
\providecommand \BibitemOpen [0]{}%
\providecommand \bibitemStop [0]{}%
\providecommand \bibitemNoStop [0]{.\EOS\space}%
\providecommand \EOS [0]{\spacefactor3000\relax}%
\providecommand \BibitemShut  [1]{\csname bibitem#1\endcsname}%
\let\auto@bib@innerbib\@empty
\bibitem [{\citenamefont {Arkani-Hamed}\ \emph
  {et~al.}(2021{\natexlab{a}})\citenamefont {Arkani-Hamed}, \citenamefont
  {Huang},\ and\ \citenamefont {Huang}}]{Arkani-Hamed:2021a}%
  \BibitemOpen
  \bibfield  {author} {\bibinfo {author} {\bibfnamefont {N.}~\bibnamefont
  {Arkani-Hamed}}, \bibinfo {author} {\bibfnamefont {T.-C.}\ \bibnamefont
  {Huang}},\ and\ \bibinfo {author} {\bibfnamefont {Y.-t.}\ \bibnamefont
  {Huang}},\ }\bibfield  {title} {\bibinfo {title} {Scattering amplitudes for
  all masses and spins},\ }\href {https://doi.org/10.1007/JHEP11(2021)070}
  {\bibfield  {journal} {\bibinfo  {journal} {JHEP}\ }\textbf {\bibinfo
  {volume} {11}},\ \bibinfo {pages} {070}}\BibitemShut {NoStop}%
\bibitem [{\citenamefont {Conde}\ and\ \citenamefont
  {Marzolla}(2016)}]{Conde:2016vxs}%
  \BibitemOpen
  \bibfield  {author} {\bibinfo {author} {\bibfnamefont {E.}~\bibnamefont
  {Conde}}\ and\ \bibinfo {author} {\bibfnamefont {A.}~\bibnamefont
  {Marzolla}},\ }\bibfield  {title} {\bibinfo {title} {{Lorentz Constraints on
  Massive Three-Point Amplitudes}},\ }\href
  {https://doi.org/10.1007/JHEP09(2016)041} {\bibfield  {journal} {\bibinfo
  {journal} {JHEP}\ }\textbf {\bibinfo {volume} {09}},\ \bibinfo {pages}
  {041}}\BibitemShut {NoStop}%
\bibitem [{\citenamefont {Dittmaier}(1998)}]{Dittmaier:1998nn}%
  \BibitemOpen
  \bibfield  {author} {\bibinfo {author} {\bibfnamefont {S.}~\bibnamefont
  {Dittmaier}},\ }\bibfield  {title} {\bibinfo {title} {{Weyl-van der Waerden
  formalism for helicity amplitudes of massive particles}},\ }\href
  {https://doi.org/10.1103/PhysRevD.59.016007} {\bibfield  {journal} {\bibinfo
  {journal} {Phys. Rev. D}\ }\textbf {\bibinfo {volume} {59}},\ \bibinfo
  {pages} {016007} (\bibinfo {year} {1998})}\BibitemShut {NoStop}%
\bibitem [{\citenamefont {Ochirov}(2018)}]{Ochirov:2018}%
  \BibitemOpen
  \bibfield  {author} {\bibinfo {author} {\bibfnamefont {A.}~\bibnamefont
  {Ochirov}},\ }\bibfield  {title} {\bibinfo {title} {{Helicity amplitudes for
  QCD with massive quarks}},\ }\href@noop {} {\bibfield  {journal} {\bibinfo
  {journal} {JHEP}\ }\textbf {\bibinfo {volume} {2018}}\bibinfo  {number} {
  (4)},\ \bibinfo {pages} {089}}\BibitemShut {NoStop}%
\bibitem [{\citenamefont {Alves}\ \emph {et~al.}(2022)\citenamefont {Alves},
  \citenamefont {Bertuzzo},\ and\ \citenamefont {Salla}}]{Alves:2022}%
  \BibitemOpen
\bibfield  {number} {  }\bibfield  {author} {\bibinfo {author} {\bibfnamefont
  {G.~F.~S.}\ \bibnamefont {Alves}}, \bibinfo {author} {\bibfnamefont
  {E.}~\bibnamefont {Bertuzzo}},\ and\ \bibinfo {author} {\bibfnamefont
  {G.~M.}\ \bibnamefont {Salla}},\ }\bibfield  {title} {\bibinfo {title}
  {{On-shell approach to neutrino oscillations}},\ }\href
  {https://doi.org/10.1103/PhysRevD.106.036028} {\bibfield  {journal} {\bibinfo
   {journal} {Phys. Rev. D}\ }\textbf {\bibinfo {volume} {106}},\ \bibinfo
  {pages} {036028} (\bibinfo {year} {2022})}\BibitemShut {NoStop}%
\bibitem [{\citenamefont {Guevara}\ \emph {et~al.}(2019)\citenamefont
  {Guevara}, \citenamefont {Ochirov},\ and\ \citenamefont
  {Vines}}]{Guevara:2019}%
  \BibitemOpen
  \bibfield  {author} {\bibinfo {author} {\bibfnamefont {A.}~\bibnamefont
  {Guevara}}, \bibinfo {author} {\bibfnamefont {A.}~\bibnamefont {Ochirov}},\
  and\ \bibinfo {author} {\bibfnamefont {J.}~\bibnamefont {Vines}},\ }\bibfield
   {title} {\bibinfo {title} {{Scattering of Spinning Black Holes from
  Exponentiated Soft Factors}},\ }\href
  {https://doi.org/10.1007/JHEP09(2019)056} {\bibfield  {journal} {\bibinfo
  {journal} {JHEP}\ }\textbf {\bibinfo {volume} {09}},\ \bibinfo {pages}
  {056}}\BibitemShut {NoStop}%
\bibitem [{\citenamefont {Herderschee}\ \emph {et~al.}(2019)\citenamefont
  {Herderschee}, \citenamefont {Koren},\ and\ \citenamefont
  {Trott}}]{Herderschee:2019}%
  \BibitemOpen
  \bibfield  {author} {\bibinfo {author} {\bibfnamefont {A.}~\bibnamefont
  {Herderschee}}, \bibinfo {author} {\bibfnamefont {S.}~\bibnamefont {Koren}},\
  and\ \bibinfo {author} {\bibfnamefont {T.}~\bibnamefont {Trott}},\ }\bibfield
   {title} {\bibinfo {title} {{Massive On-Shell Supersymmetric Scattering
  Amplitudes}},\ }\href {https://doi.org/10.1007/JHEP10(2019)092} {\bibfield
  {journal} {\bibinfo  {journal} {JHEP}\ }\textbf {\bibinfo {volume} {10}},\
  \bibinfo {pages} {092}}\BibitemShut {NoStop}%
\bibitem [{\citenamefont {Chiodaroli}\ \emph {et~al.}(2023)\citenamefont
  {Chiodaroli}, \citenamefont {Gunaydin}, \citenamefont {Johansson},\ and\
  \citenamefont {Roiban}}]{Chiodaroli:2023}%
  \BibitemOpen
  \bibfield  {author} {\bibinfo {author} {\bibfnamefont {M.}~\bibnamefont
  {Chiodaroli}}, \bibinfo {author} {\bibfnamefont {M.}~\bibnamefont
  {Gunaydin}}, \bibinfo {author} {\bibfnamefont {H.}~\bibnamefont
  {Johansson}},\ and\ \bibinfo {author} {\bibfnamefont {R.}~\bibnamefont
  {Roiban}},\ }\bibfield  {title} {\bibinfo {title} {{Spinor-helicity formalism
  for massive and massless amplitudes in five dimensions}},\ }\href
  {https://doi.org/10.1007/JHEP02(2023)040} {\bibfield  {journal} {\bibinfo
  {journal} {JHEP}\ }\textbf {\bibinfo {volume} {02}},\ \bibinfo {pages}
  {040}}\BibitemShut {NoStop}%
\bibitem [{\citenamefont {Kim}\ \emph {et~al.}(2021)\citenamefont {Kim},
  \citenamefont {Kim},\ and\ \citenamefont {Lee}}]{Kim:2021}%
  \BibitemOpen
  \bibfield  {author} {\bibinfo {author} {\bibfnamefont {J.-H.}\ \bibnamefont
  {Kim}}, \bibinfo {author} {\bibfnamefont {J.-W.}\ \bibnamefont {Kim}},\ and\
  \bibinfo {author} {\bibfnamefont {S.}~\bibnamefont {Lee}},\ }\bibfield
  {title} {\bibinfo {title} {{The relativistic spherical top as a massive
  twistor}},\ }\href {https://doi.org/10.1088/1751-8121/ac11be} {\bibfield
  {journal} {\bibinfo  {journal} {J. Phys. A}\ }\textbf {\bibinfo {volume}
  {54}},\ \bibinfo {pages} {335203} (\bibinfo {year} {2021})},\ \Eprint
  {https://arxiv.org/abs/2102.07063} {arXiv:2102.07063 [hep-th]} \BibitemShut
  {NoStop}%
\bibitem [{\citenamefont {Albonico}\ \emph {et~al.}(2024)\citenamefont
  {Albonico}, \citenamefont {Geyer},\ and\ \citenamefont
  {Mason}}]{Albonico:2024}%
  \BibitemOpen
  \bibfield  {author} {\bibinfo {author} {\bibfnamefont {G.}~\bibnamefont
  {Albonico}}, \bibinfo {author} {\bibfnamefont {Y.}~\bibnamefont {Geyer}},\
  and\ \bibinfo {author} {\bibfnamefont {L.}~\bibnamefont {Mason}},\ }\bibfield
   {title} {\bibinfo {title} {{Massive ambitwistor-strings; twistorial
  models}},\ }\href {https://doi.org/10.1007/JHEP01(2024)127} {\bibfield
  {journal} {\bibinfo  {journal} {JHEP}\ }\textbf {\bibinfo {volume} {01}},\
  \bibinfo {pages} {127}}\BibitemShut {NoStop}%
\bibitem [{\citenamefont {Kim}(2025)}]{Kim:2025}%
  \BibitemOpen
  \bibfield  {author} {\bibinfo {author} {\bibfnamefont {J.-H.}\ \bibnamefont
  {Kim}},\ }\bibfield  {title} {\bibinfo {title} {{Asymptotic spinspacetime}},\
  }\href {https://doi.org/10.1103/PhysRevD.111.105011} {\bibfield  {journal}
  {\bibinfo  {journal} {Phys. Rev. D}\ }\textbf {\bibinfo {volume} {111}},\
  \bibinfo {pages} {105011} (\bibinfo {year} {2025})},\ \Eprint
  {https://arxiv.org/abs/2309.11886} {arXiv:2309.11886 [hep-th]} \BibitemShut
  {NoStop}%
\bibitem [{\citenamefont {Elvang}\ and\ \citenamefont
  {Huang}(2015)}]{Elvang:2015}%
  \BibitemOpen
  \bibfield  {author} {\bibinfo {author} {\bibfnamefont {H.}~\bibnamefont
  {Elvang}}\ and\ \bibinfo {author} {\bibfnamefont {Y.-t.}\ \bibnamefont
  {Huang}},\ }\href@noop {} {\emph {\bibinfo {title} {{Scattering Amplitudes in
  Gauge Theory and Gravity}}}}\ (\bibinfo  {publisher} {Cambridge University
  Press},\ \bibinfo {address} {Cambridge, U.K.},\ \bibinfo {year}
  {2015})\BibitemShut {NoStop}%
\bibitem [{\citenamefont {Schwartz}(2013)}]{Schwartz:2013}%
  \BibitemOpen
  \bibfield  {author} {\bibinfo {author} {\bibfnamefont {M.~D.}\ \bibnamefont
  {Schwartz}},\ }\href@noop {} {\emph {\bibinfo {title} {{Quantum Field Theory
  and the Standard Model}}}}\ (\bibinfo  {publisher} {Cambridge University
  Press},\ \bibinfo {address} {Cambridge, U.K.},\ \bibinfo {year}
  {2013})\BibitemShut {NoStop}%
\bibitem [{\citenamefont {Dixon}(2014)}]{Dixon:2014}%
  \BibitemOpen
  \bibfield  {author} {\bibinfo {author} {\bibfnamefont {L.~J.}\ \bibnamefont
  {Dixon}},\ }\bibfield  {title} {\bibinfo {title} {{A brief introduction to
  modern amplitude methods}},\ }in\ \href@noop {} {\emph {\bibinfo {booktitle}
  {{Theoretical Advanced Study Institute in Elementary Particle Physics}:
  {Particle Physics: The Higgs Boson and Beyond}}}}\ (\bibinfo {year} {2014})\
  pp.\ \bibinfo {pages} {31--67}\BibitemShut {NoStop}%
\bibitem [{\citenamefont {Georgi}(1999)}]{Georgi:1999wka}%
  \BibitemOpen
  \bibfield  {author} {\bibinfo {author} {\bibfnamefont {H.}~\bibnamefont
  {Georgi}},\ }\href@noop {} {\emph {\bibinfo {title} {{Lie algebras in
  particle physics}}}},\ \bibinfo {edition} {2nd}\ ed.\ (\bibinfo  {publisher}
  {Perseus Books},\ \bibinfo {address} {Reading, MA},\ \bibinfo {year}
  {1999})\BibitemShut {NoStop}%
\bibitem [{\citenamefont {Penrose}(1967)}]{Penrose:1967}%
  \BibitemOpen
  \bibfield  {author} {\bibinfo {author} {\bibfnamefont {R.}~\bibnamefont
  {Penrose}},\ }\bibfield  {title} {\bibinfo {title} {{Twistor algebra}},\
  }\href {https://doi.org/10.1063/1.1705200} {\bibfield  {journal} {\bibinfo
  {journal} {J. Math. Phys.}\ }\textbf {\bibinfo {volume} {8}},\ \bibinfo
  {pages} {345} (\bibinfo {year} {1967})}\BibitemShut {NoStop}%
\bibitem [{\citenamefont {Penrose}\ and\ \citenamefont
  {Rindler}(1986)}]{Penrose:1986}%
  \BibitemOpen
  \bibfield  {author} {\bibinfo {author} {\bibfnamefont {R.}~\bibnamefont
  {Penrose}}\ and\ \bibinfo {author} {\bibfnamefont {W.}~\bibnamefont
  {Rindler}},\ }\href@noop {} {\emph {\bibinfo {title} {{Spinors and
  Space-Time: Spinor and Twistor Methods in Space-Time Geometry}}}},\
  Vol.~\bibinfo {volume} {2}\ (\bibinfo  {publisher} {Cambridge University
  Press},\ \bibinfo {address} {Cambridge, U.K.},\ \bibinfo {year}
  {1986})\BibitemShut {NoStop}%
\bibitem [{\citenamefont {Arkani-Hamed}\ \emph {et~al.}(2016)\citenamefont
  {Arkani-Hamed}, \citenamefont {Bourjaily}, \citenamefont {Cachazo},
  \citenamefont {Goncharov}, \citenamefont {Postnikov},\ and\ \citenamefont
  {Trnka}}]{Arkani-Hamed:2016}%
  \BibitemOpen
  \bibfield  {author} {\bibinfo {author} {\bibfnamefont {N.}~\bibnamefont
  {Arkani-Hamed}}, \bibinfo {author} {\bibfnamefont {J.~L.}\ \bibnamefont
  {Bourjaily}}, \bibinfo {author} {\bibfnamefont {F.}~\bibnamefont {Cachazo}},
  \bibinfo {author} {\bibfnamefont {A.~B.}\ \bibnamefont {Goncharov}}, \bibinfo
  {author} {\bibfnamefont {A.}~\bibnamefont {Postnikov}},\ and\ \bibinfo
  {author} {\bibfnamefont {J.}~\bibnamefont {Trnka}},\ }\href@noop {} {\emph
  {\bibinfo {title} {{Grassmannian Geometry of Scattering Amplitudes}}}}\
  (\bibinfo  {publisher} {Cambridge University Press},\ \bibinfo {year}
  {2016})\BibitemShut {NoStop}%
\bibitem [{\citenamefont {Britto}\ \emph {et~al.}(2005)\citenamefont {Britto},
  \citenamefont {Cachazo}, \citenamefont {Feng},\ and\ \citenamefont
  {Witten}}]{Britto:2005}%
  \BibitemOpen
  \bibfield  {author} {\bibinfo {author} {\bibfnamefont {R.}~\bibnamefont
  {Britto}}, \bibinfo {author} {\bibfnamefont {F.}~\bibnamefont {Cachazo}},
  \bibinfo {author} {\bibfnamefont {B.}~\bibnamefont {Feng}},\ and\ \bibinfo
  {author} {\bibfnamefont {E.}~\bibnamefont {Witten}},\ }\bibfield  {title}
  {\bibinfo {title} {{Direct Proof of the Tree-Level Scattering Amplitude
  Recursion Relation in Yang-Mills Theory}},\ }\href
  {https://doi.org/10.1103/physrevlett.94.181602} {\bibfield  {journal}
  {\bibinfo  {journal} {Phys. Rev. Lett.}\ }\textbf {\bibinfo {volume} {94}},\
  \bibinfo {pages} {181602} (\bibinfo {year} {2005})}\BibitemShut {NoStop}%
\bibitem [{\citenamefont {Wu}\ and\ \citenamefont {Zhu}(2022)}]{Wu:2022}%
  \BibitemOpen
  \bibfield  {author} {\bibinfo {author} {\bibfnamefont {C.}~\bibnamefont
  {Wu}}\ and\ \bibinfo {author} {\bibfnamefont {S.-H.}\ \bibnamefont {Zhu}},\
  }\bibfield  {title} {\bibinfo {title} {{Massive on-shell recursion relations
  for n-point amplitudes}},\ }\href {https://doi.org/10.1007/JHEP06(2022)117}
  {\bibfield  {journal} {\bibinfo  {journal} {JHEP}\ }\textbf {\bibinfo
  {volume} {06}},\ \bibinfo {pages} {117}},\ \Eprint
  {https://arxiv.org/abs/2112.12312} {arXiv:2112.12312 [hep-th]} \BibitemShut
  {NoStop}%
\bibitem [{\citenamefont {Christensen}\ \emph {et~al.}(2023)\citenamefont
  {Christensen}, \citenamefont {Diaz-Quiroz}, \citenamefont {Field},
  \citenamefont {Hayward},\ and\ \citenamefont {Miles}}]{Christensen:2022nja}%
  \BibitemOpen
  \bibfield  {author} {\bibinfo {author} {\bibfnamefont {N.}~\bibnamefont
  {Christensen}}, \bibinfo {author} {\bibfnamefont {H.}~\bibnamefont
  {Diaz-Quiroz}}, \bibinfo {author} {\bibfnamefont {B.}~\bibnamefont {Field}},
  \bibinfo {author} {\bibfnamefont {J.}~\bibnamefont {Hayward}},\ and\ \bibinfo
  {author} {\bibfnamefont {J.}~\bibnamefont {Miles}},\ }\bibfield  {title}
  {\bibinfo {title} {{Challenges with internal photons in constructive QED}},\
  }\href {https://doi.org/10.1016/j.nuclphysb.2023.116278} {\bibfield
  {journal} {\bibinfo  {journal} {Nucl. Phys. B}\ }\textbf {\bibinfo {volume}
  {993}},\ \bibinfo {pages} {116278} (\bibinfo {year} {2023})},\ \Eprint
  {https://arxiv.org/abs/2209.15018} {arXiv:2209.15018 [hep-ph]} \BibitemShut
  {NoStop}%
\bibitem [{\citenamefont {Cohen-Tannoudji}\ \emph
  {et~al.}(1991{\natexlab{a}})\citenamefont {Cohen-Tannoudji}, \citenamefont
  {Diu},\ and\ \citenamefont {Lalo\"e}}]{Cohen-Tannoudji:1991b}%
  \BibitemOpen
  \bibfield  {author} {\bibinfo {author} {\bibfnamefont {C.}~\bibnamefont
  {Cohen-Tannoudji}}, \bibinfo {author} {\bibfnamefont {B.}~\bibnamefont
  {Diu}},\ and\ \bibinfo {author} {\bibfnamefont {F.}~\bibnamefont {Lalo\"e}},\
  }\href@noop {} {\emph {\bibinfo {title} {{Quantum Mechanics}}}},\
  Vol.~\bibinfo {volume} {2}\ (\bibinfo  {publisher} {Wiley},\ \bibinfo
  {address} {Hoboken, NJ},\ \bibinfo {year} {1991})\ p.\ \bibinfo {pages}
  {626}\BibitemShut {NoStop}%
\bibitem [{\citenamefont {Weinberg}(2005)}]{Weinberg:1995}%
  \BibitemOpen
  \bibfield  {author} {\bibinfo {author} {\bibfnamefont {S.}~\bibnamefont
  {Weinberg}},\ }\href@noop {} {\emph {\bibinfo {title} {{The Quantum Theory of
  Fields. Vol. 1: Foundations}}}}\ (\bibinfo  {publisher} {Cambridge University
  Press},\ \bibinfo {address} {Cambridge, U.K.},\ \bibinfo {year}
  {2005})\BibitemShut {NoStop}%
\bibitem [{\citenamefont {Benincasa}\ and\ \citenamefont
  {Cachazo}(2008)}]{Benincasa:2008}%
  \BibitemOpen
  \bibfield  {author} {\bibinfo {author} {\bibfnamefont {P.}~\bibnamefont
  {Benincasa}}\ and\ \bibinfo {author} {\bibfnamefont {F.}~\bibnamefont
  {Cachazo}},\ }\bibfield  {title} {\bibinfo {title} {{Consistency Conditions
  on the S-Matrix of Massless Particles}},\ }\href@noop {} {\  (\bibinfo {year}
  {2008})},\ \Eprint {https://arxiv.org/abs/[arXiv:hep-th/0705.4305]}
  {arXiv:[arXiv:hep-th/0705.4305] [hep-th]} \BibitemShut {NoStop}%
\bibitem [{\citenamefont {Cohen-Tannoudji}\ \emph
  {et~al.}(1991{\natexlab{b}})\citenamefont {Cohen-Tannoudji}, \citenamefont
  {Diu},\ and\ \citenamefont {Lalo\"e}}]{Cohen-Tannoudji:1991a}%
  \BibitemOpen
  \bibfield  {author} {\bibinfo {author} {\bibfnamefont {C.}~\bibnamefont
  {Cohen-Tannoudji}}, \bibinfo {author} {\bibfnamefont {B.}~\bibnamefont
  {Diu}},\ and\ \bibinfo {author} {\bibfnamefont {F.}~\bibnamefont {Lalo\"e}},\
  }\href@noop {} {\emph {\bibinfo {title} {{Quantum Mechanics}}}},\ \bibinfo
  {edition} {1st}\ ed.,\ Vol.~\bibinfo {volume} {1}\ (\bibinfo  {publisher}
  {Wiley},\ \bibinfo {address} {Hoboken, NJ},\ \bibinfo {year} {1991})\ p.\
  \bibinfo {pages} {898}\BibitemShut {NoStop}%
\bibitem [{\citenamefont {Arkani-Hamed}\ \emph
  {et~al.}(2021{\natexlab{b}})\citenamefont {Arkani-Hamed}, \citenamefont
  {Huang},\ and\ \citenamefont {Huang}}]{Arkani-Hamed:2021b}%
  \BibitemOpen
  \bibfield  {author} {\bibinfo {author} {\bibfnamefont {N.}~\bibnamefont
  {Arkani-Hamed}}, \bibinfo {author} {\bibfnamefont {T.-C.}\ \bibnamefont
  {Huang}},\ and\ \bibinfo {author} {\bibfnamefont {Y.-t.}\ \bibnamefont
  {Huang}},\ }\bibfield  {title} {\bibinfo {title} {{The EFT-Hedron}},\ }\href
  {https://doi.org/10.1007/JHEP05(2021)259} {\bibfield  {journal} {\bibinfo
  {journal} {JHEP}\ }\textbf {\bibinfo {volume} {05}},\ \bibinfo {pages}
  {259}}\BibitemShut {NoStop}%
\bibitem [{\citenamefont {Coleman}\ \emph {et~al.}(1969)\citenamefont
  {Coleman}, \citenamefont {Wess},\ and\ \citenamefont
  {Zumino}}]{Coleman:1969}%
  \BibitemOpen
  \bibfield  {author} {\bibinfo {author} {\bibfnamefont {S.~R.}\ \bibnamefont
  {Coleman}}, \bibinfo {author} {\bibfnamefont {J.}~\bibnamefont {Wess}},\ and\
  \bibinfo {author} {\bibfnamefont {B.}~\bibnamefont {Zumino}},\ }\bibfield
  {title} {\bibinfo {title} {{Structure of phenomenological Lagrangians. 1.}},\
  }\href {https://doi.org/10.1103/PhysRev.177.2239} {\bibfield  {journal}
  {\bibinfo  {journal} {Phys. Rev.}\ }\textbf {\bibinfo {volume} {177}},\
  \bibinfo {pages} {2239} (\bibinfo {year} {1969})}\BibitemShut {NoStop}%
\bibitem [{\citenamefont {Callan}\ \emph {et~al.}(1969)\citenamefont {Callan},
  \citenamefont {Coleman}, \citenamefont {Wess},\ and\ \citenamefont
  {Zumino}}]{Callan:1969}%
  \BibitemOpen
  \bibfield  {author} {\bibinfo {author} {\bibfnamefont {C.~G.}\ \bibnamefont
  {Callan}, \bibfnamefont {Jr.}}, \bibinfo {author} {\bibfnamefont {S.~R.}\
  \bibnamefont {Coleman}}, \bibinfo {author} {\bibfnamefont {J.}~\bibnamefont
  {Wess}},\ and\ \bibinfo {author} {\bibfnamefont {B.}~\bibnamefont {Zumino}},\
  }\bibfield  {title} {\bibinfo {title} {{Structure of phenomenological
  Lagrangians. 2.}},\ }\href {https://doi.org/10.1103/PhysRev.177.2247}
  {\bibfield  {journal} {\bibinfo  {journal} {Phys. Rev.}\ }\textbf {\bibinfo
  {volume} {177}},\ \bibinfo {pages} {2247} (\bibinfo {year}
  {1969})}\BibitemShut {NoStop}%
\bibitem [{\citenamefont {Penrose}(2004)}]{Penrose:2004}%
  \BibitemOpen
  \bibfield  {author} {\bibinfo {author} {\bibfnamefont {R.}~\bibnamefont
  {Penrose}},\ }\href@noop {} {\emph {\bibinfo {title} {{The Road to Reality: A
  Complete Guide to the Laws of the Universe}}}}\ (\bibinfo  {publisher}
  {Jonathan Cape (Random House)},\ \bibinfo {address} {London, U.K.},\ \bibinfo
  {year} {2004})\BibitemShut {NoStop}%
\bibitem [{\citenamefont {Synge}(1965)}]{Synge:1965}%
  \BibitemOpen
  \bibfield  {author} {\bibinfo {author} {\bibfnamefont {J.~L.}\ \bibnamefont
  {Synge}},\ }\href@noop {} {\emph {\bibinfo {title} {Relativity: The Special
  Theory}}},\ \bibinfo {edition} {2nd}\ ed.\ (\bibinfo  {publisher}
  {North-Holland Publishing Co.},\ \bibinfo {year} {1965})\BibitemShut
  {NoStop}%
\bibitem [{\citenamefont {Penrose}\ and\ \citenamefont
  {MacCallum}(1972)}]{Penrose:1972}%
  \BibitemOpen
  \bibfield  {author} {\bibinfo {author} {\bibfnamefont {R.}~\bibnamefont
  {Penrose}}\ and\ \bibinfo {author} {\bibfnamefont {M.}~\bibnamefont
  {MacCallum}},\ }\bibfield  {title} {\bibinfo {title} {{Twistor theory: an
  approach to the quantisation of fields and space-time}},\ }\href@noop {}
  {\bibfield  {journal} {\bibinfo  {journal} {Phys. Rep.}\ }\textbf {\bibinfo
  {volume} {6}},\ \bibinfo {pages} {241} (\bibinfo {year} {1972})}\BibitemShut
  {NoStop}%
\bibitem [{\citenamefont {Christensen}\ \emph {et~al.}(2020)\citenamefont
  {Christensen}, \citenamefont {Field}, \citenamefont {Moore},\ and\
  \citenamefont {Pinto}}]{Christensen:2020}%
  \BibitemOpen
  \bibfield  {author} {\bibinfo {author} {\bibfnamefont {N.}~\bibnamefont
  {Christensen}}, \bibinfo {author} {\bibfnamefont {B.}~\bibnamefont {Field}},
  \bibinfo {author} {\bibfnamefont {A.}~\bibnamefont {Moore}},\ and\ \bibinfo
  {author} {\bibfnamefont {S.}~\bibnamefont {Pinto}},\ }\bibfield  {title}
  {\bibinfo {title} {{Two-, three-, and four-body decays in the constructive
  standard model}},\ }\href {https://doi.org/10.1103/PhysRevD.101.065019}
  {\bibfield  {journal} {\bibinfo  {journal} {Phys. Rev. D}\ }\textbf {\bibinfo
  {volume} {101}},\ \bibinfo {pages} {065019} (\bibinfo {year}
  {2020})}\BibitemShut {NoStop}%
\bibitem [{\citenamefont {Bachu}\ and\ \citenamefont
  {Yelleshpur}(2020)}]{Bachu:2020}%
  \BibitemOpen
  \bibfield  {author} {\bibinfo {author} {\bibfnamefont {B.}~\bibnamefont
  {Bachu}}\ and\ \bibinfo {author} {\bibfnamefont {A.}~\bibnamefont
  {Yelleshpur}},\ }\bibfield  {title} {\bibinfo {title} {{On-Shell Electroweak
  Sector and the Higgs Mechanism}},\ }\href
  {https://doi.org/10.1007/JHEP08(2020)039} {\bibfield  {journal} {\bibinfo
  {journal} {JHEP}\ }\textbf {\bibinfo {volume} {08}},\ \bibinfo {pages}
  {039}}\BibitemShut {NoStop}%
\bibitem [{\citenamefont {Durieux}\ \emph {et~al.}(2020)\citenamefont
  {Durieux}, \citenamefont {Kitahara}, \citenamefont {Shadmi},\ and\
  \citenamefont {Weiss}}]{Durieux:2020}%
  \BibitemOpen
  \bibfield  {author} {\bibinfo {author} {\bibfnamefont {G.}~\bibnamefont
  {Durieux}}, \bibinfo {author} {\bibfnamefont {T.}~\bibnamefont {Kitahara}},
  \bibinfo {author} {\bibfnamefont {Y.}~\bibnamefont {Shadmi}},\ and\ \bibinfo
  {author} {\bibfnamefont {Y.}~\bibnamefont {Weiss}},\ }\bibfield  {title}
  {\bibinfo {title} {{The electroweak effective field theory from on-shell
  amplitudes}},\ }\href {https://doi.org/10.1007/JHEP01(2020)119} {\bibfield
  {journal} {\bibinfo  {journal} {JHEP}\ }\textbf {\bibinfo {volume} {01}},\
  \bibinfo {pages} {119}}\BibitemShut {NoStop}%
\bibitem [{\citenamefont {Ballav}\ and\ \citenamefont
  {Manna}(2022)}]{Ballav:2022}%
  \BibitemOpen
  \bibfield  {author} {\bibinfo {author} {\bibfnamefont {S.}~\bibnamefont
  {Ballav}}\ and\ \bibinfo {author} {\bibfnamefont {A.}~\bibnamefont {Manna}},\
  }\bibfield  {title} {\bibinfo {title} {{Recursion relations for scattering
  amplitudes with massive particles II: Massive vector bosons}},\ }\href
  {https://doi.org/10.1016/j.nuclphysb.2022.115935} {\bibfield  {journal}
  {\bibinfo  {journal} {Nucl. Phys. B}\ }\textbf {\bibinfo {volume} {983}},\
  \bibinfo {pages} {115935} (\bibinfo {year} {2022})}\BibitemShut {NoStop}%
\bibitem [{\citenamefont {Balkin}\ \emph {et~al.}(2022)\citenamefont {Balkin},
  \citenamefont {Durieux}, \citenamefont {Kitahara}, \citenamefont {Shadmi},\
  and\ \citenamefont {Weiss}}]{Balkin:2022}%
  \BibitemOpen
  \bibfield  {author} {\bibinfo {author} {\bibfnamefont {R.}~\bibnamefont
  {Balkin}}, \bibinfo {author} {\bibfnamefont {G.}~\bibnamefont {Durieux}},
  \bibinfo {author} {\bibfnamefont {T.}~\bibnamefont {Kitahara}}, \bibinfo
  {author} {\bibfnamefont {Y.}~\bibnamefont {Shadmi}},\ and\ \bibinfo {author}
  {\bibfnamefont {Y.}~\bibnamefont {Weiss}},\ }\bibfield  {title} {\bibinfo
  {title} {{On-shell Higgsing for EFTs}},\ }\href
  {https://doi.org/10.1007/JHEP03(2022)129} {\bibfield  {journal} {\bibinfo
  {journal} {JHEP}\ }\textbf {\bibinfo {volume} {03}},\ \bibinfo {pages}
  {129}}\BibitemShut {NoStop}%
\bibitem [{\citenamefont {Bachu}(2024)}]{Bachu:2024}%
  \BibitemOpen
  \bibfield  {author} {\bibinfo {author} {\bibfnamefont {B.}~\bibnamefont
  {Bachu}},\ }\bibfield  {title} {\bibinfo {title} {{Spontaneous symmetry
  breaking from an on-shell perspective}},\ }\href
  {https://doi.org/10.1007/JHEP02(2024)098} {\bibfield  {journal} {\bibinfo
  {journal} {JHEP}\ }\textbf {\bibinfo {volume} {02}},\ \bibinfo {pages}
  {098}}\BibitemShut {NoStop}%
\bibitem [{\citenamefont {Liu}\ and\ \citenamefont {Yin}(2022)}]{Liu:2022}%
  \BibitemOpen
  \bibfield  {author} {\bibinfo {author} {\bibfnamefont {D.}~\bibnamefont
  {Liu}}\ and\ \bibinfo {author} {\bibfnamefont {Z.}~\bibnamefont {Yin}},\
  }\bibfield  {title} {\bibinfo {title} {{Gauge invariance from on-shell
  massive amplitudes and tree-level unitarity}},\ }\href
  {https://doi.org/10.1103/PhysRevD.106.076003} {\bibfield  {journal} {\bibinfo
   {journal} {Phys. Rev. D}\ }\textbf {\bibinfo {volume} {106}},\ \bibinfo
  {pages} {076003} (\bibinfo {year} {2022})}\BibitemShut {NoStop}%
\bibitem [{\citenamefont {Arkani-Hamed}\ and\ \citenamefont
  {Trnka}(2014)}]{Arkani-Hamed:2014}%
  \BibitemOpen
  \bibfield  {author} {\bibinfo {author} {\bibfnamefont {N.}~\bibnamefont
  {Arkani-Hamed}}\ and\ \bibinfo {author} {\bibfnamefont {J.}~\bibnamefont
  {Trnka}},\ }\bibfield  {title} {\bibinfo {title} {{The Amplituhedron}},\
  }\href {https://doi.org/10.1007/JHEP10(2014)030} {\bibfield  {journal}
  {\bibinfo  {journal} {JHEP}\ }\textbf {\bibinfo {volume} {10}},\ \bibinfo
  {pages} {030}},\ \Eprint {https://arxiv.org/abs/1312.2007} {arXiv:1312.2007
  [hep-th]} \BibitemShut {NoStop}%
\end{thebibliography}%

\end{document}